\documentclass[prd,onecolumn,eqsecnum]{revtex4} 
\usepackage{amssymb}
\usepackage{graphicx}
\renewcommand{\i}{{\rm i}} 
\newcommand{\e}{{\rm e}} 
\renewcommand{\r}{{\sf r}} 
\newcommand{\f}{{\sf f}} 
\DeclareFontFamily{OT1}{rsfs}{} 
\DeclareFontShape{OT1}{rsfs}{m}{n}{<-7> rsfs5 
    <7-10> rsfs7 <10-> rsfs10}{}   
\DeclareMathAlphabet{\scr}{OT1}{rsfs}{m}{n}
\begin{document}
\title{Mode-sum regularization of the scalar self-force: Formulation
  in terms of a tetrad decomposition of the singular field}  
\author{Roland Haas and Eric Poisson}
\affiliation{Department of Physics, University of Guelph, Guelph,
Ontario, Canada N1G 2W1}
\date{July 24, 2006}  
\begin{abstract}
We examine the motion in Schwarzschild spacetime of a point particle
endowed with a scalar charge. The particle produces a retarded scalar
field which interacts with the particle and influences its motion via
the action of a self-force. We assume that the magnitude of the scalar
charge is small, and that the deviations from geodesic motion produced
by the self-force are small. This problem is analogous to that of an
electric charge moving under the action of its electromagnetic
self-force, and to that of a small mass moving under the action of its
gravitational self-force. We exploit the spherical symmetry of the
Schwarzschild spacetime and decompose the scalar field in
spherical-harmonic modes. Although each mode is bounded at the
position of the particle, a mode-sum evaluation of the self-force
requires regularization because the sum does not converge: the
retarded field is infinite at the position of the particle. The 
regularization procedure involves the computation of 
{\it regularization parameters}, which are obtained from a mode
decomposition of the Detweiler-Whiting singular field; these are
subtracted from the modes of the retarded field, and the result is a 
mode-sum that converges to the actual self-force. We present such
a computation in this paper. While regularization parameters have been
presented before in the literature, there are two main aspects of our
work that are new. First, we define the regularization parameters as
{\it scalar quantities} by referring them to a {\it tetrad
decomposition} of the singular field. This is different from standard
practice, which is to define regularization parameters as vectorial
quantities. The advantage of dealing with tetrad components is that
these, unlike vector components, are naturally decomposed in scalar 
spherical harmonics. Second, we calculate, for any bound orbit around
a Schwarzschild black hole, four sets of regularization parameters
(denoted schematically by $A$, $B$, $C$, and $D$) instead of the usual
three ($A$, $B$, and $C$). While only the first three regularization
parameters are needed to produce a convergent mode-sum, the inclusion
of a fourth parameter has the practically important consequence of
accelerating the convergence. The focus of this paper is entirely on
the computation of regularization parameters for the scalar
self-force. The techniques that we introduce in this work are not,
however, restricted to this context. They will readily be exported to
the electromagnetic and gravitational cases, but we leave this
generalization for future work. As proof of principle that 
our methods are reliable, we calculate the self-force acting on a
scalar charge in circular motion around a Schwarzschild black hole,
and compare our answers with those recorded in the literature. We
leave for future work the generalization of this calculation to
generic orbits.   
\end{abstract}
\maketitle

\section{Introduction and summary}  

\subsection{Motivation} 

The capture of solar-mass compact objects by massive black holes 
residing in galactic centers has been identified as one of the most
promising sources of gravitational waves for the Laser Interferometer
Space Antenna \cite{LISA}. The need for accurate templates for signal
detection and source identification is currently motivating an
intense effort from many workers to determine the motion of a
relativistic two-body system with a small mass ratio. This is done in
a treatment that goes beyond the test-mass description in which the
small mass moves on a geodesic in the spacetime of the large black
hole. An additional complication arises from the fact that the
treatment cannot rely on a slow-motion or weak-field
approximation. The accelerated motion of the small mass in the
background spacetime of the large black hole is governed by the body's
{\it gravitational self-force} \cite{mino-etal:97, quinn-wald:97},
which encodes the influence of the body's own gravitational field on
its motion. To compute the gravitational self-force acting on a body
moving on a bound orbit around a Schwarzschild or Kerr black hole is
currently the focus of much work; for reviews, see
Refs.~\cite{poisson:04b, poisson:04c} and a special issue of 
{\it Classical and Quantum Gravity} \cite{lousto:05} devoted to this
topic.   

The complexities associated with the computation and interpretation of
the gravitational self-force have motivated the formulation of
educational toy problems. These have the advantage of being much
simpler to deal with, but they should nevertheless capture the
essential physics of self-forced motions in curved spacetime. One such
toy problem, which is based on real-world physics that is interesting 
in its own right, is the motion of an electrically charged particle in
curved spacetime, subjected to a self-force produced by its own 
electromagnetic field. The foundations for this problem were laid more
than 45 years ago by DeWitt and Brehme \cite{dewitt-brehme:60} (their
work was later corrected by Hobbs \cite{hobbs:68}).  

Another such problem is the motion of a particle endowed with a 
{\it scalar charge}; here the foundations were provided more recently
by Quinn \cite{quinn:00}. In spite of its academic nature (there are
no known macroscopic particles endowed with a scalar charge), the
scalar self-force problem is a useful toy problem because of its
relative technical simplicity, and because the motion of a particle
under the influence of its scalar self-force is expected to resemble
closely the motions obtained in the other contexts (electromagnetic
and gravitational). The self-forced motion of a scalar charge in 
Schwarzschild spacetime is the focus of the work presented in this 
paper. We exploit this simple situation to introduce computational 
techniques that are required for the concrete evaluation of the
self-force. These techniques, however, are not limited to the context 
of the scalar self-force, and they will readily be exported to the 
electromagnetic and gravitational cases. We leave this
generalization for future work.      

\subsection{The problem} 

Consider a particle of mass $m$ and scalar charge $q$ moving on a
world line $\gamma$ of the Schwarzschild spacetime. The world line is
described by the parametric relations $z^\alpha(\tau)$ in which $\tau$
is proper time. The particle produces a retarded scalar potential
$\Phi$ that satisfies the inhomogeneous wave equation \cite{quinn:00}  
\begin{equation} 
\Box \Phi(x) = -4\pi q \int_\gamma \delta_4(x,z)\, d\tau, 
\label{1.1}
\end{equation}
in which $\Box := g^{\alpha\beta} \nabla_\alpha \nabla_\beta$ is the 
covariant wave operator, and $\delta_4(x,z)$ is a scalarized Dirac
distribution with support on the world line. Except for the factor
$-4\pi$ inserted for convenience, the right-hand side of
Eq.~(\ref{1.1}) is the particle's scalar charge density.  

The retarded potential $\Phi(x)$ produces a field $\Phi_\alpha(x) :=
\nabla_\alpha \Phi(x)$ that acts on the particle and influences its 
motion. As shown by Quinn \cite{quinn:00}, the particle's acceleration
is proportional to the components of $\Phi_\alpha$ that are orthogonal 
to the particle's velocity vector $u^\alpha := dz^\alpha/d\tau$, 
\begin{equation} 
m a^\alpha = q \bigl( g^{\alpha\beta} + u^\alpha u^\beta \bigr) 
\Phi_{\beta}(z),  
\label{1.2}
\end{equation} 
where $a^\alpha := Du^\alpha/d\tau$ is the particle's acceleration
vector, the covariant derivative of the velocity vector along the
world line. Quinn also showed that the longitudinal component of the
retarded field is responsible for a change in the particle's inertial
mass,  
\begin{equation} 
\frac{d m}{d\tau} = -q u^\alpha \Phi_{\alpha}(z). 
\label{1.3}
\end{equation}  
This effect was explored in cosmological situations in
Refs.~\cite{burko-etal:02, haas-poisson:05}. 

Equations (\ref{1.2}) and (\ref{1.3}) have only formal validity
because the singular field $\Phi_\alpha(x)$ diverges as $x \to z$: the
field of a point charge is necessarily infinite at the position of the
particle. Quinn \cite{quinn:00} was able to regularize these equations
so as to produce meaningful equations of motion for the charged
particle. In breakthrough work that plays a central role in this
paper, Detweiler and Whiting \cite{detweiler-whiting:03} showed that
Quinn's regularization procedure amounts to a decomposition of the
retarded potential into uniquely defined singular and regular
potentials,  
\begin{equation} 
\Phi(x) = \Phi^{\rm S}(x) + \Phi^{\rm R}(x),  
\label{1.4}
\end{equation}
with $\Phi^{\rm S}$ denoting the singular potential and 
$\Phi^{\rm R}$ the regular potential. As was shown by Detweiler and
Whiting, the singular potential possesses the following properties:
(i) it satisfies the same wave equation as the retarded potential,
that is, it is a solution to Eq.~(\ref{1.1}); (ii) it displays the
same singularity structure as the retarded potential near the
particle's world line; and (iii) it does not exert a force on the
point charge. These properties make the decomposition of
Eq.~(\ref{1.4}) unique. The regular potential, on the other hand,
possesses the following properties: (i) it satisfies a homogeneous
version of Eq.~(\ref{1.1}), with a zero right-hand side; (ii) it is
smooth on and near the particle's world line; and (iii) it alone
determines the self-force acting on the particle. The conclusion,
therefore, is that the actual equations of motion for the particle are
Eqs.~(\ref{1.2}) and (\ref{1.3}) with the regular field 
$\Phi^{\rm R}_\alpha := \nabla_\alpha \Phi^{\rm R}$ substituted in
place of the retarded field $\Phi_\alpha$; this conclusion is in full
agreement with Quinn's earlier work. The Detweiler-Whiting
decomposition also plays an essential role in the electromagnetic and
gravitational self-force problems.   

In Schwarzschild spacetime it is computationally advantageous to solve
Eq.~(\ref{1.1}) after decomposing $\Phi(x)$ in spherical harmonics
$Y_{lm}(\theta,\phi)$. Adopting the usual Schwarzschild coordinates 
$[t,r,\theta,\phi]$, we would write 
\begin{equation} 
\Phi(x) = \sum_{lm} \Phi_{lm}(t,r) Y_{lm}(\theta,\phi) 
\label{1.5}
\end{equation} 
and call $\Phi_{lm}(t,r)$ the spherical-harmonic modes of the retarded
potential; the right-hand side involves a sum over all integers $l =
0, 1, 2, \cdots, \infty$ and a nested sum over all integers $m = -l,
-l+1, \cdots, l-1, l$. After performing the decomposition of
Eq.~(\ref{1.5}), Eq.~(\ref{1.1}) turns into a two-dimensional
wave equation for each mode function $\Phi_{lm}(t,r)$; this equation
is displayed in Eq.~(\ref{1.12}), below. 

The spherical-harmonic modes of the retarded potential give rise to
quantities $\Phi_{\alpha\, l}(x) := \nabla_\alpha \sum_m \Phi_{lm}
Y_{lm}$, such that the retarded field can be expressed as the mode-sum  
\begin{equation} 
\Phi_\alpha(x) = \sum_l \Phi_{\alpha\, l}(x). 
\label{1.6}
\end{equation} 
We call these quantities the {\it multipole coefficients} of the
retarded field. Each $\Phi_{\alpha\, l}(x)$ is bounded in the limit $x
\to z$, in spite of the fact that the retarded field is infinite on
the particle's world line. (The multipole coefficients are
discontinuous at $x=z$.) The mode-sum of Eq.~(\ref{1.6}), of course,
does not converge when the retarded field is evaluated on the world
line.  

The failure of Eq.~(\ref{1.6}) to converge when $x=z$ is exactly
compensated for by the failure of 
\begin{equation} 
\Phi^{\rm S}_\alpha(x) = \sum_l \Phi^{\rm S}_{\alpha\, l}(x)
\label{1.7}
\end{equation} 
to converge, because (as was noted previously) the retarded and
singular fields share the same singularity structure near the world
line. Here, $\Phi^{\rm S}_{\alpha\, l} := \nabla_\alpha \sum_m
\Phi^{\rm S}_{lm} Y_{lm}$ with $\Phi^{\rm S}_{lm}$ denoting the
spherical-harmonic modes of the singular potential. The regular field
can thus be expressed as  
\begin{equation} 
\Phi^{\rm R}_\alpha(z) = \lim_{x \to z} \sum_l 
\Bigl[ \Phi_{\alpha\, l}(x) - \Phi^{\rm S}_{\alpha\, l}(x) \Bigr],   
\label{1.8}
\end{equation} 
in terms of a {\it converging mode-sum}. The limiting procedure
involved in Eq.~(\ref{1.8}) is introduced to handle the (shared) 
discontinuity of the multipole coefficients $\Phi_{\alpha\, l}(x)$ and   
$\Phi^{\rm S}_{\alpha\, l}(x)$ at $x = z$. After subtraction the
multipole coefficients of the regular field are smooth at $x=z$, the
mode-sum converges, and the limit can be taken without difficulty.   

The prescription contained in Eq.~(\ref{1.8}) becomes a practical
method to evaluate $\Phi^{\rm R}_\alpha(z)$ --- and therefore the 
self-force acting on the scalar charge --- when one can actually
compute the quantities $\Phi^{\rm S}_{\alpha\, l}(x)$ for a field
point $x$ close to the world line. As we shall review below, this
computation is possible because the singular field 
$\Phi^{\rm S}_\alpha(x)$ is known in a neighborhood of the world line;
it can be expressed as a Laurent expansion in powers of the distance
to the world line. The end result for the multipole coefficients takes
the schematic form  
\begin{equation} 
\Phi^{\rm S}_{\alpha\, l}(x) = q \biggl[ 
(l + {\textstyle \frac{1}{2}}) A_{\alpha} 
+ B_{\alpha} + \frac{C_{\alpha}}{(l + {\textstyle \frac{1}{2}})}  
+ \cdots \biggr], 
\label{1.9}
\end{equation}
in which the quantities $A_\alpha$, $B_\alpha$, and $C_\alpha$, known
as {\it regularization parameters}, are independent of $l$ but depend
on the state of motion of the particle at $z$. Notice that the sum
over the $A_\alpha$ term would diverge quadratically, that the sum
over the $B_\alpha$ term would diverge linearly, and that the sum over
the $C_\alpha$ term would diverge logarithmically. The remaining
terms, those designated with $(\cdots)$, lead to {\it a converging sum
that evaluates to zero} by virtue of the Detweiler-Whiting axiom, 
according to which the singular field does not produce a force on the
particle. This ensures that after removal of the $A_\alpha$,
$B_\alpha$, and $C_\alpha$ terms, the mode-sum of Eq.~(\ref{1.8}) will 
converge to the correct value for $\Phi^{\rm R}_\alpha(z)$.  

Our central task in this paper is the computation of regularization
parameters for a scalar charge moving on a bound orbit around a
Schwarzschild black hole. We leave for future work the completion of a
calculation of the actual self-force. This would involve, over and
above the work presented here, a numerical determination of the
retarded field produced by a scalar charge moving in Schwarzschild
spacetime.   

We consider the particle's charge $q$ to be small, and we have in mind 
a perturbative implementation of Eqs.~(\ref{1.2}) and (\ref{1.3}). In 
a zeroth-order approximation, the particle is taken to have a constant
inertial mass and to move on a geodesic of the Schwarzschild
spacetime. In a first-order approximation, the regular field 
$\Phi^{\rm R}_\alpha$ is computed for this geodesic motion and
substituted on the right-hand sides of Eqs.~(\ref{1.2}) and
(\ref{1.3}). This iterative process could be continued, but we suppose 
that $q$ is sufficiently small that the process can be stopped after
a single iteration. In this regime {\it the self-force can be computed 
while assuming that the motion is geodesic}. We adopt this
small-charge approximation here, noting that it reflects the spirit 
of the small-mass-ratio approximation in the gravitational
problem. The assumption leads to much simplification; the computation
of regularization parameters for accelerated particles would be
significantly more involved.    

\subsection{Past work} 

We are not the first researchers to define and compute regularization
parameters for the mode-sum computation of self-forces in curved
spacetime. In fact, the literature is vast and the field has already 
reached a fairly mature state. But as we argue below, we believe
that this work (and our promise for extensions toward the
electromagnetic and gravitational problems) is a significant addition.     

The main ideas behind the mode-sum regularization of self-forces were 
first introduced by Barack and Ori in their pioneering work
\cite{barack-ori:00, barack:00, barack:01}, which was later perfected  
and extended by an Israeli-Japanese consortium including Barack, Ori, 
Mino, Nakano, and Sasaki \cite{barack-etal:02, barack-ori:02,
barack-ori:03a, mino-etal:02}. These authors computed regularization 
parameters for the mode-sum evaluation of the (scalar,
electromagnetic, and gravitational) self-force acting on a particle
moving on any bound geodesic of the Schwarzschild spacetime. Barack
and Ori \cite{barack-ori:03b} were then able to extend these results
to the Kerr spacetime. This early work on regularization parameters is
nicely summarized in Ref.~\cite{hikida-etal:05}.  

The (so-called early) work reviewed in the preceding paragraph was  
carried out before the discovery by Detweiler and Whiting of the
retarded field's decomposition into uniquely identified singular and
regular pieces \cite{detweiler-whiting:03}. It relied on an
alternative decomposition, in terms of ``direct'' and ``tail'' pieces,
which did not come with the same degree of mathematical elegance. For 
example, unlike the regular field which is smooth everywhere in a
neighborhood of the world line, the tail part of the retarded field is
not differentiable on the world line. The Detweiler-Whiting
decomposition provides a much sounder foundation for the definition
and computation of regularization parameters. In follow-up works,
Detweiler, Messaritaki, and Whiting \cite{detweiler-etal:03} carried
out such a computation for a scalar charge moving on a circular orbit
around a Schwarzschild black hole, and Kim \cite{kim:04} extended
these results to a generic orbit.   

A number of works present explicit computations of the self-force 
(for various charges undergoing various motions in various spacetimes) 
by mode-sum techniques. Burko computed the self-force acting on an
electric charge in circular motion in Minkowski spacetime
\cite{burko:00a}. He and his coworkers also considered scalar and
electric charges kept stationary in a Schwarzschild spacetime
\cite{burko:00b}, in a spacetime that contains a spherical matter
shell (Burko, Liu, and Soen \cite{burko-etal:01}), and in a Kerr
spacetime (Burko and Liu \cite{burko-liu:01}). In addition, Burko
computed the scalar self-force acting on a particle in circular motion
around a Schwarzschild black hole \cite{burko:00c}. This calculation 
was since revisited by Detweiler, Messaritaki, and Whiting
\cite{detweiler-etal:03}, as well as Diaz-Rivera, Messaritaki, and
Whiting \cite{diaz-rivera-etal:04}, who also considered the case of
slightly eccentric motion. Barack and Burko dealt with a particle
falling radially into a Schwarzschild black hole, and evaluated the
scalar self-force acting on this particle \cite{barack-burko:00};  
Lousto \cite{lousto:00}, and Barack and Lousto
\cite{barack-lousto:02}, computed the gravitational self-force for
radial infall.    
   
\subsection{This work} 

In this paper we present a new calculation of regularization
parameters for the mode-sum evaluation of the scalar self-force
acting on a particle moving on a bound geodesic of the Schwarzschild
spacetime. Our calculation is new in two main respects.  

First, we define the regularization parameters as 
{\it scalar quantities} by referring them to a {\it tetrad
decomposition} of the singular field. In contrast, the original
parameters of Eq.~(\ref{1.9}) are vectorial quantities that refer to 
$\Phi^{\rm S}_\alpha$, the vectorial components of the singular
field. The idea here is to introduce a basis of orthonormal vectors
$e^\alpha_{(\mu)}(x)$ at every point of the Schwarzschild spacetime; 
the superscript $\alpha$ is the usual vectorial index, and the
subscript $(\mu) = \{(0),(1),(2),(3)\}$ is a label that designates an
individual member of the tetrad. The vectors satisfy
$g_{\alpha\beta} e^\alpha_{(\mu)} e^\beta_{(\nu)} = \eta_{(\mu)(\nu)}
= \mbox{diag}[-1,1,1,1]$. The four quantities $\Phi^{\rm S}_{(\mu)} 
:= e^\alpha_{(\mu)} \Phi^{\rm S}_\alpha$ are the {\it frame
components} of the singular field, and these are {\it scalar
functions} of the spacetime coordinates. 

The advantage of introducing the tetrad and the associated
decomposition of vector fields is that each frame component 
$\Phi^{\rm S}_{(\mu)}(x)$ is a scalar function that can naturally be   
expanded in scalar spherical harmonics. This is quite unlike the
vector $\Phi^{\rm S}_{\alpha}(x)$, which could be expanded elegantly
in vectorial harmonics (a procedure that has not been adopted in the 
self-force literature) or inelegantly in scalar harmonics (as was 
done in all previous works on regularization parameters). By
introducing the tetrad we are able, in this work, to provide an
elegant definition for the regularization parameters. Our specific 
choice of tetrad will be specified below.  

Second, we calculate an additional set of regularization parameters in
order to accelerate the numerical convergence of the
mode-sum. Together with our implementation of the tetrad
decomposition, this amounts to replacing Eq.~(\ref{1.8}) by 
\begin{equation} 
\Phi^{\rm R}_{(\mu)}(z) = \lim_{x \to z} \sum_l 
\Bigl[ \Phi_{(\mu)l}(x) - \Phi^{\rm S}_{(\mu)l}(x) \Bigr],   
\label{1.10}
\end{equation} 
and Eq.~(\ref{1.9}) by 
\begin{equation} 
\Phi^{\rm S}_{(\mu)l} = q \biggl[ 
(l + {\textstyle \frac{1}{2}}) A_{(\mu)} 
+ B_{(\mu)} + \frac{C_{(\mu)}}{(l + {\textstyle \frac{1}{2}})} 
+ \frac{D_{(\mu)}}{(l - {\textstyle \frac{1}{2}}) 
(l + {\textstyle \frac{3}{2}})} + \cdots \biggr].  
\label{1.11}
\end{equation}
The regularization parameters $A_{(\mu)}$, $B_{(\mu)}$, and
$C_{(\mu)}$ have already appeared (in vectorial form) in
Eq.~(\ref{1.9}); the regularization parameters $D_{(\mu)}$ are new. We
calculate all of these for a scalar charge moving on a bound geodesic
of the Schwarzschild spacetime. 

We note that the vectorial parameters $D_\alpha$ were computed by
Detweiler, Messaritaki, and Whiting for the special case of circular
motion in Schwarzschild spacetime \cite{detweiler-etal:03}, and that
they were computed by Kim for generic orbits; Kim's results are 
recorded in his PhD dissertation \cite{kim:phd}, but they have yet to     
appear in the peer-reviewed literature. We note also that because the
vectors $e^\alpha_{(\mu)}$ contain an angular dependence, the
operations of [multiplication by a tetrad vector] and [extraction of 
multipole coefficients] do not commute:  
$(e^\alpha_{(\mu)} \Phi_\alpha)_l \neq e^\alpha_{(\mu)}
(\Phi_\alpha)_l$; as a consequence, our expressions for $A_{(\mu)}$, 
$B_{(\mu)}$, $C_{(\mu)}$, and $D_{(\mu)}$ cannot be compared directly 
to those for $A_\alpha$, $B_\alpha$, $C_\alpha$, and $D_\alpha$ that
have appeared in the literature.                 

It is important to mention that the sum over $l$ of the $D_{(\mu)}$
term in Eq.~(\ref{1.11}) is actually zero, because $\sum_{l=0}^\infty
[(l - \frac{1}{2})(l + \frac{3}{2})]^{-1} = 0$. And as we have seen,
the same statement applies to the remaining terms in Eq.~(\ref{1.11}), 
those designated by $(\cdots)$. These terms, therefore, do not
contribute to the final value of $\Phi^{\rm R}_{(\mu)}(z)$ when 
the sum is evaluated in full, from $l=0$ to $l=\infty$. Nevertheless, 
the $D_{(\mu)}$ term does play a useful role when the sum is truncated
to some finite upper bound $l = l_{\rm max}$: It produces a
significant acceleration of the sum's convergence. This property has  
practical importance, because to truncate the sum over $l$ is a 
computational necessity.   

We compute the regularization parameters $A_{(\mu)}$, $B_{(\mu)}$,
$C_{(\mu)}$, and $D_{(\mu)}$ by importing many techniques from the
literature. In fact, an advantage of arriving late into this 
business is that we can pick and choose, from a variety of sources,
the computational methods that are the most compelling. Thus, for
example, in Sec.~II we develop a covariant local expansion for  
$\Phi^{\rm S}_\alpha(x)$ that is inspired by Mino, Nakano, and Sasaki  
\cite{mino-etal:02}, but which rests on the more secure 
Detweiler-Whiting decomposition of the retarded field into singular
and regular pieces \cite{detweiler-whiting:03}. As another example, in
Sec.~V and the Appendix we employ the rotated angular coordinates
$(\alpha,\beta)$ of Barack and Ori \cite{barack-ori:02}, but we do so
without having to deal with vector components that are singular at
the point of evaluation $(\alpha = 0)$. As a third example, in Sec.~V
we routinely make substitutions such as $\alpha \to 
\sqrt{2-2\cos\alpha} + O(\alpha^3)$ to turn a function of the angles
that is well-defined only in a neighborhood of $\alpha = 0$ to another
function that is well-defined on the entire sphere. We stole this 
powerful idea from Detweiler, Messaritaki, and Whiting
\cite{detweiler-etal:03}, but we implement it without making contact   
with the Thorne-Hartle-Zhang coordinates \cite{thorne-hartle:85,
zhang:86}; these form an important part of their analysis, but they
play no role here. 

\subsection{Prescription} 

We summarize our main results in the form of a detailed prescription
for the concrete evaluation of the scalar self-force. We recall that
the force is acting on a particle moving on a generic orbit around a
Schwarzschild black hole, and that it is evaluated within the
small-charge approximation described near the end of Sec.~I B.    

{\bf First step: Integrate the wave equation.} The first task that
must be accomplished is to solve Eq.~(\ref{1.1}) for the retarded
potential $\Phi(x)$. This is best accomplished by decomposing the
potential in spherical harmonics, as in Eq.~(\ref{1.5}). Each mode 
$\Phi_{lm}(t,r)$ of the retarded potential satisfies the reduced wave  
equation  
\begin{equation} 
\biggl\{ -\frac{\partial^2}{\partial t^2} 
+ \frac{\partial^2}{\partial r_*^2} 
- f \biggl[ \frac{l(l+1)}{r^2} + \frac{2M}{r^3} \biggr] 
\biggr\} \bigl( r \Phi_{lm} \bigr) = -\frac{4\pi q f}{r u^t} 
Y_{lm}({\textstyle \frac{\pi}{2}},0) \e^{-\i m \varphi(t)}\, 
\delta\bigl( r - \r(t) \bigr), 
\label{1.12}
\end{equation} 
in which the right-hand side is the reduction of the scalar charge
density $-4\pi q \int \delta_4(x,z)\, d\tau$. This equation must be
solved while imposing ingoing-wave boundary conditions at the
black-hole event horizon, and outgoing-wave boundary conditions at
infinity. The modes are complex functions, related to each other by
$\Phi_{l,-m} = (-1)^m \bar{\Phi}_{lm}$, in which an overbar indicates
complex conjugation; this condition ensures that the scalar potential
of Eq.~(\ref{1.5}) is real. 

In Eq.~(\ref{1.12}), the reduced wave operator is written in terms of
the tortoise coordinate $r^* = \int f^{-1}\, dr 
= r + 2M \ln(r/2M - 1)$, where $M$ is the black-hole mass and 
\begin{equation} 
f := 1 - \frac{2M}{r}.  
\label{1.13} 
\end{equation} 
The reduced charge density depends on the functions $\r(t)$ and
$\varphi(t)$, which give the coordinate positions of the world
line. These are obtained by solving the geodesic equations, 
\begin{eqnarray} 
\dot{t} &=& \frac{E}{1 - 2M/\r},  
\label{1.14} \\ 
\dot{\r}^2 &=& E^2 - \biggl( 1 - \frac{2M}{\r} \biggr) 
\biggl(1 + \frac{L^2}{\r^2} \biggr),  
\label{1.15} \\ 
\dot{\varphi} &=& \frac{L}{\r^2},  
\label{1.16}
\end{eqnarray} 
in which an overdot indicates differentiation with respect to proper
time $\tau$. The constant $E$ is the particle's conserved energy per
unit rest-mass, and the constant $L$ is the conserved angular momentum
per unit rest-mass. It is assumed that the particle is moving with
$\theta = \frac{\pi}{2}$ in the black hole's equatorial plane, and the 
right-hand side of Eq.~(\ref{1.12}) also involves $u^t = \dot{t} =
E/\f$, the time component of the particle's velocity vector.  

Equations (\ref{1.12}) and (\ref{1.14})--(\ref{1.16}) can be
integrated with any reliable numerical method, and this procedure
returns $\Phi_{lm}(t,r)$ for selected values of $l$ and $m$. From this
we extract the quantities 
\begin{equation} 
\Phi_{lm}(t,\r^+), \quad 
\frac{\partial}{\partial t} \Phi_{lm}(t,\r^+), \qquad 
\frac{\partial}{\partial r} \Phi_{lm}(t,\r^+),
\label{1.17}
\end{equation} 
which are evaluated at $r = \r^+ := \r(t) + \Delta$, slightly away
from the radial position of the particle at time $t$. The symbol  
$\Delta$ represents a small radial displacement, which can be either
positive or negative. While the first two quantities listed above are
actually continuous across $r = \r(t)$, the third quantity is
discontinuous, and by evaluating it at $r = \r^+$ we make its value
unambiguous.   

{\bf Second step: Convert the modes.} The first step provides us with 
the spherical-harmonic modes of the retarded potential $\Phi$. The 
computation of the self-force, however, as implemented in
Eq.~(\ref{1.10}), requires the spherical-harmonic modes of
$\Phi_{(\mu)} := e^\alpha_{(\mu)} \nabla_\alpha \Phi$, the frame
components of the retarded field in the selected basis of orthonormal 
vectors. Our choice of tetrad is motivated in Sec.~IV; in the usual
ordering $[t,r,\theta,\phi]$ of the Schwarzschild coordinates we have 
\begin{eqnarray} 
e^\alpha_{(0)} &=& \biggl[ \frac{1}{\sqrt{f}}, 0, 0, 0 \biggr], 
\label{1.18} \\ 
e^\alpha_{(1)} &=& \biggl[ 0, \sqrt{f}\sin\theta\cos\phi, 
\frac{1}{r} \cos\theta\cos\phi, -\frac{\sin\phi}{r\sin\theta} \biggr], 
\label{1.19} \\ 
e^\alpha_{(2)} &=& \biggl[ 0, \sqrt{f}\sin\theta\sin\phi, 
\frac{1}{r} \cos\theta\sin\phi, \frac{\cos\phi}{r\sin\theta} \biggr], 
\label{1.20} \\ 
e^\alpha_{(3)} &=& \biggl[ 0, \sqrt{f}\cos\theta, 
-\frac{1}{r} \sin\theta, 0 \biggr]. 
\label{1.21}
\end{eqnarray} 
In practice it is useful to introduce, as substitutes for
$e^\alpha_{(1)}$ and $e^\alpha_{(2)}$, the complex combinations 
$e^\alpha_{(\pm)} := e^\alpha_{(1)} \pm \i e^\alpha_{(2)}$, or 
\begin{equation} 
e^\alpha_{(\pm)} = \biggl[ 0, \sqrt{f}\sin\theta \e^{\pm\i\phi},   
\frac{1}{r} \cos\theta \e^{\pm\i\phi}, 
\frac{\pm\i\e^{\pm\i\phi}}{r\sin\theta} \biggr].  
\label{1.22}
\end{equation}
As shown in Sec.~IV, the spherical-harmonic modes
$\Phi_{(\mu)lm}(t,r)$ are given in terms of $\Phi_{lm}(t,r)$ by  
\begin{eqnarray} 
\Phi_{(0)lm} &=& \frac{1}{\sqrt{f}} 
\frac{\partial}{\partial t} \Phi_{lm}, 
\label{1.23} \\ 
\Phi_{(+)lm} &=& -\sqrt{ \frac{(l+m-1)(l+m)}{(2l-1)(2l+1)} }  
\biggl( \sqrt{f} \frac{\partial}{\partial r} 
       - \frac{l-1}{r} \biggr) \Phi_{l-1,m-1} 
\nonumber \\ & & \mbox{}
+ \sqrt{ \frac{(l-m+1)(l-m+2)}{(2l+1)(2l+3)} }  
\biggl( \sqrt{f} \frac{\partial}{\partial r} 
       + \frac{l+2}{r} \biggr) \Phi_{l+1,m-1}, 
\label{1.24} \\ 
\Phi_{(-)lm} &=& \sqrt{ \frac{(l-m-1)(l-m)}{(2l-1)(2l+1)} }  
\biggl( \sqrt{f} \frac{\partial}{\partial r} 
       - \frac{l-1}{r} \biggr) \Phi_{l-1,m+1} 
\nonumber \\ & & \mbox{}
- \sqrt{ \frac{(l+m+1)(l+m+2)}{(2l+1)(2l+3)} }  
\biggl( \sqrt{f} \frac{\partial}{\partial r} 
       + \frac{l+2}{r} \biggr) \Phi_{l+1,m+1}, 
\label{1.25} \\ 
\Phi_{(3)lm} &=& \sqrt{ \frac{(l-m)(l+m)}{(2l-1)(2l+1)} }  
\biggl( \sqrt{f} \frac{\partial}{\partial r} 
       - \frac{l-1}{r} \biggr) \Phi_{l-1,m} 
\nonumber \\ & & \mbox{}
+ \sqrt{ \frac{(l-m+1)(l+m+1)}{(2l+1)(2l+3)} }  
\biggl( \sqrt{f} \frac{\partial}{\partial r} 
       + \frac{l+2}{r} \biggr) \Phi_{l+1,m}. 
\label{1.26}
\end{eqnarray} 
These can be evaluated at $r = \r^+$ by substituting the values
extracted in Eq.~(\ref{1.17}).  

{\bf Third step: Regularize the mode sum.} The spherical-harmonic
modes $\Phi_{(\mu)lm}(t,r)$ give rise to the {\it multipole
coefficients} of the retarded field, which are defined by 
\begin{equation} 
\Phi_{(\mu)l}(t,r,\theta,\phi) := \sum_{m=-l}^l \Phi_{(\mu)lm}(t,r)
Y_{lm}(\theta,\phi). 
\label{1.27} 
\end{equation} 
The frame components of the retarded field are then given by 
\begin{equation} 
\Phi_{(\mu)}(t,r,\theta,\phi) = 
\sum_l \Phi_{(\mu)l}(t,r,\theta,\phi). 
\label{1.28}
\end{equation} 
This sum can be evaluated at $r = \r^+ := \r(t) + \Delta$,
$\theta = \frac{\pi}{2}$, and $\phi = \varphi(t)$. The sum would  
diverge in the limit $\Delta \to 0$, but it is to be regularized as in  
Eq.~(\ref{1.10}), which we rewrite as 
\begin{equation} 
\Phi^{\rm R}_{(\mu)}(t,\r,{\textstyle \frac{\pi}{2}},\varphi)  
= \lim_{\Delta \to 0} \sum_l \biggl\{ 
\Phi_{(\mu)l}(t,\r^+,{\textstyle \frac{\pi}{2}},\varphi) 
- q \biggl[ (l + {\textstyle \frac{1}{2}}) A_{(\mu)} 
+ B_{(\mu)} + \frac{C_{(\mu)}}{(l + {\textstyle \frac{1}{2}})} 
+ \frac{D_{(\mu)}}{(l - {\textstyle \frac{1}{2}}) 
(l + {\textstyle \frac{3}{2}})} + \cdots \biggr] \biggr\}
\label{1.29}
\end{equation}
after also involving Eq.~(\ref{1.11}). 

We compute the regularization parameters in Sec.~V of this paper. We
find  
\begin{eqnarray} 
A_{(0)} &=& \frac{\dot{\r}}{\sqrt{\f}(\r^2+L^2)} \mbox{sign}(\Delta),  
\label{1.30} \\ 
A_{(+)} &=& -\e^{\i\varphi} \frac{E}{\sqrt{\f}(\r^2+L^2)}
\mbox{sign}(\Delta), 
\label{1.31} \\ 
A_{(3)} &=& 0,  
\label{1.32}
\end{eqnarray} 
where $\f := 1 - 2M/\r$ and $\mbox{sign}(\Delta)$ is equal to $+1$ if
$\Delta > 0$ and to $-1$ if $\Delta < 0$. We have, in addition,  
$A_{(-)} = \bar{A}_{(+)}$, $A_{(1)} = \mbox{Re}[A_{(+)}]$, and 
$A_{(2)} = \mbox{Im}[A_{(+)}]$.  

We also find 
\begin{eqnarray} 
B_{(0)} &=& -\frac{E \r \dot{\r}}{\sqrt{\f}(\r^2+L^2)^{3/2}} {\scr E}  
+ \frac{E \r \dot{\r}}{2\sqrt{\f}(\r^2+L^2)^{3/2}} {\scr K}, 
\label{1.33} \\ 
B_{(+)} &=& \e^{\i\varphi} \bigl( B^{\rm c}_{(+)} - \i B^{\rm s}_{(+)}
\bigr), 
\label{1.34} \\ 
B^{\rm c}_{(+)} &=& \biggl[ 
\frac{\r \dot{\r}^2}{\sqrt{\f}(\r^2+L^2)^{3/2}} 
+ \frac{\sqrt{\f}}{2 \r \sqrt{\r^2+L^2}} 
\biggr] {\scr E} 
\nonumber \\ & & \mbox{} 
- \biggl[ 
\frac{\r \dot{\r}^2}{2\sqrt{\f}(\r^2+L^2)^{3/2}} 
+ \frac{\sqrt{\f}-1}{\r \sqrt{\r^2+L^2}} 
\biggr] {\scr K},
\label{1.35} \\ 
B^{\rm s}_{(+)} &=&
- \frac{(2-\sqrt{\f})\dot{\r}}{2L\sqrt{\r^2+L^2}\sqrt{\f}} {\scr E}  
+ \frac{(2-\sqrt{\f})\dot{\r}}{2L\sqrt{\r^2+L^2}\sqrt{\f}} {\scr K},   
\label{1.36} \\ 
B_{(3)} &=& 0. 
\label{1.37}
\end{eqnarray} 
We have, in addition, $B_{(-)} = \bar{B}_{(+)}$, $B_{(1)} =
\mbox{Re}[B_{(+)}] = B^{\rm c}_{(+)} \cos\varphi + B^{\rm s}_{(+)} 
\sin\varphi$, and $B_{(2)} = \mbox{Im}[B_{(+)}] = B^{\rm c}_{(+)}
\sin\varphi - B^{\rm s}_{(+)} \cos\varphi$. 

We have introduced the (rescaled) elliptic integrals 
\begin{equation} 
{\scr E} := \frac{2}{\pi} \int_0^{\pi/2} (1-k\sin^2\psi)^{1/2}\, d\psi  
= F(-{\textstyle \frac{1}{2}},{\textstyle \frac{1}{2}}; 1; k)
\label{1.38}
\end{equation} 
and 
\begin{equation} 
{\scr K} := \frac{2}{\pi} \int_0^{\pi/2} (1-k\sin^2\psi)^{-1/2}\, d\psi     
= F({\textstyle \frac{1}{2}},{\textstyle \frac{1}{2}}; 1; k), 
\label{1.39}
\end{equation} 
in which $k := L^2/(\r^2 + L^2)$. As indicated in Eqs.~(\ref{1.38})
and (\ref{1.39}), the elliptic integrals can also be expressed in
terms of hypergeometric functions. 

We also find 
\begin{equation}
C_{(\mu)} = 0
\label{1.40}
\end{equation} 
and 
\begin{eqnarray} 
D_{(0)} &=& -\biggl[ 
\frac{E \r^3 (\r^2-L^2) \dot{\r}^3}
{2\sqrt{\f}(\r^2+L^2)^{7/2}} 
+ \frac{E(\r^7 + 30M \r^6 - 7 L^2 \r^5 + 114 M L^2 \r^4 
+ 104 M L^4 \r^2 + 36 M L^6) \dot{\r}}
  {16 \r^4 \sqrt{\f}(\r^2+L^2)^{5/2}} 
\biggr] {\scr E} 
\nonumber \\ & & \mbox{} 
+ \biggl[ 
\frac{E \r^3 (5 \r^2 - 3 L^2) \dot{\r}^3}
  {16\sqrt{\f}(\r^2+L^2)^{7/2}} 
+ \frac{E(\r^5 + 16M \r^4 - 3 L^2 \r^3 + 42 M L^2 \r^2 
+ 18 M L^4) \dot{\r}}{16 \r^2
  \sqrt{\f}(\r^2+L^2)^{5/2}} \biggr] {\scr K}, 
\label{1.41} \\ 
D_{(+)} &=& \e^{\i\varphi} \bigl( D^{\rm c}_{(+)} - \i D^{\rm s}_{(+)} 
\bigr), 
\label{1.42} \\ 
D^{\rm c}_{(+)} &=& 
\biggl[ \frac{\r^3 (\r^2 - L^2) \dot{\r}^4}
{2\sqrt{\f}(\r^2+L^2)^{7/2}} 
- \frac{\r \dot{\r}^2}{4(\r^2+L^2)^{3/2}} 
+ \frac{(3\r^7 + 6 M \r^6 - L^2 \r^5 + 31 M L^2 \r^4 
  + 26 M L^4 \r^2 + 9 M L^6) \dot{\r}^2}
  {4 \r^4 \sqrt{\f} (\r^2+L^2)^{5/2}} 
\nonumber \\ & & \mbox{} 
+ \frac{(3\r^7 + 8 M \r^6 + L^2 \r^5 + 26 M L^2 \r^4 
  + 22 M L^4 \r^2 + 8 M L^6) \sqrt{\f}}
  {16 \r^6 (\r^2+L^2)^{3/2}} 
- \frac{\r^3 + 2 M \r^2 + 4 M L^2}
  {8 \r^4 \sqrt{\r^2+L^2}} \bigg] {\scr E} 
\nonumber \\ & & \mbox{} 
+ \biggl[ -\frac{\r^3 (5 \r^2 - 3 L^2) \dot{\r}^4}
 {16\sqrt{\f}(\r^2+L^2)^{7/2}} 
+ \frac{\r \dot{\r}^2}{8(\r^2+L^2)^{3/2}} 
- \frac{(7 \r^5 + 12 M \r^4 - L^2 \r^3 
  + 46 M L^2 \r^2 + 18 M L^4) \dot{\r}^2}
  {16 \r^2 \sqrt{\f} (\r^2+L^2)^{5/2}} 
\nonumber \\ & & \mbox{} 
- \frac{(7\r^5 + 6 M \r^4 + 6 L^2 \r^3 + 12 M L^2 \r^2 
  + 4 M L^4) \sqrt{\f}}
  {16 \r^4 (\r^2+L^2)^{3/2}} 
+ \frac{3}{8 \r \sqrt{\r^2+L^2}} \biggr] {\scr K}, 
\label{1.43} \\ 
D^{\rm s}_{(+)} &=& 
\biggl[ 
\frac{\r^2(\r^2 - 7L^2)(\sqrt{\f} - 2)\dot{\r}^3}
  {16 L \sqrt{\f} (\r^2+L^2)^{5/2}} 
- \frac{(2\r^7 + M \r^6 + 5 L^2 \r^5 + 10 M L^2 \r^4 
  + 29 M L^4 \r^2 + 14 M L^6) \dot{\r}} 
  {8 \r^5 L (\r^2+L^2)^{3/2}}
\nonumber \\ & & \mbox{} 
+ \frac{(\r^5 - M \r^4 + 4 L^2 \r^3 - 5 M L^2 \r^2 
  + 2 M L^4) \dot{\r}}
  {4 \r^3 L \sqrt{\f}(\r^2+L^2)^{3/2}} \biggr] {\scr E} 
\nonumber \\ & & \mbox{} 
+ \biggl[ 
- \frac{\r^2(\r^2 - 3L^2)(\sqrt{\f} - 2)\dot{\r}^3}
  {16 L \sqrt{\f} (\r^2+L^2)^{5/2}} 
+ \frac{(4\r^5 + 2 M \r^4 + 7 L^2 \r^3 + 10 M L^2 \r^2 
  + 14 M L^4) \dot{\r}} 
  {16 \r^3 L (\r^2+L^2)^{3/2}}
\nonumber \\ & & \mbox{} 
- \frac{(2 \r^3 - 2 M \r^2 + 5 L^2 \r - 8 M L^2) \dot{\r}}
  {8 \r L \sqrt{\f}(\r^2+L^2)^{3/2}} \biggr] {\scr K}, 
\label{1.44} \\ 
D_{(3)} &=& 0. 
\label{1.45}
\end{eqnarray} 
We have, in addition, $D_{(-)} = \bar{D}_{(+)}$, $D_{(1)} =
\mbox{Re}[D_{(+)}] = D^{\rm c}_{(+)} \cos\varphi + D^{\rm s}_{(+)}
\sin\varphi$, and $D_{(2)} = \mbox{Im}[D_{(+)}] = D^{\rm c}_{(+)}
\sin\varphi - D^{\rm s}_{(+)} \cos\varphi$. 

{\bf Fourth step: Construct the self-force.} The mode-sum of
Eq.~(\ref{1.29}) can now be evaluated. Thanks to the presence of the
regularization parameters $D_{(\mu)}$, the sum converges quickly to a
precise estimate for the frame components 
$\Phi^{\rm R}_{(\mu)}(z)$. The vector field 
$\Phi^{\rm R}_\alpha := \nabla_\alpha \Phi^{\rm R}$ is related to
these by   
\begin{equation} 
\Phi^{\rm R}_\alpha = -\Phi^{\rm R}_{(0)} e_{(0)\alpha} 
+ \frac{1}{2} \Phi^{\rm R}_{(+)} e_{(-)\alpha} 
+ \frac{1}{2} \Phi^{\rm R}_{(-)} e_{(+)\alpha} 
+ \Phi^{\rm R}_{(3)} e_{(3)\alpha}. 
\label{1.46}
\end{equation} 
After involving Eqs.~(\ref{1.18})--(\ref{1.22}) and evaluating this at
$r = \r(t)$, $\theta = \frac{\pi}{2}$, and $\phi = \varphi(t)$, we
obtain
\begin{eqnarray} 
\Phi^{\rm R}_t &=& \sqrt{\f} \Phi^{\rm R}_{(0)}, 
\label{1.47} \\ 
\Phi^{\rm R}_r &=& \frac{1}{2\sqrt{\f}} \Bigl( 
\Phi^{\rm R}_{(+)} \e^{-\i\varphi} 
+ \Phi^{\rm R}_{(-)} \e^{\i\varphi} \Bigr),  
\label{1.48} \\ 
\Phi^{\rm R}_\theta &=& -\r \Phi^{\rm R}_{(3)}, 
\label{1.49} \\ 
\Phi^{\rm R}_\phi &=& -\frac{\i \r}{2} \Bigl( 
\Phi^{\rm R}_{(+)} \e^{-\i\varphi} 
- \Phi^{\rm R}_{(-)} \e^{\i\varphi} \Bigr). 
\label{1.50}
\end{eqnarray}
These, finally, can be substituted into Eqs.~(\ref{1.2}) and
(\ref{1.3}) for a concrete evaluation of the scalar self-force.  

\subsection{Case study: Particle on a circular orbit.} 

The prescription detailed in the preceding subsection will be fully
implemented in a future publication. To convince ourselves that it
actually works, we carried out the computations for the special case
of a scalar charge moving on a circular orbit around a Schwarzschild
black hole. Our results are not new: They reproduce some already
obtained by Burko \cite{burko:00c}, Detweiler, Messaritaki, and
Whiting \cite{detweiler-etal:03}, as well as Diaz-Rivera, Messaritaki,
and Whiting \cite{diaz-rivera-etal:04}. Nevertheless, we present them
here (without derivations) because they constitute a proof of
principle that the prescription is valid. 

We place a scalar charge on a circular, geodesic orbit at a radius
$r_0 = 6M$ (this is the innermost stable circular orbit). We go
through all the steps listed in Sec.~I E and compute 
$\Phi^{\rm R}_{(\mu)}$, the frame components of the regular field
evaluated at $r = \r(t) := r_0$, $\theta = \frac{\pi}{2}$, and $\phi = 
\varphi(t) := \Omega t$, where $\Omega = \sqrt{M/r_0^3}$ is the
particle's angular velocity. Without loss of generality we evaluate
the scalar field $\Phi^{\rm R}_\alpha$ at $t = 0$, so that 
$\varphi = 0$. This is related to the frame components by  
\begin{equation} 
\Phi^{\rm R}_t = \sqrt{f_0} \Phi^{\rm R}_{(0)}, \qquad 
\Phi^{\rm R}_r = \frac{1}{\sqrt{f_0}} \mbox{Re} \Phi^{\rm R}_{(+)},
\qquad 
\Phi^{\rm R}_\theta = 0, \qquad 
\Phi^{\rm R}_\phi = r_0 \mbox{Im} \Phi^{\rm R}_{(+)},
\label{1.51}
\end{equation}
where $f_0 = 1-2M/r_0$. The regular field satisfies the helical
condition 
\begin{equation} 
\Phi^{\rm R}_t + \Omega \Phi^{\rm R}_\phi = 0, 
\label{1.52}
\end{equation}
and we find that the real and imaginary parts of $\Phi^{\rm R}_{(+)}$
fully determine the scalar self-force.  

\begin{figure}
\includegraphics[angle=-90,scale=0.33]{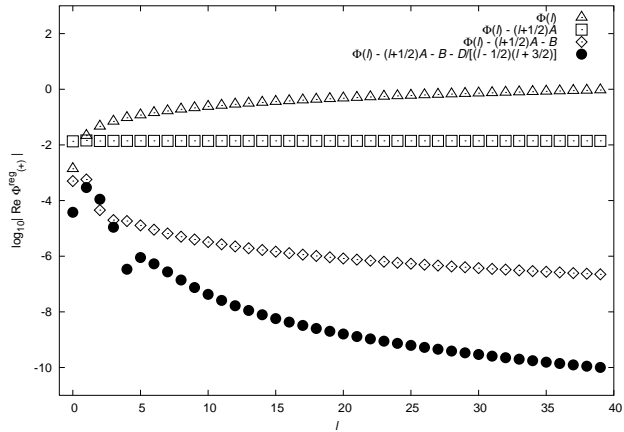}
\caption{Plot of several regularized versions of
$|\mbox{Re}\Phi_{(+)l}|$ as a function of multipole order $l$. The
different curves are described in the text.}    
\end{figure}

In Fig.~1 we plot several regularized versions of
$|\mbox{Re}\Phi_{(+)l}|$, the absolute value of the real part of 
$\Phi_{(+)l}$, the multipole coefficients of the frame component
$\Phi_{(+)}$ of the retarded field; we plot this on a logarithmic
scale as a function of $l$, in the interval $0 \leq l < 40$. The
upper curve (in open triangles) represents the unregularized
multipole coefficients of the retarded field; we see that this
function grows linearly with $l$, so that the sum of its terms  
diverges. The second curve (in open squares) is what is obtained after 
subtracting $(l+\frac{1}{2}) A_{(+)}$ from the first curve; we see
here that the curve is approximately constant, so that its sum also 
diverges. The third curve (in open diamonds) is what is obtained
after subtracting $B_{(+)}$ and $C_{(+)}/(l+\frac{1}{2}) = 0$ from the
second curve; this produces a function that decays as $1/l^2$, and
this leads to a converging sum. The convergence is accelerated,
however, with the fourth curve (in solid circles), which is obtained
after subtracting $D_{(+)}/[(l-\frac{1}{2})(l+\frac{3}{2})]$ from the
third curve; this produces a function that decays as $1/l^4$.  

Figure 1 constitutes a very robust test of our numerical and
analytical computations: Any error would give rise to a gross
violation of the properties listed above. For example, an error in the
analytical form of $D_{(+)}$ would produce a curve in full circles
that would still decay as $1/l^2$ instead of the observed
$1/l^4$. Similarly, a coding error would return a retarded field whose
singularity structure would not be compatible with the regularization
parameters, and this would again produce very visible effects in
Fig.~1. We have tested this observation by deliberately inserting
errors at various places in our numerical code. The results follow
expectations and convince us that the prescription is valid. (We have
not yet tested the $\dot{\r} \neq 0$ sector of the prescription.)  

\begin{figure}
\includegraphics[angle=-90,scale=0.33]{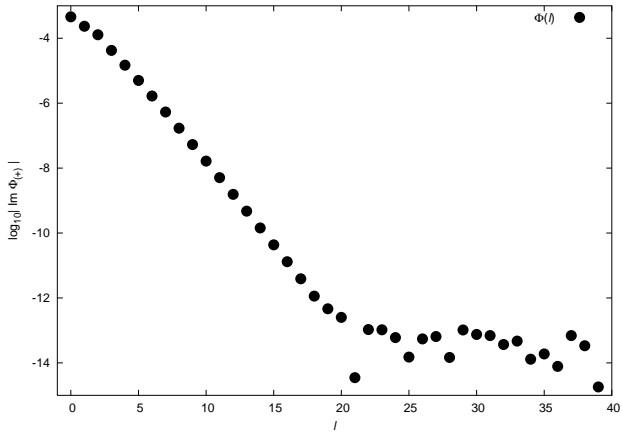}
\caption{Plot of $|\mbox{Im}\Phi_{(+)l}|$ as a function of multipole
order $l$.}     
\end{figure}

In Fig.~2 we plot $|\mbox{Im}\Phi_{(+)l}|$, the absolute value of the
imaginary part of $\Phi_{(+)l}$; again we plot this on a logarithmic 
scale as a function of $l$, in the interval $0 \leq l < 40$. This
curve requires no regularization: the imaginary parts of $A_{(+)}$,
$B_{(+)}$, and $D_{(+)}$ all vanish when $\dot{\r} = 0$. (Recall that
we evaluate the field at $t=0$ and $\varphi := \Omega t = 0$.)
Here we see that $|\mbox{Im}\Phi_{(+)l}|$ decays {\it exponentially}
as a function of $l$ (approximately as $\e^{-l/2}$) until round-off 
errors start to dominate when $l$ is approximately equal to 20. This
curve, of course, produces a rapidly converging sum.  

Our final numerical results are 
\begin{eqnarray}  
\frac{M^2}{q} \Phi^{\rm R}_t &\simeq& 3.60907254 \times 10^{-4}, 
\label{1.53} \\
\frac{M^2}{q} \Phi^{\rm R}_r &\simeq& 1.67730 \times 10^{-4}, 
\label{1.54} \\ 
\frac{M}{q} \Phi^{\rm R}_\phi &\simeq& -5.30423170 \times 10^{-3},  
\label{1.55} 
\end{eqnarray}  
with the number of significant digits reflecting our best estimation
of the code's numerical accuracy (the last digit is uncertain). The
least accurate number is for $\Phi^{\rm R}_r$, which is obtained after
several rounds of regularization. Our numbers are consistent with
results obtained by Diaz-Rivera, Messaritaki, and Whiting
\cite{diaz-rivera-etal:04}: In their Table I they list $(M^2/q)
\Phi^{\rm R}_r = 1.6772834 \times 10^{-4}$ for $r_0 = 6M$.    

\subsection{Organization of this paper} 
  
The chain of calculations that lead to the prescription detailed in
Sec.~I E is a long one, and it occupies the remaining sections of the
paper. Here is how the rest of the paper is organized. 

We begin in Sec.~II with the development of a covariant local 
expansion of the singular field $\Phi^{\rm S}_\alpha(x)$ in the
vicinity of the particle's world line. The expansion is based on the
assumption that the scalar charge follows a geodesic of a vacuum
spacetime, but it is otherwise general; we do not yet, at this stage, 
assume that the metric is given by the Schwarzschild solution. The
expansion must be carried out to a sufficient degree of accuracy to
permit the determination of all four regularization parameters. Such
an accurate expansion has never appeared in the literature, and we
present it here for the first time. 

In Sec.~III we convert the covariant expansion into an explicit
coordinate expansion that can be evaluated for any spacetime whose
metric is expressed in any coordinate system. The methods by which we 
obtain the covariant and coordinate expansions rely heavily on the 
general theory of bitensors. These were introduced by Synge
\cite{synge:60} and DeWitt and Brehme \cite{dewitt-brehme:60}, and
the theory is conveniently summarized in Poisson's contribution to    
{\it Living Reviews in Relativity} \cite{poisson:04b}. This last paper 
is an essential resource for the calculations presented in Secs.~II
and III, and we will repeatedly refer to it as LRR. 

In Sec.~IV we motivate our choice of tetrad $e^\alpha_{(\mu)}$. We
also work out the relationships listed in
Eqs.~(\ref{1.23})--(\ref{1.26}) between $\Phi_{(\mu)lm}$, the
spherical-harmonic modes of $\Phi_{(\mu)} := e^\alpha_{(\mu)}
\nabla_\alpha \Phi$, and $\Phi_{lm}$, the modes of the scalar
potential. As we explain in this section, the tetrad is selected so as
to produce a simple relationship in which $\Phi_{(\mu)lm}$ is linked
to the neighboring modes $\Phi_{l\pm 1,m}$ and $\Phi_{l\pm 1, m\pm 1}$
only; other tetrads would lead to more complicated couplings and are
best avoided.    
   
In Sec.~V we compute the regularization parameters $A_{(\mu)}$,
$B_{(\mu)}$, $C_{(\mu)} = 0$, and $D_{(\mu)}$. We do this by expanding 
our local coordinate expansion for $\Phi^{\rm S}_{(\mu)}(x)$ in
Legendre polynomials and showing that the result takes the form of
Eq.~(\ref{1.11}). To perform the Legendre decompositions we rely on 
techniques imported from Detweiler, Messaritaki, and
Whiting \cite{detweiler-etal:03}; these are summarized in the
Appendix.  

Many calculations presented in this paper are extremely tedious and
could not have been carried out with pen and paper. We relied heavily
on the symbolic manipulator {\sc GRTensorII} \cite{grtensor} working
under {\sc Maple}. Throughout the paper we use geometrized units in
which $G = c = 1$, and we adhere to the sign conventions of Misner,
Thorne, and Wheeler \cite{MTW:73}.   
   
\section{Covariant local expansion of the singular field} 

\subsection{Singular field} 

The singular part $\Phi^{\rm S}$ of the retarded potential $\Phi$
produced by a point scalar charge $q$ moving on an arbitrary world
line of an arbitrary curved spacetime was first correctly identified
by Detweiler and Whiting \cite{detweiler-whiting:03}. As they have
shown, the singular potential possesses the following properties: (i)
it satisfies the same wave equation as the retarded potential,
Eq.~(\ref{1.1}); (ii) it displays the same singularity structure as
the retarded potential near the particle's world line; and (iii) it
does not exert a force on the point charge. The scalar self-force
acting on the charge therefore results from the sole action of the
regular potential $\Phi^{\rm R} = \Phi - \Phi^{\rm S}$, which is
smooth on the world line. The self-force is proportional to 
$\Phi^{\rm R}_\alpha := \nabla_\alpha \Phi^{\rm R}$, and it can be
calculated by first computing $\Phi_\alpha := \nabla_\alpha \Phi$,
then removing from this the singular part $\Phi^{\rm S}_\alpha 
:= \nabla_\alpha \Phi^{\rm S}$, and finally evaluating the result at
the position of the particle.      

Our task in this section is to develop a covariant expansion of the   
singular field $\Phi^{\rm S}_\alpha$ in powers of $\epsilon$, a
book-keeping quantity that loosely represents the distance between
the field point and the world line. (The distance to the world line
will be defined precisely below.) As was justified near the end of
Sec.~I B, we shall restrict our attention to the case of a
scalar charge $q$ that moves on a {\it geodesic of a vacuum
spacetime}; the charge's acceleration vector and the spacetime's Ricci
tensor will therefore be set equal to zero. The expansion begins with
a term of order $\epsilon^{-2}$ and we shall keep it accurate through
order $\epsilon$, neglecting terms of order $\epsilon^2$ and higher.  

Our starting point is the expression displayed in Sec.~5.1.5
[Eq.~(413)] of LRR \cite{poisson:04b},  
\begin{eqnarray}
\Phi^{\rm S}_{\alpha}(x) &=& 
-\frac{q}{2r^2} U(x,x') \nabla_\alpha r  
- \frac{q}{2 {r_{\rm adv}}^2} U(x,x'') \nabla_\alpha r_{\rm adv} 
+ \frac{q}{2r} \Bigl[ \nabla_\alpha U(x,x') + u^{\alpha'}
  \nabla_{\alpha'} U(x,x') \nabla_\alpha u \Bigr]  
\nonumber \\  & & \mbox{} 
+ \frac{q}{2r_{\rm adv}} \Bigl[ \nabla_\alpha U(x,x'') 
+ u^{\alpha''} \nabla_{\alpha''} U(x,x'') \nabla_\alpha v \Bigr]  
+ \frac{q}{2} \Bigl[ V(x,x') \nabla_\alpha u 
- V(x,x'') \nabla_\alpha v \Bigr]  
\nonumber \\  & & \mbox{} 
- \frac{q}{2} \int_u^v \nabla_\alpha V(x,z)\, d\tau. 
\label{2.1}
\end{eqnarray}
We have introduced a large number of symbols. To begin, $x$ is the
field point at which the singular field is evaluated, and the world
line is described by parametric relations $z^\mu(\tau)$ involving the
proper-time parameter $\tau$. The points $x'$ and $x''$ on the world
line are known respectively as the {\it retarded} and {\it advanced}
points associated with $x$; these are defined such that $x$ and $x'$
are linked by a unique future-directed null geodesic originating on
the world line, while $x$ and $x''$ are linked by a past-directed null
geodesic that also originates on the world line. We define the scalar 
field $u(x)$ as the value of the proper-time parameter at $x' \equiv 
z(\tau = u)$; this is known as the {\it retarded time} function of the
field point $x$. Similarly, we define the {\it advanced time} function
$v(x)$ as the proper time at $x'' \equiv z(\tau = v)$. With
$\sigma(x,z)$ denoting Synge's world function \cite{synge:60}, equal
to half the squared geodesic distance between the field point $x$ and
the point $z$ on the world line, we have that $\sigma(x,x') =
\sigma(x,x'') = 0$. (Our subsequent developments rely heavily on a
working knowledge of the general theory of bitensors; this material is
reviewed in Sec.~2 of LRR \cite{poisson:04b}.) 

The gradient of Synge's world function is denoted
$\sigma_\alpha(x,\bar{x})$ if $\sigma(x,\bar{x})$ is differentiated
with respect to its first argument, and
$\sigma_{\bar{\alpha}}(x,\bar{x})$ if it is differentiated instead
with respect to its second argument. The {\it retarded distance}
$r(x)$ between $x$ and the world line refers to the retarded point
$x'$ and is defined by 
\begin{equation} 
r := \sigma_{\alpha'}(x,x') u^{\alpha'}, 
\label{2.2} 
\end{equation} 
where $u^{\alpha'}$ is the particle's velocity vector at $x'$; this is
an affine-parameter distance along the null geodesic that links $x$ to
$x'$. The {\it advanced distance} $r_{\rm adv}(x)$ between $x$ and the
world line refers instead to the advanced point $x''$ and is defined
by  
\begin{equation} 
r_{\rm adv} := -\sigma_{\alpha''}(x,x'') u^{\alpha''}, 
\label{2.3} 
\end{equation} 
where $u^{\alpha''}$ is the particle's velocity vector at $x''$; this
also is an affine-parameter distance. We note that the distance
functions are both nonnegative. As a consequence of their defining
relations (see Secs.~3.3.3 and 3.4.4 of LRR \cite{poisson:04b} for an
extended discussion), we have the scaling relations $r = O(\epsilon)$,
$r_{\rm adv} = O(\epsilon)$, and $v-u = O(\epsilon)$. It also follows
from the defining relations that the gradients of $u$, $v$, $r$, and
$r_{\rm adv}$ that appear in Eq.~(\ref{2.1}) are given by   
\begin{eqnarray} 
\nabla_\alpha u &=& -\sigma_{\alpha}(x,x')/r, 
\label{2.4} \\ 
\nabla_\alpha v &=& \sigma_{\alpha}(x,x'')/r_{\rm adv}, 
\label{2.5} \\ 
\nabla_\alpha r &=& \sigma_{\alpha'\beta'} u^{\alpha'} u^{\beta'}
\nabla_\alpha u + \sigma_{\alpha'\alpha} u^{\alpha'}, 
\label{2.6} \\ 
\nabla_\alpha r_{\rm adv} &=& -\sigma_{\alpha''\beta''} u^{\alpha''}
u^{\beta''} \nabla_\alpha v - \sigma_{\alpha''\alpha} u^{\alpha''}.   
\label{2.7}
\end{eqnarray} 
Equations (\ref{2.4}) and (\ref{2.5}) are always valid and follow
directly from the conditions $\sigma(x,x') = \sigma(x,x'') = 0$;
Eqs.~(\ref{2.6}) and (\ref{2.7}) are valid when the world line is a
geodesic of the curved spacetime.   

The biscalars $U(x,z)$ and $V(x,z)$ that appear in Eq.~(\ref{2.1}) are
respectively the ``direct'' and ``tail'' parts of the retarded Green's
function $G(x,z)$ associated with the scalar potential $\Phi(x)$. (The 
general theory of scalar Green's functions in curved spacetime is
reviewed in Sec.~4.3 of LRR \cite{poisson:04b}.) For our purposes in
this section, the only relevant properties of these objects are the
scaling relations    
\begin{eqnarray} 
U(x,x') &=& 1 + O(\epsilon^4), 
\label{2.8} \\ 
U(x,x'') &=& 1 + O(\epsilon^4), 
\label{2.9} \\ 
\nabla_\alpha U(x,x') &=& O(\epsilon^3), 
\label{2.10} \\
\nabla_{\alpha'} U(x,x') &=& O(\epsilon^3), 
\label{2.11} \\
\nabla_\alpha U(x,x'') &=& O(\epsilon^3), 
\label{2.12} \\
\nabla_{\alpha''} U(x,x'') &=& O(\epsilon^3), 
\label{2.13} \\ 
V(x,x') &=& O(\epsilon^2), 
\label{2.14} \\ 
V(x,x'') &=& O(\epsilon^2), 
\label{2.15} \\ 
\nabla_\alpha V(x,z) &=& O(\epsilon). 
\label{2.16}
\end{eqnarray} 
These relations are valid in a Ricci-flat spacetime only.   

Substituting Eqs.~(\ref{2.8})--(\ref{2.16}) into Eq.~(\ref{2.1})
produces the substantial simplification 
\begin{equation} 
\Phi^{\rm S}_\alpha(x) = -\frac{q}{2r^2} \nabla_\alpha r  
- \frac{q}{2r_{\rm adv}^2} \nabla_\alpha r_{\rm adv} 
+ O(\epsilon^2). 
\label{2.17} 
\end{equation} 
This will be turned into a more explicit expression in the course of
the following subsections. 

\subsection{Reference point on the world line} 

The expression of Eq.~(\ref{2.17}) refers to two separate points on
the world line, the retarded point $x' \equiv z(u)$ and the advanced
point $x'' \equiv z(v)$; each is linked to $x$ by the null conditions
$\sigma(x,x') = \sigma(x,x'') = 0$. We find it convenient to introduce
a third point $\bar{x} \equiv z(\bar{\tau})$ on the world line, and to
go through the lengthy procedure of re-expressing 
$\nabla_\alpha \Phi^{\rm S}(x)$ solely in terms of tensorial
quantities that are evaluated at $\bar{x}$. An important aspect of
this transcription is that we take the point $\bar{x}$ to be 
{\it completely arbitrary}, except for the following restriction: We
assume that $\bar{x}$ is in a spacelike relation with $x$, so that it
lies after $x'$ but before $x''$ on the world line; $\bar{x}$ is
otherwise arbitrary. The transcription produces two major
advantages: First, it consolidates the dependence of the singular
field on the world line to a single point instead of two; and second,
it eliminates the dependence of $\Phi^{\rm S}_\alpha$ on $x$
that is only implicitly contained in $x'(x)$ and $x''(x)$. Our
resulting expression for the singular field, which appears in
Eq.~(\ref{2.62}) below, contains a dependence on $x$ that is fully  
explicit.       

Having made arbitrary choices for the field point $x$ and the
reference point $\bar{x}$ on the world line, we define the quantities  
\begin{equation} 
\bar{r} := \sigma_{\bar{\alpha}}(x,\bar{x}) u^{\bar{\alpha}} 
\label{2.18}
\end{equation} 
and 
\begin{equation} 
s^2 := \bigl( g^{\bar{\alpha}\bar{\beta}} 
+ u^{\bar{\alpha}} u^{\bar{\beta}} \bigr)
\sigma_{\bar{\alpha}}(x,\bar{x}) \sigma_{\bar{\beta}}(x,\bar{x}). 
\label{2.19}
\end{equation} 
We note that $\bar{r} = O(\epsilon)$ and that its definition is
similar to that of $r$ and $r_{\rm adv}$ provided in the preceding 
subsection. In fact, when $\bar{x} \to x'$ we have that $\bar{r} \to  
r$, while $\bar{r} \to -r_{\rm adv}$ when $\bar{x} \to x''$; somewhere
between $x'$ and $x''$ we have $\bar{r}$ changing sign, from a
positive value to a negative value. The quantity $s^2 = O(\epsilon^2)$ 
is the squared distance between $\bar{x}$ and $x$ as measured by an
observer at $\bar{x}$ that is momentarily comoving with the charged
particle. We note the useful identity $s^2 = 2\sigma(x,\bar{x}) +
\bar{r}^2$, which shows that $s^2$ is necessarily positive when $x$
and $\bar{x}$ are in a spacelike relation.  

Our remaining task is to carry out the procedure described in the
first paragraph of this subsection. This involves many steps, and
lengthy calculations. The first step is to determine the positions of 
the retarded and advanced points relative to the reference point
$\bar{x}$; this we do in Sec.~II C. The next steps involve computing
the various pieces of the singular field of Eq.~(\ref{2.17}) and
expressing them in terms of tensorial quantities defined at
$\bar{x}$. We begin in Sec.~II D with a computation of $r$ and 
$r_{\rm adv}$. We continue in Sec.~II E with a computation of
$\sigma_{\alpha}(x,x')$ and $\sigma_{\alpha}(x,x'')$, which appear in
the expressions for $\nabla_\alpha u$ and $\nabla_\alpha v$ --- see
Eqs.~(\ref{2.4}) and (\ref{2.5}). In Secs.~II F and G we compute the
quantities $\sigma_{\alpha\alpha'}u^{\alpha'}$, 
$\sigma_{\alpha\alpha''}u^{\alpha''}$, $\sigma_{\alpha'\beta'}
u^{\alpha'} u^{\beta'}$, and $\sigma_{\alpha''\beta''}u^{\alpha''}
u^{\beta''}$, which are involved in the expressions for 
$\nabla_\alpha r$ and $\nabla_\alpha r_{\rm adv}$ --- see
Eqs.~(\ref{2.6}) and (\ref{2.7}). In Secs.~II H and I we collect our
results and compute $\nabla_\alpha u$, $\nabla_\alpha v$, 
$\nabla_\alpha r$, and $\nabla_\alpha r_{\rm adv}$. And finally, in
Sec.~II J we produce our final expression for $\nabla_\alpha
\Phi^{\rm S}(x)$ --- see Eq.~(\ref{2.62}) below. The reader who does
not wish to go through these computations may simply jump to Sec.~II J
for the punch line.    

\subsection{Retarded and advanced points} 

We wish to determine the positions of the retarded point $x' = z(u)$
and the advanced point $x'' = z(v)$ relative to the arbitrary
reference point $\bar{x} = z(\bar{\tau})$ on the world line. We will
achieve this by obtaining expressions for   
\begin{equation} 
\Delta_+ := v - \bar{\tau} > 0
\label{2.20}
\end{equation} 
and 
\begin{equation} 
\Delta_- := u-\bar{\tau} < 0. 
\label{2.21}
\end{equation} 
These will be given in the form of expansions in powers of
$\epsilon$. We will rely on Taylor-expansion techniques reviewed in
Sec.~3.4 of LRR \cite{poisson:04b}, as well as the standard
bitensorial expansion (see, for example, Ref.~\cite{christensen:76})  
\begin{equation} 
\sigma_{\bar{\alpha}\bar{\beta}} =
g_{\bar{\alpha}\bar{\beta}} - \frac{1}{3}
R_{\bar{\alpha}\bar{\gamma}\bar{\beta}\bar{\delta}}
  \sigma^{\bar{\gamma}} \sigma^{\bar{\delta}} + \frac{1}{12}  
R_{\bar{\alpha}\bar{\gamma}\bar{\beta}\bar{\delta};\bar{\epsilon}} 
  \sigma^{\bar{\gamma}} \sigma^{\bar{\delta}} \sigma^{\bar{\epsilon}}
  + O(\epsilon^4),  
\label{2.22}
\end{equation} 
in which the metric, as well as the Riemann tensor and its covariant
derivative, is evaluated at the reference point $\bar{x}$.  

We keep $x$ fixed and introduce the function 
\begin{equation} 
\sigma(\tau) := \sigma\bigl(x,z(\tau)\bigr) 
\label{2.23}
\end{equation} 
of the proper-time parameter $\tau$ on the world line. We note the
special values $\sigma(u) = \sigma(v) = 0$, and that $\sigma(\tau)$ is
positive in the interval $u < \tau < v$. We express $\sigma(\tau)$ as
a Taylor expansion around the reference point $\tau = \bar{\tau}$, and 
evaluate it at $\tau = w$, which stands collectively for either $u$
or $v$. With $\Delta := w - \bar{\tau}$ (and therefore equal to either
$\Delta_+$ or $\Delta_-$), the result is
\begin{equation} 
0 = \bar{\sigma} + \dot{\sigma} \Delta + \frac{1}{2} \ddot{\sigma}
\Delta^2 + \frac{1}{6} \sigma^{(3)} \Delta^3 + \frac{1}{24}
\sigma^{(4)} \Delta^4 + \cdots, 
\label{2.24}
\end{equation} 
where $\bar{\sigma} := \sigma(\bar{\tau})$ and all derivatives of
$\sigma(\tau)$, which are indicated by overdots or a number within
brackets, are evaluated at $\tau = \bar{\tau}$. The computation of 
the derivatives is simplified by the fact that the motion is
geodesic. From Eq.~(\ref{2.23}) we have 
$\dot{\sigma} = \sigma_{\bar{\alpha}} u^{\bar{\alpha}}$, 
$\ddot{\sigma} = \sigma_{\bar{\alpha}\bar{\beta}}
u^{\bar{\alpha}} u^{\bar{\beta}}$, 
${\sigma}^{(3)} = \sigma_{\bar{\alpha}\bar{\beta}\bar{\gamma}} 
u^{\bar{\alpha}} u^{\bar{\beta}} u^{\bar{\gamma}}$, and 
${\sigma}^{(4)} =
\sigma_{\bar{\alpha}\bar{\beta}\bar{\gamma}\bar{\delta}} 
u^{\bar{\alpha}} u^{\bar{\beta}} u^{\bar{\gamma}} u^{\bar{\delta}}$.  
To evaluate the first derivative we simply involve Eq.~(\ref{2.18})
and get $\dot{\sigma} = \bar{r}$. For the second derivative we involve
Eq.~(\ref{2.22}) and obtain 
\begin{equation} 
\ddot{\sigma} = -1 - \frac{1}{3} R_{u\sigma u\sigma} 
+ \frac{1}{12} R_{u\sigma u\sigma | \sigma} + O(\epsilon^4), 
\label{2.25}
\end{equation} 
where we have introduced the notation $R_{u\sigma u\sigma} :=
R_{\bar{\alpha}\bar{\mu}\bar{\beta}\bar{\nu}} u^{\bar{\alpha}}
\sigma^{\bar{\mu}} u^{\bar{\beta}} \sigma^{\bar{\nu}}$ and 
$R_{u\sigma u\sigma | \sigma} := 
R_{\bar{\alpha}\bar{\mu}\bar{\beta}\bar{\nu};\bar{\lambda}}
u^{\bar{\alpha}} \sigma^{\bar{\mu}} u^{\bar{\beta}} \sigma^{\bar{\nu}}
\sigma^{\bar{\lambda}}$; many variants of this notation will appear
below. To evaluate the third derivative we begin by differentiating
Eq.~(\ref{2.22}) to obtain an expansion for
$\sigma_{\bar{\alpha}\bar{\beta}\bar{\gamma}}$, which we then contract
with the velocity vector. The result is
\begin{equation} 
\sigma^{(3)} = -\frac{1}{4} R_{u\sigma u\sigma | u} +
O(\epsilon^3). 
\label{2.26}
\end{equation} 
We proceed similarly for the fourth derivative and obtain 
\begin{equation} 
\sigma^{(4)} = O(\epsilon^2).
\label{2.27}
\end{equation} 
Gathering the results, Eq.~(\ref{2.24}) becomes 
\begin{equation}
0 = \bar{\sigma} + \bar{r} \Delta - \frac{1}{2} \biggl( 1 
+ \frac{1}{3} R_{u\sigma u\sigma} 
- \frac{1}{12} R_{u\sigma u\sigma | \sigma} \biggr) \Delta^2 
- \frac{1}{24} R_{u\sigma u\sigma | u} \Delta^3 
+ O(\epsilon^6). 
\label{2.28}
\end{equation} 
This equation must now be solved for $\Delta$. 

We assume that $\Delta$ can be expressed as an expansion in powers of
$\epsilon$, in the form 
\begin{equation} 
\Delta = \Delta_1 + \Delta_2 + \Delta_3 + \Delta_4 + O(\epsilon^5)
\label{2.29}
\end{equation} 
with $\Delta_n = O(\epsilon^n)$. Substituting Eq.~(\ref{2.29}) into
Eq.~(\ref{2.28}) and equating each coefficient of $\epsilon^n$ to zero
returns a hierarchy of equations to be solved. The first equation
determines $\Delta_1$:   
\begin{equation} 
\Delta_1^2 - 2\bar{r} \Delta_1 - 2\bar{\sigma} = 0.
\label{2.30}
\end{equation} 
In view of Eq.~(\ref{2.19}) we have that $2\bar{\sigma} = s^2 
- \bar{r}^2$, and the two solutions to Eq.~(\ref{2.30}) are
$\Delta^+_{1} = \bar{r} + s$ and $\Delta^-_{1} = \bar{r} - s$. The 
remaining equations produce $\Delta_2 = 0$, 
\begin{equation} 
\Delta_3 = \frac{R_{u\sigma u\sigma} \Delta_1^2}
{6 (\bar{r} - \Delta_1)}, 
\label{2.31}
\end{equation} 
and 
\begin{equation} 
\Delta_4 = \frac{ \bigl( \Delta_1 R_{u\sigma u\sigma | u} 
- R_{u\sigma u\sigma | \sigma} \bigr) \Delta_1^2} 
{24 (\bar{r} - \Delta_1)}. 
\label{2.32}
\end{equation} 
The series expansion for $\Delta$ is now determined to the required 
degree of accuracy.  

Collecting our results, we conclude that $\Delta_\pm$ is given by   
\begin{equation} 
\Delta_\pm = (\bar{r} \pm s) \mp \frac{(\bar{r} \pm s)^2}{6s} 
R_{u\sigma u\sigma} \mp \frac{(\bar{r} \pm s)^2}{24s}
\Bigl[ (\bar{r} \pm s) R_{u\sigma u\sigma | u} 
- R_{u\sigma u\sigma | \sigma} \Bigr] + O(\epsilon^5). 
\label{2.33}
\end{equation} 
This determines the positions of the retarded point $x' = z(u) = 
z(\bar{\tau} + \Delta_-)$ and the advanced point $x'' = z(v) = 
z(\bar{\tau} + \Delta_+)$ relative to the reference point
$\bar{x} = z(\bar{\tau})$ on the world line. We note that in
Eq.~(\ref{2.33}), the first term on the right-hand side is of order
$\epsilon$, the second term is of order $\epsilon^3$, and the third
term (involving the square brackets) is of order $\epsilon^4$. We
recall the notation introduced below Eq.~(\ref{2.25}): In
Eq.~(\ref{2.33}),  
\begin{equation} 
R_{u\sigma u\sigma} := 
R_{\bar{\alpha}\bar{\mu}\bar{\beta}\bar{\nu}} u^{\bar{\alpha}}
\sigma^{\bar{\mu}} u^{\bar{\beta}} \sigma^{\bar{\nu}}
\label{2.34}
\end{equation} 
and 
\begin{equation} 
R_{u\sigma u\sigma | \sigma} := 
R_{\bar{\alpha}\bar{\mu}\bar{\beta}\bar{\nu};\bar{\lambda}}
u^{\bar{\alpha}} \sigma^{\bar{\mu}} u^{\bar{\beta}} \sigma^{\bar{\nu}} 
\sigma^{\bar{\lambda}}.  
\label{2.35}
\end{equation} 
The notation is unambiguous and easily adaptable to other projections 
of the Riemann tensor; for example, $R_{u\sigma u\sigma | u} := 
R_{\bar{\alpha}\bar{\mu}\bar{\beta}\bar{\nu};\bar{\gamma}}
u^{\bar{\alpha}} \sigma^{\bar{\mu}} u^{\bar{\beta}} \sigma^{\bar{\nu}} 
u^{\bar{\gamma}}$ also appears in Eq.~(\ref{2.33}).      

\subsection{Calculation of $r$ and $r_{\rm adv}$} 

We now wish to express $r$ and $r_{\rm adv}$ in terms of tensorial
quantities that are evaluated at $\bar{x}$. Once more our strategy 
is to perform a Taylor expansion around $\tau = \bar{\tau}$. From  
Eqs.~(\ref{2.2}) and (\ref{2.23}) we obtain $r = \dot{\sigma}(u)$,
which may be expanded as 
\begin{equation} 
r = \dot{\sigma} + \ddot{\sigma} \Delta_- 
+ \frac{1}{2} \sigma^{(3)} \Delta_-^2  
+ \frac{1}{6} \sigma^{(4)} \Delta_-^3 + \cdots, 
\label{2.36}
\end{equation}
where $\Delta_- = u - \bar{\tau}$ and where all derivatives of
$\sigma(\tau)$ are evaluated at $\tau = \bar{\tau}$. Similarly,   
Eqs.~(\ref{2.3}) and (\ref{2.23}) give $r_{\rm adv} = 
-\dot{\sigma}(v)$ and 
\begin{equation} 
r_{\rm adv} = -\dot{\sigma} - \ddot{\sigma} \Delta_+ 
- \frac{1}{2} \sigma^{(3)} \Delta_+^2  
- \frac{1}{6} \sigma^{(4)} \Delta_+^3 + \cdots, 
\label{2.37}
\end{equation}
where $\Delta_+ = v - \bar{\tau}$. We recall that $\dot{\sigma} =
\bar{r}$, and that expressions for the higher derivatives of
$\sigma(\tau)$ were obtained in Eqs.~(\ref{2.25})--(\ref{2.27}). 
Furthermore, Eq.~(\ref{2.33}) gives $\Delta_\pm$ as an expansion in
powers of $\epsilon$. Making these substitutions into
Eqs.~(\ref{2.36}) and (\ref{2.37}) produces, after some
simplification,  
\begin{equation} 
r = s - \frac{\bar{r}^2-s^2}{6s} R_{u\sigma u\sigma} 
- \frac{\bar{r}-s}{24s} \Bigl[ (\bar{r}-s)(\bar{r}+2s)
  R_{u\sigma u\sigma | u} 
- (\bar{r}+s) R_{u\sigma u\sigma | \sigma} \Bigr] 
+ O(\epsilon^5) 
\label{2.38}
\end{equation} 
and 
\begin{equation} 
r_{\rm adv} = s - \frac{\bar{r}^2-s^2}{6s} R_{u\sigma u\sigma}  
- \frac{\bar{r}+s}{24s} \Bigl[ (\bar{r}+s)(\bar{r}-2s)
  R_{u\sigma u\sigma | u} 
- (\bar{r}-s) R_{u\sigma u\sigma | \sigma} \Bigr] 
+ O(\epsilon^5).  
\label{2.39}
\end{equation} 
We note that in Eqs.~(\ref{2.38}) and (\ref{2.39}), the first term on
the right-hand side is of order $\epsilon$, the second term is of
order $\epsilon^3$, and the third term (involving the square brackets)
is of order $\epsilon^4$. Notice also that the difference between $r$
and $r_{\rm adv}$ is of order $\epsilon^4$.  

\subsection{Calculation of $\sigma_{\alpha}(x,x')$ and
  $\sigma_{\alpha}(x,x'')$} 
 
We continue to keep $x$ fixed and introduce the vector-valued function  
\begin{equation} 
\sigma_\alpha(\tau) := \sigma_\alpha\bigl(x,z(\tau)\bigr) 
\label{2.40}
\end{equation} 
on the world line; the vectorial index refers to the fixed point $x$,
and $\sigma_\alpha(\tau)$ is a set of four {\it scalar functions} of
the argument $\tau$. In terms of this we have $\sigma_\alpha(x,x') =
\sigma_\alpha(u)$ and $\sigma_\alpha(x,x'') = \sigma_\alpha(v)$, and
we wish to express these in terms of tensorial quantities evaluated at
$\bar{x}$. Letting $w$ stand for either $u$ or $v$, and re-introducing
$\Delta := w - \bar{\tau}$, Taylor expansion gives 
\begin{equation} 
\sigma_\alpha(w) = \bar{\sigma}_\alpha + \dot{\sigma}_\alpha \Delta 
+ \frac{1}{2} \ddot{\sigma}_\alpha \Delta^2 
+ \frac{1}{6} {\sigma}^{(3)}_\alpha \Delta^3
+ \frac{1}{24} {\sigma}^{(4)}_\alpha \Delta^4 + \cdots, 
\label{2.41} 
\end{equation} 
where $\bar{\sigma}_\alpha = \sigma_\alpha(\bar{\tau})$, and where
all derivatives are evaluated at $\tau = \bar{\tau}$. We shall
evaluate each term on the right-hand side of Eq.~(\ref{2.41}). We 
rely on results obtained in Sec.~II C, as well as the standard
bitensorial expansion (see, for example, Ref.~\cite{christensen:76}) 
\begin{equation} 
\sigma_{\bar{\alpha}\beta}(x,\bar{x}) = g^{\bar{\beta}}_{\ \beta}
\Bigl[ - g_{\bar{\alpha}\bar{\beta}} - \frac{1}{6}
R_{\bar{\alpha}\bar{\gamma}\bar{\beta}\bar{\delta}}
  \sigma^{\bar{\gamma}} \sigma^{\bar{\delta}} + \frac{1}{12}  
R_{\bar{\alpha}\bar{\gamma}\bar{\beta}\bar{\delta};\bar{\epsilon}} 
  \sigma^{\bar{\gamma}} \sigma^{\bar{\delta}} \sigma^{\bar{\epsilon}}
  + O(\epsilon^4) \Bigr], 
\label{2.42}
\end{equation} 
where $g^{\bar{\beta}}_{\ \beta}(x,\bar{x})$ is the parallel
propagator, which takes a vector at $x$ and carries it to $\bar{x}$  
by parallel transport. 

We begin by recalling the identity $\sigma_\alpha(x,\bar{x}) = 
-g^{\bar{\alpha}}_{\ \alpha}(x,\bar{x}) 
\sigma_{\bar{\alpha}}(x,\bar{x})$, which follows from the geometrical
interpretation of $\sigma_\alpha$ and $-\sigma_{\bar{\alpha}}$ as
tangent vectors on the spacelike geodesic that links the points $x$
and $\bar{x}$. The identity allows us to write $\bar{\sigma}_\alpha$
as   
\begin{equation} 
\bar{\sigma}_\alpha = -g^{\bar{\alpha}}_{\ \alpha} 
\sigma_{\bar{\alpha}}. 
\label{2.43}
\end{equation} 
The derivative of $\sigma_\alpha(\tau)$ is given by
$\dot{\sigma}_\alpha = \sigma_{\alpha \bar{\alpha}} u^{\bar{\alpha}}$,
and involving Eq.~(\ref{2.42}) produces 
\begin{equation} 
\dot{\sigma}_\alpha = g^{\bar{\alpha}}_{\ \alpha} \biggl[ 
- u_{\bar{\alpha}} - \frac{1}{6} R_{\bar{\alpha}\sigma u\sigma} 
+ \frac{1}{12} R_{\bar{\alpha}\sigma u\sigma | \sigma} 
+ O(\epsilon^4) \biggr], 
\label{2.44}
\end{equation} 
where we have introduced a notation similar to that of
Eqs.~(\ref{2.34}) and (\ref{2.35}): $R_{\bar{\alpha}\sigma u \sigma}
:= R_{\bar{\alpha}\bar{\mu}\bar{\beta}\bar{\nu}} \sigma^{\bar{\mu}} 
u^{\bar{\beta}} \sigma^{\bar{\nu}}$ and 
$R_{\bar{\alpha}\sigma u\sigma | \sigma} := 
R_{\bar{\alpha}\bar{\mu}\bar{\beta}\bar{\nu};\bar{\lambda}}
\sigma^{\bar{\mu}} u^{\bar{\beta}} \sigma^{\bar{\nu}} 
\sigma^{\bar{\lambda}}$. 

The second derivative of $\sigma_\alpha(\tau)$ is 
$\ddot{\sigma}_\alpha = \sigma_{\alpha \bar{\alpha}\bar{\beta}}
u^{\bar{\alpha}} u^{\bar{\beta}}$, and an expression for
$\sigma_{\alpha \bar{\alpha}\bar{\beta}}$ can be obtained by
differentiating Eq.~(\ref{2.22}) with respect to $x^\alpha$. This
expression is simplified with the help of Eq.~(\ref{2.42}), which
can be truncated to $\sigma^{\bar{\mu}}_{\ \alpha} 
= -g^{\bar{\mu}}_{\ \alpha} + O(\epsilon^2)$ for the purposes of this
computation. The end result, after contracting with $u^{\bar{\alpha}}
u^{\bar{\beta}}$ and invoking the symmetries of the Riemann tensor, is  
\begin{equation} 
\ddot{\sigma}_\alpha = g^{\bar{\alpha}}_{\ \alpha} \biggl[ 
\frac{2}{3} R_{\bar{\alpha}u\sigma u} 
- \frac{1}{12} \bigl( 3R_{\bar{\alpha}u\sigma u | \sigma}  
+ R_{\bar{\alpha}\sigma u \sigma | u} \bigr) 
+ O(\epsilon^3) \biggr]. 
\label{2.45}
\end{equation} 
Similar computations for the third and fourth derivatives produce 
\begin{equation} 
{\sigma}^{(3)}_\alpha = g^{\bar{\alpha}}_{\ \alpha} \biggl[ 
\frac{1}{2} R_{\bar{\alpha}u\sigma u | u} 
+ O(\epsilon^2) \biggr] 
\label{2.46}
\end{equation} 
and 
\begin{equation} 
{\sigma}^{(4)}_\alpha = O(\epsilon).  
\label{2.47} 
\end{equation} 

Equations (\ref{2.43})--(\ref{2.47}) can now be incorporated into 
Eq.~(\ref{2.41}), in which we also substitute Eq.~(\ref{2.33}). After
simplification, the final results are 
\begin{eqnarray} 
\sigma_\alpha(x,x') &=& g^{\bar{\alpha}}_{\ \alpha} \biggl\{  
- \Bigl[ \sigma_{\bar{\alpha}} + (\bar{r}-s) u_{\bar{\alpha}} \Bigr]  
- \biggl[ \frac{1}{6} (\bar{r}-s) R_{\bar{\alpha}\sigma u\sigma} 
+ \frac{(\bar{r}-s)^2}{6s} R_{u\sigma u\sigma} u_{\bar{\alpha}} 
- \frac{1}{3} (\bar{r}-s)^2 R_{\bar{\alpha}u\sigma u} \biggr] 
\nonumber \\ & & \mbox{} 
+ \biggl[ 
\frac{1}{12} (\bar{r}-s) R_{\bar{\alpha}\sigma u \sigma|\sigma}  
- \frac{(\bar{r}-s)^2}{24s} \Bigl(
  (\bar{r}-s) R_{u\sigma u\sigma | u} 
  - R_{u\sigma u\sigma | \sigma} \Bigr) u_{\bar{\alpha}} 
\nonumber \\ & & \mbox{} 
- \frac{1}{24} (\bar{r}-s)^2 \bigl(3R_{\bar{\alpha}u\sigma u | \sigma}  
  + R_{\bar{\alpha}\sigma u \sigma | u} \bigr) 
+ \frac{1}{12} (\bar{r}-s)^3 R_{\bar{\alpha}u\sigma u | u} \biggr]  
+ O(\epsilon^5) \biggr\} 
\label{2.48} 
\end{eqnarray} 
and 
\begin{eqnarray} 
\sigma_\alpha(x,x'') &=& g^{\bar{\alpha}}_{\ \alpha} \biggl\{  
- \Bigl[ \sigma_{\bar{\alpha}} + (\bar{r}+s) u_{\bar{\alpha}} \Bigr]  
- \biggl[ \frac{1}{6} (\bar{r}+s) R_{\bar{\alpha}\sigma u\sigma} 
- \frac{(\bar{r}+s)^2}{6s} R_{u\sigma u\sigma} u_{\bar{\alpha}} 
- \frac{1}{3} (\bar{r}+s)^2 R_{\bar{\alpha}u\sigma u} \biggr]  
\nonumber \\ & & \mbox{} 
+ \biggl[
\frac{1}{12} (\bar{r}+s) R_{\bar{\alpha}\sigma u \sigma|\sigma}  
+ \frac{(\bar{r}+s)^2}{24s} \Bigl( 
  (\bar{r}+s) R_{u\sigma u\sigma | u} 
  - R_{u\sigma u\sigma | \sigma} \Bigr) u_{\bar{\alpha}} 
\nonumber \\ & & \mbox{} 
- \frac{1}{24} (\bar{r}+s)^2 \bigl(3R_{\bar{\alpha}u\sigma u | \sigma}  
  + R_{\bar{\alpha}\sigma u \sigma | u} \bigr) 
+ \frac{1}{12} (\bar{r}+s)^3 R_{\bar{\alpha}u\sigma u | u} \biggr]  
+ O(\epsilon^5) \biggr\}.  
\label{2.49} 
\end{eqnarray} 
In these equations, terms grouped within square brackets are of the
same order of magnitude: The first group of terms is of order 
$\epsilon$, the second group is of order $\epsilon^3$, and the third
group is of order $\epsilon^4$. We recall the notation introduced
below Eq.~(\ref{2.44}): In Eqs.~(\ref{2.48}) and (\ref{2.49}), the
various partial projections of the Riemann tensor are given by
equations of the form   
\begin{equation} 
R_{\bar{\alpha}\sigma u \sigma} := 
R_{\bar{\alpha}\bar{\mu}\bar{\beta}\bar{\nu}} \sigma^{\bar{\mu}} 
u^{\bar{\beta}} \sigma^{\bar{\nu}}
\label{2.50}
\end{equation} 
and 
\begin{equation} 
R_{\bar{\alpha}\sigma u\sigma | \sigma} := 
R_{\bar{\alpha}\bar{\mu}\bar{\beta}\bar{\nu};\bar{\lambda}}
\sigma^{\bar{\mu}} u^{\bar{\beta}} \sigma^{\bar{\nu}} 
\sigma^{\bar{\lambda}}. 
\label{2.51}
\end{equation} 
The full projections of the Riemann tensor have already been
introduced in Eqs.~(\ref{2.34}) and (\ref{2.35}). 

\subsection{Calculation of $\sigma_{\alpha\alpha'}u^{\alpha'}$ and
  $\sigma_{\alpha\alpha''}u^{\alpha''}$} 
 
In terms of the vector-valued function $\sigma_\alpha(\tau)$
introduced in Eq.~(\ref{2.40}), we have that 
$\sigma_{\alpha\alpha'}u^{\alpha'} =
\dot{\sigma}_\alpha(u)$ and $\sigma_{\alpha\alpha''}u^{\alpha''} =
\dot{\sigma}_\alpha(v)$. With $w$ standing for either $u$ or $v$, and
with $\Delta := w - \bar{\tau}$, Taylor expansion gives 
\begin{equation} 
\dot{\sigma}_\alpha(w) = \dot{\sigma}_\alpha  
+ \ddot{\sigma}_\alpha \Delta 
+ \frac{1}{2} {\sigma}^{(3)}_\alpha \Delta^2
+ \frac{1}{6} {\sigma}^{(4)}_\alpha \Delta^3 + \cdots, 
\label{2.52} 
\end{equation} 
in which all derivatives of $\sigma_\alpha(\tau)$ are evaluated at
$\tau = \bar{\tau}$. These quantities are displayed in
Eqs.~(\ref{2.44})--(\ref{2.47}), and Eq.~(\ref{2.33}) provides an
expression for $\Delta$. Substitution and simplification yields 
\begin{eqnarray} 
\sigma_{\alpha \alpha'} u^{\alpha'} &=& 
g^{\bar{\alpha}}_{\ \alpha} \biggl\{  
- u_{\bar{\alpha}} 
- \biggl[ \frac{1}{6} R_{\bar{\alpha}\sigma u\sigma} 
- \frac{2}{3} (\bar{r}-s) R_{\bar{\alpha}u\sigma u} \biggr]  
+ \biggl[ \frac{1}{12} R_{\bar{\alpha}\sigma u \sigma | \sigma}   
\nonumber \\ & & \mbox{} 
- \frac{1}{12} (\bar{r}-s) \bigl(3R_{\bar{\alpha}u\sigma u | \sigma}  
  + R_{\bar{\alpha}\sigma u \sigma | u} \bigr) 
+ \frac{1}{4} (\bar{r}-s)^2 R_{\bar{\alpha}u\sigma u | u} \biggr] 
+ O(\epsilon^4) \biggr\} 
\label{2.53} 
\end{eqnarray} 
and 
\begin{eqnarray} 
\sigma_{\alpha \alpha''} u^{\alpha''} &=& 
g^{\bar{\alpha}}_{\ \alpha} \biggl\{  
- u_{\bar{\alpha}} 
- \biggl[ \frac{1}{6} R_{\bar{\alpha}\sigma u\sigma} 
- \frac{2}{3} (\bar{r}+s) R_{\bar{\alpha}u\sigma u} \biggr] 
+ \biggl[ \frac{1}{12} R_{\bar{\alpha}\sigma u \sigma | \sigma}  
\nonumber \\ & & \mbox{} 
- \frac{1}{12} (\bar{r}+s) \bigl(3R_{\bar{\alpha}u\sigma u | \sigma}  
  + R_{\bar{\alpha}\sigma u \sigma | u} \bigr) 
+ \frac{1}{4} (\bar{r}+s)^2 R_{\bar{\alpha}u\sigma u | u} \biggr] 
+ O(\epsilon^4) \biggr\}.  
\label{2.54} 
\end{eqnarray} 
In these equations, the first term on the right-hand side is of order
$\epsilon^0$, the second group of terms is of order $\epsilon^2$, and
the third group is of order $\epsilon^3$. 

\subsection{Calculation of $\sigma_{\alpha'\beta'}u^{\alpha'}
  u^{\beta'}$ and $\sigma_{\alpha''\beta''}u^{\alpha''}
  u^{\beta''}$}  

Returning to the function $\sigma(\tau)$ defined by Eq.~(\ref{2.23}),
we have that $\sigma_{\alpha'\beta'}u^{\alpha'} u^{\beta'} =
\ddot{\sigma}(u)$ and $\sigma_{\alpha''\beta''}u^{\alpha''}
u^{\beta''} = \ddot{\sigma}(v)$. Taylor expansion gives 
\begin{equation} 
\ddot{\sigma}(w) = \ddot{\sigma} + {\sigma}^{(3)} \Delta
+ \frac{1}{2} {\sigma}^{(4)} \Delta^2 + \cdots.  
\label{2.55} 
\end{equation} 
Substitution of Eqs.~(\ref{2.33}) and (\ref{2.25})--(\ref{2.27}) 
produces 
\begin{equation} 
\sigma_{\alpha'\beta'}u^{\alpha'} u^{\beta'} = -1 
- \frac{1}{3} R_{u\sigma u\sigma} 
+ \frac{1}{12} \Bigl[ R_{u\sigma u\sigma | \sigma} 
  - 3(\bar{r}-s) R_{u\sigma u\sigma | u} \Bigr]  
+ O(\epsilon^4) 
\label{2.56}
\end{equation} 
and 
\begin{equation} 
\sigma_{\alpha''\beta''}u^{\alpha''} u^{\beta''} = -1 
- \frac{1}{3} R_{u\sigma u\sigma} 
+ \frac{1}{12} \Bigl[ R_{u\sigma u\sigma | \sigma} 
  - 3(\bar{r}+s) R_{u\sigma u\sigma | u} \Bigr]  
+ O(\epsilon^4). 
\label{2.57}
\end{equation} 
In these equations, the first term on the right-hand side is of order
$\epsilon^0$, the second term is of order $\epsilon^2$, and the
bracketed terms are of order $\epsilon^3$.  

\subsection{Calculation of $\nabla_\alpha u$ and $\nabla_\alpha v$} 

According to Eq.~(\ref{2.4}), the gradient of the retarded time
function $u(x)$ is $\nabla_\alpha u = -\sigma_\alpha(x,x')/r$, and 
according to Eq.~(\ref{2.5}), the gradient of the advanced time
function $v(x)$ is $\nabla_\alpha v 
= \sigma_\alpha(x,x'')/r_{\rm adv}$. Expansions of $r$ and 
$r_{\rm adv}$ in powers of $\epsilon$ were obtained in Sec.~II
D and presented in Eqs.~(\ref{2.38}) and (\ref{2.39}); these can
easily be converted into expansions for $1/r$ and 
$1/r_{\rm adv}$. Expansions for $\sigma_\alpha(x,x')$ and 
$\sigma_\alpha(x,x'')$ were developed in Sec.~II E and presented in
Eqs.~(\ref{2.48}) and (\ref{2.49}). Combining these expansions
produces  
\begin{eqnarray} 
\nabla_\alpha u &=& \frac{1}{s} g^{\bar{\alpha}}_{\ \alpha} \biggl\{   
\Bigl[ \sigma_{\bar{\alpha}} + (\bar{r}-s) u_{\bar{\alpha}} \Bigr] 
+ \biggl[ \frac{1}{6} (\bar{r}-s) R_{\bar{\alpha} \sigma u \sigma}  
- \frac{1}{3} (\bar{r}-s)^2 R_{\bar{\alpha} u \sigma u} 
+ \frac{\bar{r}^2-s^2}{6s^2} R_{u\sigma u\sigma} \sigma_{\bar{\alpha}} 
\nonumber \\ & & \mbox{}
+ \frac{(\bar{r}-s)^2(\bar{r}+2s)}{6s^2} R_{u\sigma u\sigma}
   u_{\bar{\alpha}} \biggr] 
+ \biggl[ -\frac{1}{12} (\bar{r}-s) 
          R_{\bar{\alpha}\sigma u \sigma|\sigma}    
+ \frac{1}{8} (\bar{r}-s)^2 R_{\bar{\alpha}u \sigma u |\sigma}  
+ \frac{1}{24} (\bar{r}-s)^2 R_{\bar{\alpha}\sigma u \sigma | u}  
\nonumber \\ & & \mbox{}
- \frac{1}{12} (\bar{r}-s)^3 R_{\bar{\alpha}u \sigma u | u}  
+ \frac{\bar{r}-s}{24s^2} \Bigl( 
  (\bar{r}-s)(\bar{r}+2s) R_{u\sigma u\sigma | u} 
  - (\bar{r}+s) R_{u\sigma u\sigma | \sigma} \Bigr)
  \sigma_{\bar{\alpha}}  
\nonumber \\ & & \mbox{}
+ \frac{(\bar{r}-s)^2}{24s^2} \Bigl( 
  (\bar{r}-s)(\bar{r}+3s) R_{u\sigma u\sigma | u} 
  - (\bar{r}+2s) R_{u\sigma u\sigma | \sigma} \Bigr)
  u_{\bar{\alpha}} \biggr] 
+ O(\epsilon^5) \biggr\} 
\label{2.58} 
\end{eqnarray} 
and 
\begin{eqnarray} 
\nabla_\alpha v &=& -\frac{1}{s} g^{\bar{\alpha}}_{\ \alpha} \biggl\{   
\Bigl[ \sigma_{\bar{\alpha}} + (\bar{r}+s) u_{\bar{\alpha}} \Bigr] 
+ \biggl[ \frac{1}{6} (\bar{r}+s) R_{\bar{\alpha} \sigma u \sigma} 
- \frac{1}{3} (\bar{r}+s)^2 R_{\bar{\alpha} u \sigma u} 
+ \frac{\bar{r}^2-s^2}{6s^2} R_{u\sigma u\sigma} \sigma_{\bar{\alpha}} 
\nonumber \\ & & \mbox{}
+ \frac{(\bar{r}+s)^2(\bar{r}-2s)}{6s^2} R_{u\sigma u\sigma}
   u_{\bar{\alpha}} \biggr]  
+ \biggl[ -\frac{1}{12} (\bar{r}+s) 
          R_{\bar{\alpha}\sigma u \sigma|\sigma}  
+ \frac{1}{8} (\bar{r}+s)^2 R_{\bar{\alpha}u \sigma u |\sigma}  
+ \frac{1}{24} (\bar{r}+s)^2 R_{\bar{\alpha}\sigma u \sigma | u}  
\nonumber \\ & & \mbox{}
- \frac{1}{12} (\bar{r}+s)^3 R_{\bar{\alpha}u \sigma u | u}  
+ \frac{\bar{r}+s}{24s^2} \Bigl( 
  (\bar{r}+s)(\bar{r}-2s) R_{u\sigma u\sigma | u} 
  - (\bar{r}-s) R_{u\sigma u\sigma | \sigma} \Bigr)
  \sigma_{\bar{\alpha}}  
\nonumber \\ & & \mbox{}
+ \frac{(\bar{r}+s)^2}{24s^2} \Bigl( 
  (\bar{r}+s)(\bar{r}-3s) R_{u\sigma u\sigma | u} 
  - (\bar{r}-2s) R_{u\sigma u\sigma | \sigma} \Bigr)
  u_{\bar{\alpha}} \biggr]  
+ O(\epsilon^5) \biggr\}. 
\label{2.59} 
\end{eqnarray} 
Once more we have grouped terms of the same order of magnitude. The
first group of (square-bracketed) terms within the curly brackets is
of order $\epsilon$, the second group is of order $\epsilon^3$, and
the third group is of order $\epsilon^4$. Noticing the factor $s^{-1}$
in front of the curly brackets, this means that $\nabla_\alpha u$ and
$\nabla_\alpha v$ contain terms of order $\epsilon^0$, $\epsilon^2$,
and $\epsilon^3$; the neglected terms are $O(\epsilon^4)$.   

\subsection{Calculation of $\nabla_\alpha r$ and 
$\nabla_\alpha r_{\rm adv}$} 

The gradients of the retarded and advanced distance functions were 
defined in Eqs.~(\ref{2.6}) and (\ref{2.7}); for example, 
$\nabla_\alpha r = \sigma_{\alpha'\beta'} u^{\alpha'} u^{\beta'}
\nabla_\alpha u + \sigma_{\alpha'\alpha} u^{\alpha'}$. Each 
piece of this expression was calculated separately in the preceding   
subsections: $\sigma_{\alpha'\beta'} u^{\alpha'} u^{\beta'}$ was
computed in Sec.~II G, $\nabla_\alpha u$ was computed in Sec.~II H,
and $\sigma_{\alpha'\alpha} u^{\alpha'}$ was computed in Sec.~II
F. Combining all these results, we arrive at  
\begin{eqnarray} 
\nabla_\alpha r &=& -\frac{1}{s} g^{\bar{\alpha}}_{\ \alpha} \biggl\{   
\Bigl[ \sigma_{\bar{\alpha}} + \bar{r} u_{\bar{\alpha}} \Bigr] 
+ \biggl[ \frac{1}{6} \bar{r} R_{\bar{\alpha} \sigma u \sigma} 
- \frac{1}{3} (\bar{r}^2-s^2) R_{\bar{\alpha} u \sigma u} 
+ \frac{\bar{r}^2+s^2}{6s^2} R_{u\sigma u\sigma} \sigma_{\bar{\alpha}} 
+ \frac{\bar{r}(\bar{r}^2-s^2)}{6s^2} R_{u\sigma u\sigma}
   u_{\bar{\alpha}} \biggr]  
\nonumber \\ & & \mbox{}
+ \biggl[ -\frac{1}{12} \bar{r} 
          R_{\bar{\alpha}\sigma u \sigma|\sigma}  
+ \frac{1}{8} (\bar{r}^2-s^2) R_{\bar{\alpha}u \sigma u |\sigma}  
+ \frac{1}{24} (\bar{r}^2-s^2) R_{\bar{\alpha}\sigma u \sigma | u}  
- \frac{1}{12} (\bar{r}-s)^2(\bar{r}+2s) 
  R_{\bar{\alpha}u \sigma u | u}   
\nonumber \\ & & \mbox{}
+ \frac{1}{24s^2} \Bigl( 
  (\bar{r}-s)(\bar{r}^2+\bar{r}s+4s^2) R_{u\sigma u\sigma | u}  
  - (\bar{r}^2+s^2) R_{u\sigma u\sigma | \sigma} \Bigr)
  \sigma_{\bar{\alpha}}  
\nonumber \\ & & \mbox{}
+ \frac{\bar{r}-s}{24s^2} \Bigl(
  (\bar{r}-s)(\bar{r}^2+2\bar{r}s+3s^2) R_{u\sigma u\sigma | u}  
  - \bar{r}(\bar{r}+s) R_{u\sigma u\sigma | \sigma} \Bigr)
  u_{\bar{\alpha}} \biggr]  
+ O(\epsilon^5) \biggr\} 
\label{2.60} 
\end{eqnarray} 
and 
\begin{eqnarray} 
\nabla_\alpha r_{\rm adv} &=& -\frac{1}{s} g^{\bar{\alpha}}_{\ \alpha}
\biggl\{ \Bigl[ \sigma_{\bar{\alpha}} + \bar{r} u_{\bar{\alpha}}
  \Bigr] 
+ \biggl[ \frac{1}{6} \bar{r} R_{\bar{\alpha} \sigma u \sigma}  
- \frac{1}{3} (\bar{r}^2-s^2) R_{\bar{\alpha} u \sigma u} 
+ \frac{\bar{r}^2+s^2}{6s^2} R_{u\sigma u\sigma} \sigma_{\bar{\alpha}} 
+ \frac{\bar{r}(\bar{r}^2-s^2)}{6s^2} R_{u\sigma u\sigma}
   u_{\bar{\alpha}} \biggr] 
\nonumber \\ & & \mbox{}
+ \biggl[ -\frac{1}{12} \bar{r} 
          R_{\bar{\alpha}\sigma u \sigma|\sigma}  
+ \frac{1}{8} (\bar{r}^2-s^2) R_{\bar{\alpha}u \sigma u |\sigma}  
+ \frac{1}{24} (\bar{r}^2-s^2) R_{\bar{\alpha}\sigma u \sigma | u}  
- \frac{1}{12} (\bar{r}+s)^2(\bar{r}-2s) 
  R_{\bar{\alpha}u \sigma u | u}   
\nonumber \\ & & \mbox{}
+ \frac{1}{24s^2} \Bigl( 
  (\bar{r}+s)(\bar{r}^2-\bar{r}s+4s^2) R_{u\sigma u\sigma | u}  
  - (\bar{r}^2+s^2) R_{u\sigma u\sigma | \sigma} \Bigr)
  \sigma_{\bar{\alpha}}  
\nonumber \\ & & \mbox{}
+ \frac{\bar{r}+s}{24s^2} \Bigl(
  (\bar{r}+s)(\bar{r}^2-2\bar{r}s+3s^2) R_{u\sigma u\sigma | u}  
  - \bar{r}(\bar{r}-s) R_{u\sigma u\sigma | \sigma} \Bigr)
  u_{\bar{\alpha}} \biggr] 
+ O(\epsilon^5) \biggr\}.  
\label{2.61} 
\end{eqnarray} 
In these equations, the first group of (square-bracketed) terms within
the curly brackets is of order $\epsilon$, the second group is of
order $\epsilon^3$, and the third group is of order
$\epsilon^4$. Noticing the common factor $s^{-1}$, this means that
$\nabla_\alpha r$ and $\nabla_\alpha r_{\rm adv}$ contain terms of
order $\epsilon^0$, $\epsilon^2$, and $\epsilon^3$; the neglected
terms are $O(\epsilon^4)$.  

\subsection{Final result: The singular field} 

Our final expression for the singular field is obtained by
substituting the expansions of Eqs.~(\ref{2.38}), (\ref{2.39}),
(\ref{2.60}), and (\ref{2.61}) into Eq.~(\ref{2.17}). After a long
computation and much simplification, we obtain 
\begin{eqnarray} 
\Phi^{\rm S}_\alpha(x) &=& \frac{q}{s^3} 
g^{\bar{\alpha}}_{\ \alpha} \biggl\{   
\Bigl[ \sigma_{\bar{\alpha}} + \bar{r} u_{\bar{\alpha}} \Bigr] 
+ \biggl[ \frac{1}{6} \bar{r} R_{\bar{\alpha} \sigma u \sigma} 
- \frac{1}{3} (\bar{r}^2-s^2) R_{\bar{\alpha} u \sigma u} 
+ \frac{3\bar{r}^2-s^2}{6s^2} R_{u\sigma u\sigma}
  \sigma_{\bar{\alpha}}  
+ \frac{\bar{r}(\bar{r}^2-s^2)}{2s^2} R_{u\sigma u\sigma}
   u_{\bar{\alpha}} \biggr] 
\nonumber \\ & & \mbox{}
+ \biggl[ -\frac{1}{12} \bar{r} 
          R_{\bar{\alpha}\sigma u \sigma|\sigma}  
+ \frac{1}{8} (\bar{r}^2-s^2) R_{\bar{\alpha}u \sigma u |\sigma}  
+ \frac{1}{24} (\bar{r}^2-s^2) R_{\bar{\alpha}\sigma u \sigma | u}  
- \frac{1}{12} \bar{r}(\bar{r}^2-3s^2) 
  R_{\bar{\alpha}u \sigma u | u}   
\nonumber \\ & & \mbox{}
+ \frac{1}{24s^2} \Bigl( 
  3\bar{r}(\bar{r}^2-s^2) R_{u\sigma u\sigma | u}  
  - (3\bar{r}^2-s^2) R_{u\sigma u\sigma | \sigma} \Bigr)
  \sigma_{\bar{\alpha}}  
\nonumber \\ & & \mbox{}
+ \frac{\bar{r}^2-s^2}{8s^2} \Bigl(
  (\bar{r}^2-s^2) R_{u\sigma u\sigma | u}  
  - \bar{r}R_{u\sigma u\sigma | \sigma} \Bigr)
  u_{\bar{\alpha}} \biggr]  
+ O(\epsilon^5) \biggr\}.  
\label{2.62} 
\end{eqnarray} 
In this equation, terms grouped within square brackets are of the same
order of magnitude. The first group of terms is of order $\epsilon$,
the second group is of order $\epsilon^3$, and the third group is of
order $\epsilon^4$. Noticing the common factor $s^{-3}$, this means
that $\Phi_\alpha^{\rm S}$ contains terms of order $\epsilon^{-2}$, 
$\epsilon^0$, and $\epsilon^1$; the neglected terms are
$O(\epsilon^2)$.   

The dependence of $\Phi^{\rm S}_\alpha := \nabla_\alpha \Phi^{\rm S}$
on the field point $x$ is contained in the common factor 
$g^{\bar{\alpha}}_{\ \alpha}(x,\bar{x})$, and also within the many 
occurrences of $\sigma_{\bar{\alpha}}(x,\bar{x})$. This quantity
appears explicitly in Eq.~(\ref{2.62}), and it is involved in the
definitions of 
\[
\bar{r} := \sigma_{\bar{\alpha}} u^{\bar{\alpha}}
\] 
and  
\[
s^2 := (g^{\bar{\alpha}\bar{\beta}} + u^{\bar{\alpha}}
u^{\bar{\beta}}) \sigma_{\bar{\alpha}} \sigma_{\bar{\beta}}, 
\]
which were introduced in Eqs.~(\ref{2.18}) and (\ref{2.19}),
respectively. The gradient of the world function is also involved 
in the various projections of the Riemann tensor (and its covariant
derivative) that appear in Eq.~(\ref{2.62}). We recall the notation
introduced in the preceding subsections: A subscript $u$ indicates a
projection along $u^{\bar{\alpha}}$, a subscript $\sigma$ indicates a
projection along $\sigma^{\bar{\alpha}}(x,\bar{x})$, and a vertical
bar indicates that the projection involves the covariant derivative of
the Riemann tensor. For example, 
\[
R_{\bar{\alpha}\sigma u \sigma} := 
R_{\bar{\alpha}{\bar{\mu}}\bar{\beta}\bar{\nu}} \sigma^{\bar{\mu}}
u^{\bar{\beta}} \sigma^{\bar{\nu}}
\]
and 
\[
R_{\bar{\alpha}\sigma u \sigma | \sigma} := 
R_{\bar{\alpha}{\bar{\mu}}\bar{\beta}\bar{\nu};\bar{\lambda}}
\sigma^{\bar{\mu}} u^{\bar{\beta}} \sigma^{\bar{\nu}}
\sigma^{\bar{\lambda}}.
\]
The Riemann tensor, its derivatives, and the particle's velocity
vector $u^{\bar{\alpha}}$ are all evaluated at the reference
point $\bar{x}$ on the world line. We recall that $\bar{x}$ is chosen
to be in a spacelike relation with $x$, but that it is otherwise
arbitrary.    

\section{Coordinate expansions of bitensors} 

The calculations presented in Sec.~II culminated into an expansion of
the singular field $\Phi^{\rm S}_\alpha(x)$ in powers of $\epsilon$,
the distance between the field point $x$ and the reference point
$\bar{x}$ on the world line. This expansion is fully covariant, and
the dependence on $x$ is explicitly contained in 
$\sigma_{\bar{\alpha}}(x,\bar{x})$ and 
$g^{\bar{\alpha}}_{\ \alpha}(x,\bar{x})$. Our task in this section 
is to develop {\it coordinate expansions} for these two bitensors, in 
powers of 
\begin{equation} 
w^\alpha := x^\alpha - \bar{x}^\alpha, 
\label{3.1}
\end{equation}
the difference in the coordinate positions of the points $x$ and
$\bar{x}$. These expansions, of course, will not be covariant; they
will depend on the choice of coordinate system. When the coordinate
expansions to be obtained here are substituted into Eq.~(\ref{2.62})
for the singular field, the result will be an explicit expression for
the expanded $\Phi^{\rm S}_\alpha(x)$ in the adopted system of 
coordinates.   

\subsection{Description of the geodesic linking $x$ to $\bar{x}$} 

We assume that there exists a unique geodesic segment that
begins at $\bar{x}$ and ends at $x$. This geodesic segment is denoted
$\beta$, and it is described, in the adopted coordinate system, by the
parametric relations $p^\alpha(\lambda)$. We assume that the parameter
$\lambda$ is an affine parameter, and that it is limited to the
interval $0 \leq \lambda \leq 1$. We have that $p^\alpha(0) =
\bar{x}^\alpha$ and $p^\alpha(1) = x^\alpha$, where $\bar{x}^\alpha$
are the coordinates assigned to $\bar{x}$, while $x^\alpha$ are the
coordinates assigned to $x$. 

The functions $p^\alpha(\lambda)$ may be expressed as Taylor
expansions about $\lambda = 0$:   
\begin{equation} 
p^\alpha(\lambda) = p^\alpha(0) + \dot{p}^\alpha(0) \lambda 
+ \frac{1}{2} \ddot{p}^\alpha(0) \lambda^2 
+ \frac{1}{6} p^{\alpha(3)}(0) \lambda^3 
+ \frac{1}{24} p^{\alpha(4)}(0) \lambda^4 + \cdots, 
\label{3.2}
\end{equation}
in which overdots, or a number within brackets, indicate repeated
differentiation with respect to $\lambda$. Equation (\ref{3.2})
implies  
\begin{equation} 
\dot{p}^\alpha(\lambda) = \dot{p}^\alpha(0)  
+ \ddot{p}^\alpha(0) \lambda 
+ \frac{1}{2} p^{\alpha(3)}(0) \lambda^2 
+ \frac{1}{6} p^{\alpha(4)}(0) \lambda^3 + \cdots  
\label{3.3}
\end{equation}
and 
\begin{equation} 
\ddot{p}^\alpha(\lambda) = \ddot{p}^\alpha(0)  
+ p^{\alpha(3)}(0) \lambda 
+ \frac{1}{2} p^{\alpha(4)}(0) \lambda^2 + \cdots.   
\label{3.4}
\end{equation}
These quantities are linked by the geodesic equation, 
\begin{equation} 
\ddot{p}^\alpha(\lambda) + \Gamma^\alpha_{\ \beta\gamma}(\lambda)
\dot{p}^\beta(\lambda) \dot{p}^\gamma(\lambda) = 0, 
\label{3.5}
\end{equation} 
in which $\Gamma^\alpha_{\ \beta\gamma}(\lambda)$ is the Christoffel 
connection evaluated on $\beta$. 

The range of the affine parameter $\lambda$ was chosen to ensure   
that the tangent vector $\dot{p}^\alpha(\lambda)$ is intimately
related to the gradient of Synge's world function. In fact, as
reviewed in Sec.~2.1.3 of LRR \cite{poisson:04b} --- see in particular
Eqs.~(55) and (56) --- we have that $\sigma_\alpha(x,\bar{x}) =
g_{\alpha\beta}(x) \dot{p}^\alpha(1)$ and 
\begin{equation} 
\sigma_{\bar{\alpha}}(x,\bar{x}) =
-g_{\alpha\beta}(\bar{x}) \dot{p}^\alpha(0). 
\label{3.6}
\end{equation} 

\subsection{Calculational strategy} 

Our first goal in this section is to obtain an expansion of  
$\sigma_{\bar{\alpha}}(x,\bar{x})$ in powers of the coordinate
difference of Eq.~(\ref{3.1}). This is given by $w^\alpha =
p^\alpha(1) - p^\alpha(0)$, or   
\begin{equation} 
w^\alpha = \dot{p}^\alpha(0) + \frac{1}{2} \ddot{p}^\alpha(0)  
+ \frac{1}{6} p^{\alpha(3)}(0) + \frac{1}{24} p^{\alpha(4)}(0) +
\cdots.
\label{3.7}
\end{equation}
We will achieve this in four steps. First (Sec.~II C), we substitute 
Eqs.~(\ref{3.3}) and (\ref{3.4}) into Eq.~(\ref{3.5}) and solve for 
$\ddot{p}^\alpha(0)$, $p^{\alpha(3)}(0)$, and $p^{\alpha(4)}(0)$ in
terms of $\dot{p}^\alpha(0)$. Second (Sec.~II D), we incorporate these
results into Eq.~(\ref{3.7}) and obtain $w^\alpha$ as an expansion 
in powers of $\dot{p}^\alpha(0)$. Third (also Sec.~II D), we invert
this series to obtain $\dot{p}^\alpha(0)$ expanded in powers of
$w^\alpha$. Finally (Sec.~II E), we substitute the result into
Eq.~(\ref{3.6}); our final expression for
$\sigma_{\bar{\alpha}}(x,\bar{x})$ is displayed 
in Eq.~(\ref{3.19}) below.   

Our second goal in this section is to obtain an expression for 
$g^{\bar{\alpha}}_{\ \alpha}(x,\bar{x})$ expanded in powers of
$w^\alpha$. This calculation is carried out in Secs.~II F and II G and
it follows a very similar strategy; our final expression for the
parallel propagator is displayed in Eq.~(\ref{3.30}) below. 

The calculations that follow rely on the expansions displayed
in Eqs.~(\ref{3.2})--(\ref{3.4}) and (\ref{3.7}). We also will need
the Taylor expansion of $\Gamma^\alpha_{\beta\gamma}(\lambda)$ about
$\lambda = 0$. Starting with 
\[
\Gamma^\alpha_{\ \beta\gamma}(\lambda) 
= \Gamma^\alpha_{\ \beta\gamma}(0) 
+ \Gamma^\alpha_{\ \beta\gamma,\mu}(0) 
  \bigl[ p^\mu(\lambda) - p^\mu(0) \bigr] 
+ \frac{1}{2} \Gamma^\alpha_{\ \beta\gamma,\mu\nu}(0) 
  \bigl[ p^\mu(\lambda) - p^\mu(0) \bigr] 
  \bigl[ p^\nu(\lambda) - p^\nu(0) \bigr] + \cdots
\]
and involving Eq.~(\ref{3.2}), we arrive at 
\begin{equation} 
\Gamma^\alpha_{\ \beta\gamma}(\lambda) 
= \Gamma^\alpha_{\ \beta\gamma}(0) 
+ \Bigl[ \Gamma^\alpha_{\ \beta\gamma,\mu}(0) \dot{p}^\mu(0) \Bigr] 
  \lambda 
+ \frac{1}{2} \Bigl[ \Gamma^\alpha_{\ \beta\gamma,\mu\nu}(0) 
  \dot{p}^\mu(0) \dot{p}^\nu(0) 
  + \Gamma^\alpha_{\ \beta\gamma,\mu}(0) \ddot{p}^\mu(0) \Bigr] 
  \lambda^2 + \cdots,   
\label{3.8}
\end{equation} 
in which the connection and its derivatives are evaluated at
$\bar{x}$ on the right-hand side of the equation.    

\subsection{Calculation of $\ddot{p}^\alpha(0)$, $p^{\alpha(3)}(0)$,
  and $p^{\alpha(4)}(0)$} 

We substitute Eqs.~(\ref{3.3}), (\ref{3.4}), and (\ref{3.8}) into
Eq.~(\ref{3.5}) and collect terms that share the same power of
$\lambda$. Setting the coefficient of the $\lambda^0$ term to zero
yields the condition $0 = \ddot{p}^\alpha(0) 
+ \Gamma^\alpha_{\ \beta\gamma}(0) \dot{p}^\beta(0)
\dot{p}^\gamma(0)$, or  
\begin{equation} 
\ddot{p}^\alpha(0) = -\Gamma^\alpha_{\ \beta\gamma}\, \dot{p}^\beta(0) 
\dot{p}^\gamma(0).  
\label{3.9}
\end{equation}
It is understood that here, the connection is evaluated at the 
reference point $\bar{x}$. The same comment will apply below to
all derivatives of the connection. 

Setting the coefficient of the $\lambda^1$ term to zero yields  
\[
0 = p^{\alpha(3)}(0) + 2\Gamma^\alpha_{\ \beta\gamma} \dot{p}^\beta(0)   
\ddot{p}^\gamma(0) + \Gamma^\alpha_{\ \beta\gamma,\mu} 
\dot{p}^\beta(0) \dot{p}^\gamma(0) \dot{p}^\mu(0), 
\]
and involving Eq.~(\ref{3.9}) gives 
\begin{equation} 
p^{\alpha(3)}(0) = -\Gamma^\alpha_{\ \beta\gamma\delta}\, 
\dot{p}^\beta(0) \dot{p}^\gamma(0) \dot{p}^\delta(0)
\label{3.10}
\end{equation}
with 
\begin{equation} 
\Gamma^\alpha_{\ \beta\gamma\delta} := 
\Gamma^\alpha_{\ \beta\gamma,\delta} 
- 2 \Gamma^\alpha_{\ \beta\mu} \Gamma^\mu_{\ \gamma\delta}. 
\label{3.11}
\end{equation} 

Setting the coefficient of the $\lambda^2$ term to zero yields 
\begin{eqnarray*} 
0 &=& p^{\alpha(4)}(0) + 2\Gamma^\alpha_{\ \beta\gamma} \bigl[  
\dot{p}^\beta(0) p^{\gamma(3)}(0)  
+ \ddot{p}^\beta(0) \ddot{p}^\gamma(0) \bigr] 
+ \Gamma^\alpha_{\ \beta\gamma,\mu} \bigl[ 
  4 \dot{p}^\beta(0) \ddot{p}^\gamma(0) \dot{p}^\mu(0)
  + \dot{p}^\beta(0) \dot{p}^\gamma(0) \ddot{p}^\mu(0) \bigr] 
\\ & & \mbox{} 
+ \Gamma^\alpha_{\ \beta\gamma,\mu\nu} \dot{p}^\beta(0)
  \dot{p}^\gamma(0) \dot{p}^\mu(0) \dot{p}^\nu(0), 
\end{eqnarray*}
and involving Eqs.~(\ref{3.9})--(\ref{3.11}) gives 
\begin{equation} 
p^{\alpha(4)}(0) = -\Gamma^\alpha_{\ \beta\gamma\delta\epsilon}\,  
\dot{p}^\beta(0) \dot{p}^\gamma(0) \dot{p}^\delta(0)
\dot{p}^\epsilon(0)
\label{3.12}
\end{equation} 
with 
\begin{equation} 
\Gamma^\alpha_{\ \beta\gamma\delta\epsilon} := 
\Gamma^\alpha_{\ \beta\gamma,\delta\epsilon} 
- 4 \Gamma^\alpha_{\ \beta\mu,\gamma} \Gamma^\mu_{\ \delta\epsilon} 
- \Gamma^\alpha_{\ \beta\gamma,\mu} \Gamma^\mu_{\ \delta\epsilon}
- 2\Gamma^\alpha_{\ \beta\mu} \Gamma^\mu_{\ \gamma\delta,\epsilon} 
+ 4 \Gamma^\alpha_{\ \beta\mu} \Gamma^\mu_{\ \gamma\nu}
  \Gamma^\nu_{\ \delta\epsilon} 
+ 2\Gamma^\alpha_{\ \mu\nu} \Gamma^\mu_{\ \beta\gamma} 
  \Gamma^\nu_{\ \delta\epsilon}. 
\label{3.13}
\end{equation} 

It should be noted that $\Gamma^\alpha_{\ \beta\gamma\delta}$ and 
$\Gamma^\alpha_{\ \beta\gamma\delta\epsilon}$, as defined by
Eqs.~(\ref{3.10}) and (\ref{3.12}), are both fully symmetric in their    
lower indices. This symmetry has not, however, been implemented on the
right-hand sides of Eqs.~(\ref{3.11}) and (\ref{3.13}). Although this
could easily be achieved, this operation is not necessary and we opt to
leave these expressions as they are.    

\subsection{Calculation of $\dot{p}^\alpha(0)$} 

Combining Eqs.~(\ref{3.7}), (\ref{3.9}), (\ref{3.10}), and
(\ref{3.12}) gives 
\begin{equation} 
w^\alpha = \dot{p}^\alpha(0) 
- \frac{1}{2} \Gamma^\alpha_{\ \beta\gamma}\, 
  \dot{p}^\beta(0) \dot{p}^\gamma(0)  
- \frac{1}{6} \Gamma^\alpha_{\ \beta\gamma\delta}\, 
  \dot{p}^\beta(0) \dot{p}^\gamma(0) \dot{p}^\delta(0)  
- \frac{1}{24} \Gamma^\alpha_{\ \beta\gamma\delta\epsilon}\, 
  \dot{p}^\beta(0) \dot{p}^\gamma(0) \dot{p}^\delta(0)
  \dot{p}^\epsilon(0) + \cdots. 
\label{3.14}
\end{equation} 
This is an expansion of $w^\alpha$ in powers of $\dot{p}^\alpha(0)$. 
The inverted series will take the form of 
\begin{equation} 
\dot{p}^\alpha(0) = w^\alpha 
+ A^\alpha_{\ \beta\gamma} w^\beta w^\gamma   
+ A^\alpha_{\ \beta\gamma\delta} w^\beta w^\gamma w^\delta 
+ A^\alpha_{\ \beta\gamma\delta\epsilon} w^\beta w^\gamma w^\delta
  w^\epsilon + \cdots, 
\label{3.15}
\end{equation} 
and the coefficients $A^\alpha_{\ \beta\gamma}$, 
$A^\alpha_{\ \beta\gamma\delta}$, and 
$A^\alpha_{\ \beta\gamma\delta\epsilon}$ can be determined by
inserting Eq.~(\ref{3.15}) into Eq.~(\ref{3.14}) and demanding that 
the substitution returns the identity $\dot{p}^\alpha(0) =
\dot{p}^\alpha(0)$. 

Elimination of the quadratic terms gives rise to the condition  
\begin{equation} 
A^\alpha_{\ \beta\gamma} := \frac{1}{2} \Gamma^\alpha_{\ \beta\gamma}.  
\label{3.16}
\end{equation} 
Elimination of the cubic terms produces 
$A^\alpha_{\ \beta\gamma\delta} = \frac{1}{6} 
\Gamma^\alpha_{\ \beta\gamma\delta} 
+ A^\alpha_{\ \beta\mu} \Gamma^\mu_{\ \gamma\delta}$. This becomes 
\begin{equation} 
A^\alpha_{\ \beta\gamma\delta} := \frac{1}{6} \Bigl( 
\Gamma^\alpha_{\ \beta\gamma,\delta} 
+ \Gamma^\alpha_{\ \beta\mu} \Gamma^\mu_{\ \gamma\delta} \Bigr) 
\label{3.17}
\end{equation} 
after involving Eqs.~(\ref{3.11}) and (\ref{3.16}). Elimination of the
quartic terms produces 
\[
A^\alpha_{\ \beta\gamma\delta\epsilon} = 
\frac{1}{24} \Gamma^\alpha_{\ \beta\gamma\delta\epsilon}
+ \frac{1}{3} A^\alpha_{\ \beta\mu} 
  \Gamma^\mu_{\ \gamma\delta\epsilon} 
- \frac{1}{4} A^\alpha_{\ \mu\nu} \Gamma^\mu_{\ \beta\gamma}   
  \Gamma^\nu_{\ \delta\epsilon} 
+ \frac{1}{2} A^\alpha_{\ \beta\gamma\mu} 
  \Gamma^\mu_{\ \delta\epsilon}  
+ \frac{1}{2} A^\alpha_{\ \beta\mu\epsilon} 
  \Gamma^\mu_{\ \gamma\delta}  
+ \frac{1}{2} A^\alpha_{\ \mu\beta\gamma} 
  \Gamma^\mu_{\ \delta\epsilon}.
\]
This becomes 
\begin{equation} 
A^\alpha_{\ \beta\gamma\delta\epsilon} := \frac{1}{24} \Bigl( 
\Gamma^\alpha_{\ \beta\gamma,\delta\epsilon} 
+ \Gamma^\alpha_{\ \beta\gamma,\mu} \Gamma^\mu_{\ \delta\epsilon} 
+ 2 \Gamma^\alpha_{\ \beta\mu} \Gamma^\mu_{\ \gamma\delta,\epsilon} 
+ \Gamma^\alpha_{\ \mu\nu} \Gamma^\mu_{\ \beta\gamma} 
  \Gamma^\nu_{\ \delta\epsilon} \Bigr) 
\label{3.18}
\end{equation}
after involving Eqs.~(\ref{3.11}), (\ref{3.13}), (\ref{3.16}), and
(\ref{3.17}). 

Equation (\ref{3.15}), with the coefficients of
Eqs.~(\ref{3.16})--(\ref{3.18}), gives the expansion of
$\dot{p}^\alpha(0)$ in powers of $w^\alpha = x^\alpha -
\bar{x}^\alpha$. We recall that the coefficients involve the
Christoffel connection and its partial derivatives evaluated at the
reference point $\bar{x}$. We also remark that while 
$A^\alpha_{\ \beta\gamma\delta}$ and 
$A^\alpha_{\ \beta\gamma\delta\epsilon}$ have been defined in
Eq.~(\ref{3.15}) as being fully symmetric in their lower indices, the  
right-hand sides of Eqs.~(\ref{3.17}) and (\ref{3.18}) have been left
in a non-symmetric form.    

\subsection{Final result: $\sigma_{\bar{\alpha}}$ expanded in
  powers of $w^\alpha$}

The gradient of Synge's function is obtained by substituting
Eq.~(\ref{3.15}) into Eq.~(\ref{3.6}). The result is 
\begin{equation} 
-\sigma_{\bar{\alpha}}(x,\bar{x}) = g_{\alpha\beta} w^\beta  
+ A_{\alpha\beta\gamma} w^\beta w^\gamma   
+ A_{\alpha\beta\gamma\delta} w^\beta w^\gamma w^\delta 
+ A_{\alpha\beta\gamma\delta\epsilon} w^\beta w^\gamma w^\delta
  w^\epsilon + \cdots, 
\label{3.19}
\end{equation} 
where the coefficients $A_{\alpha\beta\gamma}$,
$A_{\alpha\beta\gamma\delta}$, and
$A_{\alpha\beta\gamma\delta\epsilon}$ are obtained from
Eqs.~(\ref{3.16})--(\ref{3.18}) by lowering the first index with the
spacetime metric (as if these quantities were tensors). We recall that
the metric $g_{\alpha\beta}$, as well as the connection 
$\Gamma^\alpha_{\ \beta\gamma}$ and its partial derivatives, is
evaluated at the reference point $\bar{x}$. Equation (\ref{3.19}) is
the required coordinate expansion of $\sigma_{\bar{\alpha}}$ in powers
of $w^\alpha = x^\alpha - \bar{x}^\alpha$, the difference in the
coordinate positions of the points $x$ and $\bar{x}$. 

\subsection{Parallel transport on the spacelike geodesic} 

We next turn to our second task, the development of a coordinate
expansion for the parallel propagator 
$g^{\bar{\alpha}}_{\ \alpha}(x,\bar{x})$. We begin by introducing an
arbitrary dual vector $q_\alpha(\lambda)$ that we take to parallel
transported on $\beta$, the geodesic segment that links $x$ to
$\bar{x}$. The definition of the parallel propagator implies that
$q_\alpha(1) \equiv q_\alpha(x)$ and $q_\alpha(0) \equiv
q_{\bar{\alpha}}(\bar{x})$ are related by  
\begin{equation} 
q_\alpha(x) = g^{\bar{\alpha}}_{\ \alpha}(x,\bar{x})
q_{\bar{\alpha}}(\bar{x}). 
\label{3.20}
\end{equation} 
We shall calculate the parallel propagator by expanding
$q_\alpha(\lambda)$ in a Taylor series about $\lambda = 0$, evaluating  
this at $\lambda = 1$, and comparing the result with
Eq.~(\ref{3.20}). 

The Taylor expansion is 
\begin{equation} 
q_\alpha(\lambda) = q_\alpha(0) 
+\dot{q}_\alpha(0) \lambda  
+ \frac{1}{2} \ddot{q}^\alpha(0) \lambda^2  
+ \frac{1}{6} q_\alpha^{(3)}(0) \lambda^3 + \cdots  
\label{3.21}
\end{equation}
and it implies 
\begin{equation} 
\dot{q}_\alpha(\lambda) = \dot{q}_\alpha(0)   
+ \ddot{q}^\alpha(0) \lambda   
+ \frac{1}{2} q_\alpha^{(3)}(0) \lambda^2 + \cdots.   
\label{3.22}
\end{equation}
The dual vector is parallel transported on $\beta$ if 
\begin{equation} 
\dot{q}_\alpha(\lambda) - \Gamma^\mu_{\ \alpha\beta}(\lambda) 
q_\mu(\lambda) \dot{p}^\beta(\lambda) = 0. 
\label{3.23}
\end{equation} 
We shall work on this equation, using the expansions for
$\dot{p}^\beta(\lambda)$ and 
$\Gamma^\mu_{\ \alpha\beta}(\lambda)$ that are displayed in 
Eqs.~(\ref{3.3}) and (\ref{3.8}), respectively.  

We make the substitutions in Eq.~(\ref{3.23}) and collect terms that 
share the same power of $\lambda$. Setting the coefficient of the
$\lambda^0$ term to zero yields the condition 
\begin{equation} 
\dot{q}_\alpha(0) = \Gamma^\mu_{\ \alpha\beta}\, q_\mu(0)
\dot{p}^\beta(0). 
\label{3.24}
\end{equation} 
We recall that the Christoffel symbols are evaluated at the
reference point $\bar{x}$; the same remark applies to their
derivatives, which will appear in expressions below.    

Setting the coefficient of the $\lambda^1$ term to zero yields   
\[
\ddot{q}_\alpha(0) = \Gamma^\mu_{\ \alpha\beta} \bigl[ 
q_\mu(0) \ddot{p}^\beta(0) + \dot{q}_\mu(0) \dot{p}^\beta(0) \bigr]  
+ \Gamma^\mu_{\ \alpha\beta,\nu} \dot{p}^\nu(0) q_\mu(0)
\dot{p}^\beta(0). 
\]
This becomes 
\begin{equation} 
\ddot{q}_\alpha(0) = Q^\mu_{\ \alpha\beta\gamma}\,  
q_\mu(0) \dot{p}^\beta(0) \dot{p}^\gamma(0)
\label{3.25}
\end{equation} 
with 
\begin{equation} 
Q^\mu_{\ \alpha\beta\gamma} =  
\Gamma^\mu_{\ \alpha\beta,\gamma} 
- \Gamma^\mu_{\ \alpha\nu} \Gamma^\nu_{\ \beta\gamma} 
+ \Gamma^\mu_{\ \gamma\nu} \Gamma^\nu_{\ \alpha\beta}, 
 \label{3.26}
\end{equation} 
after involving Eqs.~(\ref{3.9}) and (\ref{3.24}). 

Setting the coefficient of the $\lambda^2$ term to zero yields 
\begin{eqnarray*} 
q^{(3)}_\alpha(0) &=& \Gamma^\mu_{\ \alpha\beta} \bigl[ 
q_\mu(0) p^{\beta(3)}(0) + 2\dot{q}_\mu(0) \ddot{p}^\beta(0) 
+ \ddot{q}_\mu(0) \dot{p}^\beta \bigr] 
+ 2\Gamma^\mu_{\ \alpha\beta,\nu} \dot{p}^\nu(0) \bigl[ 
q_\mu(0) \ddot{p}^\beta(0) + \dot{q}_\mu(0) \dot{p}^\beta(0) \bigr] 
\\ & & \mbox{} 
+ \bigl[ \Gamma^\mu_{\ \alpha\beta,\nu} \ddot{p}^\nu(0) 
+ \Gamma^\mu_{\ \alpha\beta,\nu\lambda} \dot{p}^\nu(0)
  \dot{p}^\lambda(0) \bigr] q_\mu(0) \dot{p}^\beta(0). 
\end{eqnarray*} 
This becomes 
\begin{equation} 
q^{(3)}_\alpha(0) = Q^\mu_{\ \alpha\beta\gamma\delta}\,
q_\mu(0) \dot{p}^\beta(0) \dot{p}^\gamma(0) \dot{p}^\delta(0) 
\label{3.27}
\end{equation} 
with 
\begin{eqnarray} 
Q^\mu_{\ \alpha\beta\gamma\delta} &=& 
\Gamma^\mu_{\ \alpha\beta,\gamma\delta} 
- \Gamma^\mu_{\ \alpha\nu} \Gamma^\nu_{\ \beta\gamma,\delta}
+ \Gamma^\nu_{\ \alpha\beta} \Gamma^\mu_{\ \nu\gamma,\delta}
- 2 \Gamma^\nu_{\ \beta\gamma} \Gamma^\mu_{\ \alpha\nu,\delta} 
+ 2 \Gamma^\mu_{\ \beta\nu} \Gamma^\nu_{\ \alpha\gamma,\delta}
- \Gamma^\nu_{\ \beta\gamma} \Gamma^\mu_{\ \alpha\delta,\nu} 
\nonumber \\ & & \mbox{} 
+ 2 \Gamma^\mu_{\ \alpha\nu} \Gamma^\nu_{\ \beta\lambda} 
  \Gamma^\lambda_{\ \gamma\delta} 
- 2 \Gamma^\mu_{\ \beta\nu} \Gamma^\nu_{\ \alpha\lambda} 
  \Gamma^\lambda_{\ \gamma\delta}
- \Gamma^\mu_{\ \nu\lambda} \Gamma^\nu_{\ \alpha\beta} 
  \Gamma^\lambda_{\ \gamma\delta} 
+ \Gamma^\mu_{\ \beta\nu} \Gamma^\nu_{\ \gamma\lambda} 
  \Gamma^\lambda_{\ \alpha\delta},  
\label{3.28}
\end{eqnarray} 
after involving Eqs.~(\ref{3.9})--(\ref{3.11}), as well as
Eqs.~(\ref{3.24})--(\ref{3.26}).  

Equations (\ref{3.21}), (\ref{3.24}), (\ref{3.25}), and (\ref{3.27})
combine to give  
\[
q_\alpha(1) = q_\alpha(0) 
+ \Gamma^\mu_{\ \alpha\beta}\, q_\mu(0) \dot{p}^\beta(0) 
+ \frac{1}{2} Q^\mu_{\ \alpha\beta\gamma}\, 
  q_\mu(0) \dot{p}^\beta(0) \dot{p}^\gamma(0)
+ \frac{1}{6} Q^\mu_{\ \alpha\beta\gamma\delta}\, q_\mu(0)
  \dot{p}^\beta(0) \dot{p}^\gamma(0) \dot{p}^\delta(0) + \cdots. 
\]
There is a common factor of $q_\mu(0)$, and this equation has the same
form as Eq.~(\ref{3.20}). The parallel propagator is therefore
identified as  
\begin{equation} 
g^{\bar{\mu}}_{\ \alpha}(x,\bar{x}) = \delta^\mu_{\ \alpha} 
+ \Gamma^\mu_{\ \alpha\beta}\, \dot{p}^\beta(0) 
+ \frac{1}{2} Q^\mu_{\ \alpha\beta\gamma}\, \dot{p}^\beta(0) 
  \dot{p}^\gamma(0)
+ \frac{1}{6} Q^\mu_{\ \alpha\beta\gamma\delta}\,  
  \dot{p}^\beta(0) \dot{p}^\gamma(0) \dot{p}^\delta(0) 
+ \cdots. 
\label{3.29}
\end{equation} 
The coefficients of this expansion in powers of $\dot{p}^\alpha(0)$
are given by Eqs.~(\ref{3.26}) and (\ref{3.28}). 

\subsection{Final result: $g^{\bar{\mu}}_{\ \alpha}$ expanded in
  powers of $w^\alpha$}

Our final expression for the parallel propagator is obtained by
substituting Eq.~(\ref{3.15}) into Eq.~(\ref{3.29}) and expanding the
result in powers of $w^\alpha$. After a lengthy calculation that
involves Eqs.~(\ref{3.16}), (\ref{3.17}), and (\ref{3.18}), we obtain  
\begin{equation} 
g^{\bar{\mu}}_{\ \alpha}(x,\bar{x}) = \delta^\mu_{\ \alpha} 
+ B^\mu_{\ \alpha\beta} w^\beta  
+ B^\mu_{\ \alpha\beta\gamma} w^\beta w^\gamma 
+ B^\mu_{\ \alpha\beta\gamma\delta} w^\beta w^\gamma w^\delta  
+ \cdots
\label{3.30}
\end{equation} 
with 
\begin{eqnarray} 
B^\mu_{\ \alpha\beta} &:=& \Gamma^\mu_{\ \alpha\beta}, 
\label{3.31} \\ 
B^\mu_{\ \alpha\beta\gamma} &:=& \frac{1}{2} \Bigl( 
\Gamma^\mu_{\ \alpha\beta,\gamma} 
+ \Gamma^\mu_{\ \beta\nu} \Gamma^\nu_{\ \alpha\gamma} \Bigr), 
\label{3.32} \\ 
B^\mu_{\ \alpha\beta\gamma\delta} &:=& \frac{1}{12} \Bigl( 
2\Gamma^\mu_{\ \alpha\beta,\gamma\delta} 
+ 2\Gamma^\nu_{\ \alpha\beta} \Gamma^\mu_{\ \nu\gamma,\delta}
- \Gamma^\nu_{\ \beta\gamma} \Gamma^\mu_{\ \alpha\nu,\delta} 
+ 4 \Gamma^\mu_{\ \beta\nu} \Gamma^\nu_{\ \alpha\gamma,\delta}
+ \Gamma^\nu_{\ \beta\gamma} \Gamma^\mu_{\ \alpha\delta,\nu} 
\nonumber \\ & & \mbox{} 
- \Gamma^\mu_{\ \beta\nu} \Gamma^\nu_{\ \alpha\lambda} 
  \Gamma^\lambda_{\ \gamma\delta}
+ \Gamma^\mu_{\ \nu\lambda} \Gamma^\nu_{\ \alpha\beta} 
  \Gamma^\lambda_{\ \gamma\delta} 
+ 2 \Gamma^\mu_{\ \beta\nu} \Gamma^\nu_{\ \gamma\lambda} 
  \Gamma^\lambda_{\ \alpha\delta} \Bigr). 
\label{3.33}
\end{eqnarray} 
The Christoffel connection and its partial derivatives are all 
evaluated at the reference point $\bar{x}$. 

\section{Spherical-harmonic decomposition of the scalar field} 

Our task in this section is to identify a useful way of relating a 
spherical-harmonic decomposition of the field $\Phi_\alpha :=
\nabla_\alpha \Phi$ to a spherical-harmonic decomposition of the
potential $\Phi$. As we shall see, the selected relation involves a 
decomposition of the vector $\Phi_\alpha$ in terms of a tetrad of
orthonormal vectors $e^\alpha_{(\mu)}$. So instead of decomposing
$\Phi_\alpha$ in a set of vectorial harmonics, we shall decompose each 
frame component $\Phi_{(\mu)} := \Phi_\alpha e^\alpha_{(\mu)}$ of the
vector field --- a scalar function of the spacetime coordinates --- in
scalar spherical harmonics. The decomposition of $\Phi_\alpha$ in
vectorial harmonics would make a viable alternative strategy, but one
which would prove less convenient for the purposes of calculating
regularization parameters --- see Sec.~V and the Appendix. Our scheme
leaves open the choice of tetrad, which is a priori arbitrary; our
particular choice is motivated by a desire to keep the
spherical-harmonic decompositions of $\Phi$ and $\Phi_{(\mu)}$ as
closely linked as possible.     

\subsection{Spherical-harmonic decompositions} 

Let $\Phi(x^a,\theta^A)$ be a scalar field on a
spherically-symmetric spacetime. The spacetime manifold has the
product structure ${\cal M}^2 \times S^2$, in which ${\cal M}^2$ is a 
two-dimensional submanifold that is orthogonal to the two-spheres
$S^2$. We let $x^a$ stand for any coordinate system that charts an
open domain of the submanifold ${\cal M}^2$; the lower-case Latin
index $a$ runs from 0 to 1. We let $\theta^A$ be angular coordinates
on the two-spheres; the upper-case Latin index $A$ runs from 2 to
3. For a Schwarzschild spacetime charted with the usual coordinates
$[t,r,\theta,\phi]$, we have $x^a = [t,r]$ and $\theta^A =
[\theta,\phi]$. We shall leave the coordinates $x^a$ arbitrary for the
time being, but we adopt the canonical angular coordinates $\theta^A =
[\theta,\phi]$.   

We suppose that the scalar field is expressed as a decomposition in 
spherical-harmonic functions $Y^{lm}$, so that 
\begin{equation} 
\Phi(x^a,\theta^A) = \sum_{lm} \Phi^{lm}(x^a) Y^{lm}(\theta^A). 
\label{4.1}
\end{equation}
The sum over the integer $l$ extends from $l = 0$ to $l = \infty$,
while the sum over the integer $m$ ranges from $m = -l$ to 
$m = l$. To keep $\Phi$ real the mode functions $\Phi^{lm}$ must
satisfy $\Phi^{l,-m} = (-1)^m \bar{\Phi}^{lm}$, in which an overbar 
indicates complex conjugation. The gradient $\nabla_\alpha \Phi$ of
the scalar field possesses the components 
\begin{equation} 
\partial_a \Phi = \sum_{lm} \partial_a \Phi^{lm} Y^{lm}  
\label{4.2}
\end{equation}
and 
\begin{equation} 
\partial_A \Phi = \sum_{lm} \Phi^{lm} \partial_A Y^{lm}.   
\label{4.3}
\end{equation}
These relations show that a natural basis of expansion for
$\nabla_\alpha \Phi$ would involve the scalar harmonics $Y^{lm}$
in the ${\cal M}^2$ sector, and the vectorial harmonics 
$\partial_A Y^{lm}$ in the $S^2$ sector. Following this route,
however, would introduce complications at a later stage (refer to the
last paragraph of Sec.~2 in the Appendix), and we shall adopt an
alternative strategy.  

We introduce a tetrad of orthonormal vectors $e^\alpha_{(\mu)}$ at
every point in the spherically-symmetric spacetime. The superscript
$\alpha$ is the usual vectorial index, and the subscript $(\mu) =
\{(0),(1),(2),(3)\}$ is a label that designates an individual member
of the tetrad. These vectors satisfy 
\begin{equation} 
g_{\alpha\beta} e^\alpha_{(\mu)} e^\beta_{(\nu)} = 
\eta_{(\mu)(\nu)}
\label{4.4}
\end{equation} 
with $\eta_{(\mu)(\nu)} = \mbox{diag}[-1,1,1,1]$. The four quantities 
\begin{equation} 
\Phi_{(\mu)} := e^\alpha_{(\mu)} \nabla_\alpha \Phi 
\label{4.5}
\end{equation}
are the {\it frame components} of the vector $\nabla_\alpha \Phi$;
these are {\it scalar functions} of the spacetime coordinates. The
vector field can be reconstructed from its frame components by
involving the dual tetrad $e^{(\mu)}_\alpha$, which is defined by  
\begin{equation} 
e^{(\mu)}_\alpha := \eta^{(\mu)(\nu)} g_{\alpha\beta} e^\beta_{(\nu)}, 
\label{4.6}
\end{equation} 
where $\eta^{(\mu)(\nu)} = \mbox{diag}[-1,1,1,1]$ is the matrix
inverse of $\eta_{(\mu)(\nu)}$. It is easy to show that  
\begin{equation} 
\nabla_\alpha \Phi = \Phi_{(\mu)} e^{(\mu)}_\alpha. 
\label{4.7}
\end{equation} 

Because each frame component $\Phi_{(\mu)}$ is a scalar quantity, it
is natural to decompose it in scalar harmonics. We therefore write 
\begin{equation} 
\Phi_{(\mu)}(x^a,\theta^A) = \sum_{lm} \Phi_{(\mu)}^{lm}(x^a)
Y^{lm}(\theta^A), 
\label{4.8}
\end{equation}
and seek to determine the relation between $\Phi_{(\mu)}^{lm}$,
the modes of the frame components, and $\Phi^{lm}$, the modes of the
original scalar field. 

The answer is provided by substituting Eq.~(\ref{4.5}) into the
equations $\Phi_{(\mu)}^{lm} = \int \Phi_{(\mu)} \bar{Y}^{lm}\,
d\Omega$, where $d\Omega = \sin\theta\, d\theta d\phi$ is an element
of solid angle. After involving Eqs.~(\ref{4.2}) and (\ref{4.3}), we
obtain  
\begin{equation} 
\Phi_{(\mu)}^{lm} = \sum_{l'm'} \Bigl[ C^a_{(\mu)}(l' m' | l m) 
\partial_a \Phi^{l'm'} + C_{(\mu)}(l' m' | l m) \Phi^{l'm'} \Bigr],  
\label{4.9}
\end{equation}
where the {\it coupling coefficients} are given by 
\begin{equation} 
C^a_{(\mu)}(l' m' | l m) := \int e^a_{(\mu)} Y^{l'm'} \bar{Y}^{lm}\,
d\Omega 
\label{4.10}
\end{equation} 
and 
\begin{equation} 
C_{(\mu)}(l' m' | l m) := \int e^A_{(\mu)} \partial_A Y^{l'm'}
\bar{Y}^{lm}\, d\Omega.  
\label{4.11}
\end{equation} 
Here, $e^a_{(\mu)}$ denotes the $(0,1)$ components of each basis
vector, while $e^A_{(\mu)}$ represents the angular components. The
computation of the coupling coefficients requires the specification of
the tetrad. 

In the rest of this section we will make a specific choice of tetrad
(Sec.~II B), compute the coupling coefficients for this tetrad 
(Secs.~II C and II D), and give an explicit form to
Eq.~(\ref{4.9}). The tetrad is displayed in
Eqs.~(\ref{4.21})--(\ref{4.25}) below, and the resulting relation
between $\Phi^{lm}_{(\mu)}$ and $\Phi^{lm}$ was already displayed in  
Eqs.~(\ref{1.23})--(\ref{1.26}).  

\subsection{Choice of tetrad} 

The choice of tetrad is in principle free, but we wish to find a
tetrad that leads to a simple structure for the coupling coefficients.  
Specializing to Schwarzschild spacetime and the usual coordinates
$[t,r,\theta,\phi]$, a possible choice of tetrad would be the usual
orthonormal frame 
\begin{eqnarray} 
e^\alpha_{(t)} &=& \bigl[f^{-1/2},0,0,0\bigr], 
\label{4.12} \\ 
e^\alpha_{(r)} &=& \bigl[0,f^{1/2},0,0\bigr], 
\label{4.13} \\ 
e^\alpha_{(\theta)} &=& \bigl[0,0,r^{-1},0\bigr], 
\label{4.14} \\ 
e^\alpha_{(\phi)} &=& \bigl[0,0,0,(r\sin\theta)^{-1}\bigr], 
\label{4.15}
\end{eqnarray}
where 
\begin{equation} 
f := 1 - \frac{2M}{r}.  
\label{4.16}
\end{equation} 
It is easy to show, however, that while this tetrad would lead to
simple (diagonal) coupling coefficients $C^a_{(\mu)}$, it would also
lead to coefficients $C_{(\mu)}$ that couple each $(lm)$ mode to 
{\it an infinite number} of $(l'm')$ modes. We shall not, therefore,
make this choice of tetrad.  

We shall instead introduce a ``Cartesian frame'' that is linked to
the ``spherical frame'' of Eqs.~(\ref{4.12})--(\ref{4.15}) by the same
relations that would hold in flat spacetime. Explicitly, our choice of 
tetrad is 
\begin{eqnarray} 
e^\alpha_{(0)} &:=& e^\alpha_{(t)}, 
\label{4.17} \\ 
e^\alpha_{(1)} &:=& \sin\theta\cos\phi\, e^\alpha_{(r)} 
+ \cos\theta\cos\phi\, e^\alpha_{(\theta)} 
- \sin\phi\, e^\alpha_{(\phi)}, 
\label{4.18} \\ 
e^\alpha_{(2)} &:=& \sin\theta\sin\phi\, e^\alpha_{(r)} 
+ \cos\theta\sin\phi\, e^\alpha_{(\theta)} 
+ \cos\phi\, e^\alpha_{(\phi)}, 
\label{4.19} \\ 
e^\alpha_{(3)} &:=& \cos\theta\, e^\alpha_{(r)} 
- \sin\theta\, e^\alpha_{(\theta)}. 
\label{4.20}
\end{eqnarray} 
We may loosely think of $e^\alpha_{(1)}$ as pointing in the ``$x$
direction,'' of $e^\alpha_{(2)}$ as pointing in the ``$y$ direction,'' 
and of $e^\alpha_{(3)}$ as pointing in the ``$z$ direction,'' with 
$[x,y,z]$ representing a quasi-Cartesian frame related in the usual way
to the quasi-spherical coordinates $[r,\theta,\phi]$. Independently of
this heuristic interpretation, we note that the transformation of 
Eqs.~(\ref{4.17})--(\ref{4.20}) defines a legitimate set of
orthonormal vectors. And as we shall see, this tetrad has the
desirable property of leading to a simple structure for the coupling
coefficients.  

Combining Eqs.~(\ref{4.17})--(\ref{4.20}) with
Eqs.~(\ref{4.12})--(\ref{4.15}) produces the following explicit
expressions for the basis vectors:  
\begin{eqnarray} 
e^\alpha_{(0)} &=& \biggl[ \frac{1}{\sqrt{f}}, 0, 0, 0 \biggr], 
\label{4.21} \\ 
e^\alpha_{(1)} &=& \biggl[ 0, \sqrt{f}\sin\theta\cos\phi, 
\frac{1}{r} \cos\theta\cos\phi, -\frac{\sin\phi}{r\sin\theta} \biggr], 
\label{4.22} \\ 
e^\alpha_{(2)} &=& \biggl[ 0, \sqrt{f}\sin\theta\sin\phi, 
\frac{1}{r} \cos\theta\sin\phi, \frac{\cos\phi}{r\sin\theta} \biggr], 
\label{4.23} \\ 
e^\alpha_{(3)} &=& \biggl[ 0, \sqrt{f}\cos\theta, 
-\frac{1}{r} \sin\theta, 0 \biggr]. 
\label{4.24}
\end{eqnarray} 
It is useful to introduce, as substitutes for $e^\alpha_{(1)}$ and 
$e^\alpha_{(2)}$, the complex combinations 
\begin{equation} 
e^\alpha_{(\pm)} := e^\alpha_{(1)} \pm \i e^\alpha_{(2)} 
= \biggl[ 0, \sqrt{f}\sin\theta \e^{\pm\i\phi},  
\frac{1}{r} \cos\theta \e^{\pm\i\phi}, 
\frac{\pm\i\e^{\pm\i\phi}}{r\sin\theta} \biggr].  
\label{4.25}
\end{equation}
In terms of the complex vectors we have $e^\alpha_{(1)} =
[e^\alpha_{(+)} + e^\alpha_{(-)}]/2 = \mbox{Re}[e^\alpha_{(+)}]$ and  
$e^\alpha_{(2)} = [e^\alpha_{(+)} - e^\alpha_{(-)}]/(2\i) = 
\mbox{Im}[e^\alpha_{(+)}]$. In the sequel we will work
primarily in terms of the complex tetrad $e^\alpha_{(0)}$, 
$e^\alpha_{(+)}$, $e^\alpha_{(-)}$, and $e^\alpha_{(3)}$.  

\subsection{Calculation of $C^a_{(\mu)}(l'm' | l m)$} 

We may now substitute the tetrad of Eqs.~(\ref{4.21})--(\ref{4.25})
into Eq.~(\ref{4.10}) and calculate $C^a_{(\mu)}(l'm'| lm)$, the first
set of coupling coefficients. Our computations will rely on
the standard identities (see, for example, Sec.~12.9 of
Ref.~\cite{arfken-weber:05})  
\begin{eqnarray} 
\cos\theta Y^{lm} &=& 
\sqrt{ \frac{(l-m+1)(l+m+1)}{(2l+1)(2l+3)} } Y^{l+1,m} 
+ \sqrt{ \frac{(l-m)(l+m)}{(2l-1)(2l+1)} } Y^{l-1,m}, 
\label{4.27} \\ 
\sin\theta \e^{\i\phi} Y^{lm} &=& 
-\sqrt{ \frac{(l+m+1)(l+m+2)}{(2l+1)(2l+3)} } Y^{l+1,m+1} 
+ \sqrt{ \frac{(l-m)(l-m-1)}{(2l-1)(2l+1)} } Y^{l-1,m+1}, 
\label{4.28} \\ 
\sin\theta \e^{-\i\phi} Y^{lm} &=& 
\sqrt{ \frac{(l-m+1)(l-m+2)}{(2l+1)(2l+3)} } Y^{l+1,m-1} 
- \sqrt{ \frac{(l+m)(l+m-1)}{(2l-1)(2l+1)} } Y^{l-1,m-1},   
\label{4.29}
\end{eqnarray} 
involving spherical-harmonic functions. 

Substituting Eq.~(\ref{4.21}) into Eq.~(\ref{4.10}) and invoking the 
orthonormality relations of the spherical harmonics reveals that the
only nonvanishing component of $C^a_{(0)}(l'm'| lm)$ is 
\begin{equation} 
C^t_{(0)} = \frac{1}{\sqrt{f}}\, \delta_{ll'} \delta_{mm'}. 
\label{4.30}
\end{equation} 
Substituting Eq.~(\ref{4.25}) into Eq.~(\ref{4.10}) and involving
Eq.~(\ref{4.28}) shows that 
\begin{equation} 
C^r_{(+)} = -\sqrt{ \frac{(l+m-1)(l+m)}{(2l-1)(2l+1)} } 
\sqrt{f}\, \delta_{l',l-1} \delta_{m',m-1} 
+ \sqrt{ \frac{(l-m+1)(l-m+2)}{(2l+1)(2l+3)} } 
\sqrt{f}\, \delta_{l',l+1} \delta_{m',m-1}
\label{4.31}
\end{equation} 
is the only nonvanishing component of 
$C^a_{(+)}(l'm'| lm)$. Similarly, 
\begin{equation} 
C^r_{(-)} = \sqrt{ \frac{(l-m-1)(l-m)}{(2l-1)(2l+1)} } 
\sqrt{f}\, \delta_{l',l-1} \delta_{m',m+1} 
- \sqrt{ \frac{(l+m+1)(l+m+2)}{(2l+1)(2l+3)} } 
\sqrt{f}\, \delta_{l',l+1} \delta_{m',m+1}
\label{4.32}
\end{equation} 
is the only nonvanishing component of 
$C^a_{(-)}(l'm'| lm)$. Finally, substituting Eq.~(\ref{4.24}) into
Eq.~(\ref{4.10}) and involving Eq.~(\ref{4.27}) reveals that 
\begin{equation} 
C^r_{(3)} = \sqrt{ \frac{(l-m)(l+m)}{(2l-1)(2l+1)} } 
\sqrt{f}\, \delta_{l',l-1} \delta_{m'm} 
+ \sqrt{ \frac{(l-m+1)(l+m+1)}{(2l+1)(2l+3)} } 
\sqrt{f}\, \delta_{l',l+1} \delta_{m'm}
\label{4.33}
\end{equation} 
is the only nonvanishing component of 
$C^a_{(3)}(l'm'| lm)$. 

\subsection{Calculation of $C_{(\mu)}(l'm' | l m)$} 

We now substitute the tetrad of Eqs.~(\ref{4.21})--(\ref{4.25})
into Eq.~(\ref{4.11}) and calculate $C_{(\mu)}(l'm'| lm)$, the second 
set of coupling coefficients. These computations also will rely on the 
standard identities listed in Eqs.~(\ref{4.27})--(\ref{4.29}), as well
as (see, for example, Sec.~12.7 of Ref.~\cite{arfken-weber:05}) 
\begin{equation} 
\e^{\i\phi} (\partial_\theta + \i\cot\theta \partial_\phi) Y^{lm} 
= \sqrt{(l-m)(l+m+1)} Y^{l,m+1} 
\label{4.34}
\end{equation} 
and 
\begin{equation} 
\e^{-\i\phi} (\partial_\theta - \i\cot\theta \partial_\phi) Y^{lm} 
= -\sqrt{(l+m)(l-m+1)} Y^{l,m-1}.  
\label{4.35}
\end{equation} 

Substituting Eq.~(\ref{4.21}) into Eq.~(\ref{4.11}) immediately gives 
\begin{equation} 
C_{(0)} = 0. 
\label{4.36}
\end{equation} 

Substituting Eq.~(\ref{4.25}) into Eq.~(\ref{4.11}) produces an
integral for $C_{(+)}$ that involves the factor 
\[
\cos\theta \e^{\i\phi} \partial_\theta Y^{l'm'} 
+ \frac{\i \e^{\i\phi}}{\sin\theta} \partial_\phi Y^{l'm'}. 
\]
The first term can be expressed as 
\[
\e^{\i\phi} \partial_\theta \bigl( \cos\theta Y^{l'm'} \bigr) 
+ \sin\theta \e^{\i\phi} Y^{l'm'}. 
\]
The second term, on the other hand, can be expressed as 
\[
\i \e^{\i\phi} \cot\theta \partial_\phi \bigl( \cos\theta Y^{l'm'} 
\bigr) + \i\sin\theta \e^{\i\phi} \partial_\phi Y^{l'm'}. 
\]
The sum becomes 
\[
\e^{\i\phi} (\partial_\theta + \i\cot\theta \partial_\phi) 
\bigl( \cos\theta Y^{l'm'} \bigr) 
- (m'-1) \sin\theta \e^{\i\phi} Y^{l'm'}, 
\]
and this is in such a form that Eqs.~(\ref{4.27}), (\ref{4.28}), and
(\ref{4.34}) can now be involved. Multiplying by $\bar{Y}^{lm}$ and
evaluating the integrals returns 
\begin{equation}
C_{(+)} = \sqrt{ \frac{(l+m-1)(l+m)}{(2l-1)(2l+1)} } 
\frac{l-1}{r}\, \delta_{l',l-1} \delta_{m',m-1} 
+ \sqrt{ \frac{(l-m+1)(l-m+2)}{(2l+1)(2l+3)} } 
\frac{l+2}{r}\,  \delta_{l',l+1} \delta_{m',m-1}. 
\label{4.37}
\end{equation} 
We similarly obtain 
\begin{equation}
C_{(-)} = -\sqrt{ \frac{(l-m-1)(l-m)}{(2l-1)(2l+1)} } 
\frac{l-1}{r}\, \delta_{l',l-1} \delta_{m',m+1} 
- \sqrt{ \frac{(l+m+1)(l+m+2)}{(2l+1)(2l+3)} } 
\frac{l+2}{r}\,  \delta_{l',l+1} \delta_{m',m+1}. 
\label{4.38}
\end{equation} 

Substituting Eq.~(\ref{4.24}) into Eq.~(\ref{4.11}) produces 
an integral that involves the factor 
\[
\sin\theta \partial_\theta Y^{l'm'} = 
\partial_\theta \bigl( \sin\theta Y^{l'm'} \bigr) 
- \cos\theta Y^{l'm'}.
\]
This can be expressed as 
\[
\e^{-\i\phi} (\partial_\theta - \i\cot\theta \partial_\phi) 
\bigl( \sin\theta \e^{\i\phi} Y^{l'm'} \bigr)
+ \i\e^{-\i\phi} \cot\theta \partial_\phi 
\bigl( \sin\theta \e^{\i\phi} Y^{l'm'} \bigr)  
- \cos\theta Y^{l'm'}, 
\]
or as 
\[
\e^{-\i\phi} (\partial_\theta - \i\cot\theta \partial_\phi) 
\bigl( \sin\theta \e^{\i\phi} Y^{l'm'} \bigr)
- (m'+2)\cos\theta Y^{l'm'},
\]
which is now in a useful form. After involving Eqs.~(\ref{4.27}),
(\ref{4.28}), and (\ref{4.35}), then multiplying by $\bar{Y}^{lm}$,
and finally evaluating the integrals, we arrive at  
\begin{equation}
C_{(3)} = -\sqrt{ \frac{(l-m)(l+m)}{(2l-1)(2l+1)} } 
\frac{l-1}{r}\, \delta_{l',l-1} \delta_{m'm} 
+ \sqrt{ \frac{(l-m+1)(l+m+1)}{(2l+1)(2l+3)} } 
\frac{l+2}{r}\, \delta_{l',l+1} \delta_{m'm}. 
\label{4.39}
\end{equation} 

\subsection{Final result: $\Phi^{lm}_{(\mu)}$ in terms of $\Phi^{lm}$} 

We substitute Eqs.~(\ref{4.30})--(\ref{4.33}), 
(\ref{4.36})--(\ref{4.39}) into Eq.~(\ref{4.9}), which gives 
$\Phi^{lm}_{(\mu)}$, the spherical-harmonic modes of the frame
components $\Phi_{(\mu)}$, in terms of $\Phi^{lm}$, the modes of the
original scalar field $\Phi$. After evaluating the sums over $l'$
and $m'$, we find that the relationship is given by
Eqs.~(\ref{1.23})--(\ref{1.26}), which are displayed back in Sec.~I E.  
Inspection of these equations reveals that the relationship is simple:
The structure of the coupling coefficients is such that
$\Phi^{lm}_{(\mu)}$ is linked to the neighboring modes 
$\Phi^{l\pm 1,m}$ and $\Phi^{l\pm 1, m\pm 1}$ only. This simplicity is
a benefit of the choice of tetrad made in Sec.~IV B; as was
discussed in that subsection, other choices would produce more   
complicated relationships.   

\section{Regularization parameters} 

The self-force acting on a scalar charge $q$ moving on a geodesic of
the Schwarzschild spacetime is proportional to $\Phi^{\rm R}_\alpha :=
\nabla_\alpha \Phi^{\rm R}$, where $\Phi^{\rm R}$ is the regular
potential that remains after the singular potential $\Phi^{\rm S}$ is
subtracted from the retarded potential $\Phi$. A local covariant
expansion for $\Phi^{\rm S}_\alpha := \nabla_\alpha \Phi^{\rm S}$ was 
worked out in Sec.~II, and with the help of the results presented in
Sec.~III, this can be turned into an explicit expansion in
Schwarzschild coordinates. In Sec.~IV we introduced a tetrad of
orthonormal vectors $e^\alpha_{(\mu)}$ to resolve the vector fields
$\Phi_\alpha$, $\Phi^{\rm S}_\alpha$, and $\Phi^{\rm R}_\alpha$ in
terms of their frame components $\Phi_{(\mu)} := \Phi_\alpha
e^\alpha_{(\mu)}$, $\Phi^{\rm R}_{(\mu)} := \Phi^{\rm R}_\alpha 
e^\alpha_{(\mu)}$, and $\Phi^{\rm S}_{(\mu)} := \Phi^{\rm S}_\alpha  
e^\alpha_{(\mu)}$, respectively. We have  
\begin{equation} 
\Phi^{\rm R}_{(\mu)} := \Phi_{(\mu)} - \Phi^{\rm S}_{(\mu)}. 
\label{5.1}
\end{equation} 
Also in Sec.~IV we showed how the spherical-harmonic modes of the
frame components $\Phi_{(\mu)}$ can be related to those of the scalar
potential $\Phi$. Our task in this section is to compute the
spherical-harmonic modes of the singular field 
$\Phi^{\rm S}_{(\mu)}$ and to extract from them the quantities known 
as {\it regularization parameters}. We will rely on the results
obtained in Sec.~II, III, and IV, as well as multipole-decomposition 
techniques reviewed in the Appendix. Most of the computations that are
described below were carried out with the symbolic manipulation
software {\sc Maple}, with the help of the tensor package 
{\sc GRTensorII} \cite{grtensor}. We will describe how these
calculations were performed, but space considerations compel us to
leave most details hidden.   

\subsection{Definition of the regularization parameters}

The scalar charge $q$ moves on an arbitrary geodesic of the
Schwarzschild spacetime. We place this geodesic in the equatorial
plane, and we assign to the particle the coordinates 
$t = {\sf t}(\tau)$, $r = \r(\tau)$, $\theta = \frac{\pi}{2}$, and
$\phi = \varphi(\tau)$, in which $\tau$ is proper time on the
geodesic. These functions are determined by integrating the geodesic
equations, which take the form   
$\dot{{\sf t}} = E/\f$, $\dot{\r}^2 = E^2 - \f (1 + L^2/\r^2)$, and  
$\dot{\varphi} = L/\r^2$, in which an overdot indicates
differentiation with respect to $\tau$. The constant $E$ is the
particle's conserved energy per unit rest-mass, and the constant $L$
is the conserved angular momentum per unit rest-mass. We also have
introduced the metric function $f := 1 - 2M/r$ and its value $\f$ at
$r = \r(\tau)$. At some instant $\tau = \tau_0$, the particle is found
at the point $\bar{x} = [t_0,r_0,\frac{\pi}{2},\phi_0]$ on its world
line. We let $f_0 := 1-2M/r_0$, and we wish to evaluate the scalar
self-force at that instant.    

We are interested in the value of $\Phi^{\rm R}_{(\mu)}$ at the point
$\bar{x}$. We calculate this by first decomposing the retarded field
$\Phi_{(\mu)}$ into spherical-harmonic modes, then subtracting the
modes associated with the singular field $\Phi^{\rm S}_{(\mu)}$, and
finally summing over all modes. This mode-sum converges because the
regular field is smooth on the world line; the mode-sums associated
with the retarded and singular fields do not converge on their own,
because both fields diverge on the world line. In spite of this
singular behavior, each spherical-harmonic mode $\Phi_{(\mu)lm}$ and 
$\Phi^{\rm S}_{(\mu)lm}$ is bounded as $x \to \bar{x}$; the
singularity merely gives rise to a {\it jump discontinuity} of each
mode at $\bar{x}$. The value of each mode is therefore ambiguous at
$\bar{x}$, but this ambiguity is of no consequence because the
discontinuity disappears when the mode of the singular field is
subtracted from the mode of the retarded field. The modes of the
regular field are continuous (and differentiable) at $\bar{x}$ because
$\Phi^{\rm R}_{(\mu)}$ is smooth on the world line. In practice the
discontinuity can be handled by evaluating each mode of the retarded
and singular fields slightly away from $\bar{x}$, performing the
subtraction, and then taking the limit $x \to \bar{x}$.    

A practical implementation of this prescription is contained in  
\begin{equation} 
\Phi^{\rm R}_{(\mu)}(t_0,r_0,{\textstyle \frac{\pi}{2}},\phi_0) 
= \lim_{\Delta \to 0} \sum_l \Bigl[ 
\Phi_{(\mu)l} - \Phi^{\rm S}_{(\mu)l} \Bigr], 
\label{5.6}
\end{equation} 
where 
\begin{equation} 
\Phi_{(\mu)l} := \sum_{m=-l}^l \Phi_{(\mu)lm}(t_0,r_0+\Delta)
Y^{lm}({\textstyle \frac{\pi}{2}},\phi_0') 
\label{5.7}
\end{equation} 
are the {\it multipole coefficients} of the retarded field, while 
\begin{equation} 
\Phi^{\rm S}_{(\mu)l} := \sum_{m=-l}^l 
\Phi^{\rm S}_{(\mu)lm}(t_0,r_0+\Delta)
Y^{lm}({\textstyle \frac{\pi}{2}},\phi_0') 
\label{5.8}
\end{equation} 
are the multipole coefficients of the singular field. The quantities
$\Phi_{(\mu)lm}$ and $\Phi^{\rm S}_{(\mu)lm}$ are the
spherical-harmonic modes of the retarded and singular fields,
respectively. (Details regarding the spherical-harmonic decomposition 
of a scalar function on $S^2$ are provided in the Appendix.) In
Eqs.~(\ref{5.6})--(\ref{5.8}) we choose the displaced point
to be at the same time coordinate $t_0$ as $\bar{x}$, but at a
displaced radius $r_0 + \Delta$ and a displaced azimuthal angle    
\begin{equation} 
\phi_0' := \phi_0 - c \Delta; 
\label{5.9}
\end{equation} 
the constant $c$ will be selected for convenience below, in Sec.~V D
--- see Eq.~(\ref{5.24}). (The idea of introducing a displacement in
the $\phi$ direction goes back to Mino, Nakano, and Sasaki
\cite{mino-etal:02}.)   

Equations (\ref{5.6})--(\ref{5.8}) are at the core of the procedure to  
calculate the self-force. We imagine that the modes $\Phi_{(\mu)lm}$
of the retarded field can be computed with a convenient numerical
method. From these we obtain the multipole coefficients
$\Phi_{(\mu)l}$, from which we subtract $\Phi^{\rm S}_{(\mu)l}$, the
multipole coefficients of the singular field. These can be computed
analytically, which is our task in this section. As we shall see, they
have the form  
\begin{equation} 
\Phi^{\rm S}_{(\mu)l} = q \biggl[ 
(l + {\textstyle \frac{1}{2}}) A_{(\mu)} 
+ B_{(\mu)} + \frac{C_{(\mu)}}{(l + {\textstyle \frac{1}{2}})} 
+ \frac{D_{(\mu)}}{(l - {\textstyle \frac{1}{2}}) 
(l + {\textstyle \frac{3}{2}})} + \cdots \biggr], 
\label{5.10}
\end{equation}
in which the coefficients $A_{(\mu)}$, $B_{(\mu)}$, $C_{(\mu)}$, and  
$D_{(\mu)}$, known as {\it regularization parameters}, are independent
of $l$ but depend on the state of motion of the particle at
$\bar{x}$. The subtraction produces $\Phi^{\rm R}_{(\mu)l}$, which we
sum over all values of $l$ to get $\Phi^{\rm R}_{(\mu)}$ and
eventually the self-force.  

\subsection{Rotation of the angular coordinates} 

Efficient techniques to compute multipole coefficients were devised 
by Detweiler, Messaritaki, and Whiting \cite{detweiler-etal:03}; these
are reviewed in the Appendix. They rely on a rotation of the angular
coordinates that maps the special point $(\theta=\frac{\pi}{2},\phi 
= \phi_0')$ to the North pole of the new angular coordinates. (This
idea goes back to Barack and Ori \cite{barack-ori:02,
barack-ori:03a}.) This rotation is described by  
\begin{eqnarray} 
\sin\theta\cos(\phi-\phi_0') &=& \cos\alpha, 
\label{5.11} \\ 
\sin\theta\sin(\phi-\phi_0') &=& \sin\alpha \cos\beta,  
\label{5.12} \\ 
\cos\theta &=& \sin\alpha \sin\beta, 
\label{5.13}
\end{eqnarray} 
which is a slightly modified version of the rotation described by
Eqs.~(\ref{A.5})--(\ref{A.7}). The new angles are $\alpha$ and
$\beta$, and it is easy to see that the rotation does indeed map the
point $(\theta=\frac{\pi}{2},\phi = \phi_0')$ to 
$(\alpha = 0, \beta = ?)$, with $\beta$ undetermined.  

As reviewed in the Appendix, each multipole coefficient 
$\Phi^{\rm S}_{(\mu)l}$ is computed by first expressing the singular 
field $\Phi^{\rm S}_{(\mu)}$ as a function of the angles $\alpha$ and
$\beta$, then extracting the Legendre projection 
\begin{equation} 
\Phi^{\rm S}_{(\mu)l}(\beta) := \frac{1}{2} (2l+1) \int_{-1}^1 
\Phi^{\rm S}_{(\mu)}(t_0,r_0+\Delta,\alpha,\beta) P_l(\cos\alpha)\, 
d\cos\alpha, 
\label{5.14}
\end{equation} 
and finally averaging over the angles $\beta$, 
\begin{equation} 
\Phi^{\rm S}_{(\mu)l} = \frac{1}{2\pi} \int_0^{2\pi} 
\Phi^{\rm S}_{(\mu)l}(\beta)\, d\beta. 
\label{5.15}
\end{equation} 
The decomposition of 
$\Phi^{\rm S}_{(\mu)}(t_0,r_0+\Delta,\alpha,\beta)$ in Legendre
polynomials relies on the techniques reviewed in Sec.~4 of the 
Appendix. The averaging over all angles $\beta$ relies on the
techniques reviewed in Sec.~5 of the Appendix. A summary of the
key results is provided in Sec.~6 of the Appendix.   

\subsection{Calculation of 
$\Phi^{\rm S}_{(\mu)}(t_0,r_0+\Delta,\alpha,\beta)$}   

The starting point of these computations is the calculation of the
singular field $\Phi^{\rm S}_{(\mu)}$ at a position $x$, to which we 
assign the (unrotated) coordinates $[t_0, r_0 + \Delta, \theta,
\phi]$. This point is slightly displaced from $\bar{x}$, to which we
have assigned the coordinates $[t_0, r_0, \frac{\pi}{2}, \phi_0]$. The
displacement vector is  
\begin{equation} 
w^\alpha := x^\alpha - \bar{x}^\alpha 
= [0,\Delta,\theta-{\textstyle \frac{\pi}{2}}, \phi - \phi_0], 
\label{5.16} 
\end{equation}
and an expression for $\Phi^{\rm S}_{(\mu)}(x)$ expanded in powers of    
$w^\alpha$ can be obtained by combining the results displayed in
Sec.~II J [Eq.~(\ref{2.62}) and following], Sec.~III E
[Eq.~(\ref{3.19})], Sec.~III G [Eq.~(\ref{3.30})], and
Sec.~IV B [Eqs.~(\ref{4.21})--(\ref{4.25})]. The computation of the
singular field is extremely tedious, and was handled by the symbolic 
manipulator {\sc GRTensorII} \cite{grtensor} operating under 
{\sc Maple}.     

The singular field is now expressed in terms of $w^\theta$ and
$w^\phi$ (in addition to $\Delta$), but these components of the
displacement vector are functions of $\alpha$ and $\beta$ that can be
determined from Eqs.~(\ref{5.11})--(\ref{5.13}). We have 
\begin{equation} 
w^\theta = -\mbox{arcsin}(\sin\alpha\sin\beta)
\label{5.17}
\end{equation}
and 
\begin{equation}
w^\phi = \mbox{arcsin}\biggl( \frac{\sin\alpha\cos\beta}{\sqrt{1 
- \sin^2\alpha\sin^2\beta}}\biggr) - c\Delta, 
\label{5.18}
\end{equation}
where the (as yet unassigned) constant $c$ was introduced back in 
Eq.~(\ref{5.9}). These quantities are small wherever $\alpha$ and
$\Delta$ are small, and the expansion of $\Phi^{\rm S}_{(\mu)}$ in
powers of $w^\alpha$ is valid in a small neighborhood of the North
pole $(\alpha = 0, \beta = ?)$. 

Defining  
\begin{equation} 
Q := \sqrt{1 - \cos\alpha}, 
\label{5.19}
\end{equation} 
we observe that $w^\theta$ and $w^\phi$ can each be expressed as an
expansion in powers of $Q$: 
\begin{eqnarray}
w^\theta &=& -\sqrt{2} Q \sin\beta - \frac{\sqrt{2}}{12} Q^3 
   (1 - 4\cos^2\beta) \sin\beta + O(Q^5), 
\label{5.20} \\ 
w^\phi &=& -c\Delta + \sqrt{2} Q \cos\beta + \frac{\sqrt{2}}{12} Q^3  
   (9 - 8\cos^2\beta) \cos\beta + O(Q^5). 
\label{5.21} 
\end{eqnarray} 
We note that $Q$ is formally of the same order of magnitude as
$\alpha$. Making these substitutions within the singular field returns
a double expansion in powers of $\Delta$ and $Q$, which are formally
considered to be of the same order of magnitude. This new
representation of $\Phi^{\rm S}_{(\mu)}$ involves the rotated angles
$\alpha$ and $\beta$, which is required for its substitution into
Eqs.~(\ref{5.14}) and (\ref{5.15}). Moreover, the singular 
field is expressed in terms of $\sin\beta$, $\cos\beta$, and $Q$,
{\it functions that are globally well-defined on the sphere}. This
property is critical for the successful decomposition of 
$\Phi^{\rm S}_{(\mu)}$ in Legendre polynomials, and the subsequent
averaging over $\beta$. We stole this powerful idea from Detweiler, 
Messaritaki, and Whiting \cite{detweiler-etal:03}. 

It should be acknowledged that the global extension of the singular 
field beyond the neighborhood of the North Pole is not unique. This
ambiguity, however, is of no consequence, because a different
extension $\Phi^{\prime {\rm S}}_{(\mu)}$ that continues to respect
the local expansion through order $\epsilon^1$ will be such that  
$\Phi^{\prime {\rm S}}_{(\mu)} - \Phi^{{\rm S}}_{(\mu)} 
= O(\epsilon^2)$. Because the difference vanishes at the position of
the particle, the value of the self-force, and the value of the four  
regularization parameters, will not be affected.      
  
\subsection{Squared-distance function} 

An important piece of $\Phi^{\rm S}_{(\mu)}$, as can be seen from
Eq.~(\ref{2.62}), is $s^2$, the squared distance between the points
$x$ and $\bar{x}$. This is defined by Eq.~(\ref{2.19}), and its
leading term in an expansion in powers of $w^\alpha$ is obtained by
substituting $\sigma_{\bar{\alpha}}(x,\bar{x}) = -g_{\alpha\beta}
w^\beta$, a truncated version of Eq.~(\ref{3.19}), into
Eq.~(\ref{2.19}). The result is 
\begin{equation} 
\tilde{\rho}^2 := (g_{\alpha\beta} + u_\alpha u_\beta) 
w^\alpha w^\beta, 
\label{5.22}
\end{equation} 
in which the metric and the velocity vector are evaluated at $\bar{x}
= [t_0,r_0,\frac{\pi}{2},\phi_0]$. With $u^\alpha :=
[E/f_0,\dot{r}_0,0,L/r_0^2]$ we obtain 
\begin{equation} 
\tilde{\rho}^2 = \frac{r_0(r_0-2M+r_0 \dot{r}^2_0)}{(r_0 - 2M)^2} 
\Delta^2 + r_0^2 (w^\theta)^2 + (r_0^2 + L^2) (w^\phi)^2 
+ \frac{2r_0 L \dot{r}_0}{r_0-2M} \Delta w^\phi. 
\label{5.23}
\end{equation} 
We re-express this result in terms of $w^{\prime \phi} := \phi -
\phi_0' = w^\phi + c\Delta$ and choose $c$ in order to eliminate the
term proportional to $\Delta w^\phi$. With \cite{mino-etal:02} 
\begin{equation} 
c := \frac{r_0 L \dot{r}_0}{(r_0-2M)(r_0^2+L^2)} 
\label{5.24}
\end{equation}
we find that Eq.~(\ref{5.23}) becomes 
\begin{equation} 
\tilde{\rho}^2 = \frac{r_0^4 E^2}{(r_0 - 2M)^2(r_0^2 + L^2)} 
\Delta^2 + r_0^2 (w^\theta)^2 + (r_0^2 + L^2) (w^{\prime\phi})^2, 
\label{5.25}
\end{equation} 
where we have also used the geodesic equation to eliminate
$\dot{r}^2_0$, the square of the radial velocity at $\bar{x}$, in
favor of $E^2$.   

We define a ``squared-distance function'' $\rho^2$ by making the
substitutions $w^\theta = -\sqrt{2} Q \sin\beta$ and 
$w^{\prime \phi} = \sqrt{2} Q \cos\beta$ into Eq.~(\ref{5.25}); these
are truncated versions of Eq.~(\ref{5.20}) and (\ref{5.21}),
respectively. We find that this is given by  
\begin{equation} 
\rho^2 := \frac{r_0^4 E^2}{(r_0 - 2M)^2(r_0^2 + L^2)} \Delta^2 
+ 2(r_0^2 + L^2) \chi Q^2, 
\label{5.26}
\end{equation} 
in which 
\begin{equation} 
\chi := 1 - k\sin^2\beta, \qquad 
k := \frac{L^2}{r_0^2 + L^2} 
\label{5.27}
\end{equation}
contains the dependence of $\rho^2$ on $\beta$. Recalling the
definition of $Q$ from Eq.~(\ref{5.19}), the squared-distance function 
can also be written as 
\begin{equation} 
\rho^2 = 2 (r_0^2 + L^2) \chi (\delta^2 + 1 - \cos\alpha), 
\label{5.28} 
\end{equation}
with 
\begin{equation}
\delta^2 := \frac{E^2 r_0^4}{2(r_0^2+L^2)^2 (r_0-2M)^2} 
\frac{\Delta^2}{\chi}.  
\label{5.29}
\end{equation} 

\subsection{Calculation of $\Phi^{\rm S}_{(\mu)}$ --- continued} 

At the end of Sec.~V C we had computed the singular field 
$\Phi^{\rm S}_{(\mu)}(t_0,r_0+\Delta,\alpha,\beta)$ and expressed it
in terms of $\sin\beta$, $\cos\beta$, $Q := \sqrt{1-\cos\alpha}$, and
$\Delta$; the expression also involves world-line quantities such as
$r_0$, $E$, $L$, and $\dot{r}_0$ which describe the state of motion of
the particle at $\bar{x}$. We interrupted this computation in Sec.~V D
to introduce the ``squared-distance function'' $\rho^2$, which is the
leading term in an expansion of $s^2 :=
(g^{\bar{\alpha}\bar{\beta}} + u^{\bar{\alpha}} u^{\bar{\beta}})
\sigma_{\bar{\alpha}} \sigma_{\bar{\beta}}$ in powers of $Q$ and
$\Delta$; this is itself a function of $\sin\beta$, $Q$ and $\Delta$,
as well as world-line quantities. 

The squared-distance function is introduced to eliminate the
dependence of $\Phi^{\rm S}_{(\mu)}$ on all even powers of $Q$, and to
reduce all odd powers of $Q$ to something linear in $Q$. The idea is
to solve Eq.~(\ref{5.26}) for $Q^2$ and to substitute the resulting
expression $Q^2(\rho^2,\Delta^2)$ into our current representation of
the singular field. We thus systematically replace each factor
$Q^{2n}$ in $\Phi^{\rm S}_{(\mu)}$ by $[Q^2(\rho^2,\Delta^2)]^n$, and
each factor $Q^{2n+1}$ by $[Q^2(\rho^2,\Delta^2)]^n Q$. This yields a
representation of the singular field which separates into a first set
of terms that is independent of $Q$, and a second set of terms that is
proportional to $Q$; each set contains a dependence on $\sin\beta$,
$\cos\beta$, $\rho$, and $\Delta$, as well as on the world-line
quantities.  

Our final expression for 
$\Phi^{\rm S}_{(\mu)}(t_0,r_0+\Delta,\alpha,\beta)$ is too long to be
displayed here. In fact, it is too long to be displayed anywhere, and 
we have taken measures to keep it hidden within the depths of our  
{\sc Maple} worksheets. Its basic structure is as follows. The
singular field admits an expansion in powers of $\epsilon$ of the form  
\begin{equation} 
\Phi^{\rm S}_{(\mu)} = \Phi^{\rm S}_{(\mu),-2} 
+ \Phi^{\rm S}_{(\mu),-1} + \Phi^{\rm S}_{(\mu),0} 
+ \Phi^{\rm S}_{(\mu),+1} + O(\epsilon^2), 
\label{5.30}
\end{equation} 
in which $\Phi^{\rm S}_{(\mu),-2}$ is of order $\epsilon^{-2}$,
$\Phi^{\rm S}_{(\mu),-1}$ of order $\epsilon^{-1}$, and so
on. (Recall from Sec.~II that $\epsilon$ loosely measures the
distance between $x$ and $\bar{x}$; we have that $\rho$ and $\Delta$
are both of order $\epsilon$.) The terms that appear on the right-hand
side of Eq.~(\ref{5.30}) possess the schematic form   
\begin{eqnarray}
\Phi^{\rm S}_{(\mu),-2} &=& M_{(\mu),-2} (\Delta/\rho^3) 
+ O(Q\cos\beta/\rho^3), 
\label{5.31} \\ 
\Phi^{\rm S}_{(\mu),-1} &=& M_{(\mu),-1} (1/\rho) 
+ O(Q\cos\beta \Delta/\rho^3) + O(\Delta^2/\rho^3) 
+ O(Q\cos\beta \Delta^3/\rho^5) + O(\Delta^4/\rho^5),  
\label{5.32} \\ 
\Phi^{\rm S}_{(\mu),0} &=& O(Q\cos\beta/\rho) + O(\Delta/\rho) 
+ O(Q\cos\beta \Delta^2/\rho^3) + O(\Delta^3/\rho^3) 
+ O(Q\cos\beta \Delta^4/\rho^5) 
\nonumber \\ & & 
+ O(\Delta^5/\rho^5)  
+ O(Q\cos\beta \Delta^6/\rho^7) + O(\Delta^7/\rho^7),   
\label{5.33} \\ 
\Phi^{\rm S}_{(\mu),+1} &=& M_{(\mu),+1} \rho 
+ O(Q\cos\beta \Delta/\rho) + O(\Delta^2/\rho) 
+ O(Q\cos\beta \Delta^3/\rho^3) + O(\Delta^4/\rho^3) 
\nonumber \\ & & 
+ O(Q\cos\beta \Delta^5/\rho^5) + O(\Delta^6/\rho^5)  
+ O(Q\cos\beta \Delta^7/\rho^7) + O(\Delta^8/\rho^7)   
\nonumber \\ & & 
+ O(Q\cos\beta \Delta^9/\rho^9) + O(\Delta^{10}/\rho^9).    
\label{5.34} 
\end{eqnarray} 
These equations display the dependence of each term on $Q$,
$\cos\beta$, $\Delta$, and $\rho$; the remaining dependence on $\beta$
is contained entirely in the function $\chi := 1 - k\sin^2\beta$
defined by Eq.~(\ref{5.27}). This dependence, as well as
the dependence on the world-line quantities, is left implicit in  
Eqs.~(\ref{5.31})--(\ref{5.34}). The terms in these equations that
involve the coefficients $M_{(\mu),-2}$, $M_{(\mu),-1}$, and 
$M_{(\mu),+1}$ are important: As we shall see, only these terms
actually contribute to the regularization parameters. The coefficients
depend on $\chi$ and the world-line quantities. The remaining terms in
Eqs.~(\ref{5.31})--(\ref{5.34}), those represented by the various
symbols $O(\ )$, are unimportant: They do not contribute to the
regularization parameters.  

\subsection{Final results: Regularization parameters} 

The singular field of Eqs.~(\ref{5.30})--(\ref{5.34}) can now be
substituted into Eqs.~(\ref{5.14}) and (\ref{5.15}) to compute the
multipole coefficients $\Phi^{\rm S}_{(\mu)l}$. The techniques
reviewed in the Appendix provide us with efficient calculational
rules. Equation (\ref{A.34}), for example, implies that all terms
involving $\cos\beta$ in Eqs.~(\ref{5.31})--(\ref{5.34})
average to zero and do not contribute to the multipole
coefficients. Equations (\ref{A.38})--(\ref{A.42}), on the other hand, 
imply that all remaining $O(\ )$ terms in
Eqs.~(\ref{5.31})--(\ref{5.34}) vanish in the limit $\Delta \to
0$; recall that this limiting procedure was introduced back in
Eq.~(\ref{5.6}). 

The only surviving contributions come from the terms involving the
coefficients $M_{(\mu),-2}$, $M_{(\mu),-1}$, and $M_{(\mu),+1}$. These
are handled with the help of Eqs.~(\ref{A.35})--(\ref{A.37}), and this
shows that $\Phi^{\rm S}_{(\mu)l}$ does indeed take the form of
Eq.~(\ref{5.10}). We remark that the neglected $O(\epsilon^2)$ terms
in Eq.~(\ref{5.30}) are those which would produce the neglected
$(\cdots)$ terms in Eq.~(\ref{5.10}); the $(\cdots)$ terms sum to
zero because the $O(\epsilon^2)$ terms in the singular field vanish at  
the position of the particle.      

The regularization parameters $A_{(\mu)}$ are produced by the term
involving the coefficient $M_{(\mu),-2}$ in Eq.~(\ref{5.31}). Our
results are listed back in Sec.~I E, in
Eqs.~(\ref{1.30})--(\ref{1.32}). Notice that in these equations, we
changed our notation from $r_0$ to $\r(t)$, and from $\phi_0$ to
$\varphi(t)$. The regularization parameters $B_{(\mu)}$ are produced
by the term involving the coefficient $M_{(\mu),-1}$ in
Eq.~(\ref{5.32}). Our results are listed back in Sec.~I E, in
Eqs.~(\ref{1.33})--(\ref{1.37}). The regularization parameters
$C_{(\mu)}$ would normally have originated from
Eq.~(\ref{5.33}). Because, however, there are no surviving
contributions from $\Phi^{\rm S}_{(\mu),0}$, we conclude that
$C_{(\mu)} = 0$. The regularization parameters $D_{(\mu)}$ are
produced by the term involving the coefficient $M_{(\mu),+1}$ in
Eq.~(\ref{5.34}). Our results are listed back in Sec.~I E, in 
Eqs.~(\ref{1.41})--(\ref{1.45}).  

The regularization parameters of Eqs.~(\ref{1.30})--(\ref{1.45}) are  
expressed in terms the (rescaled) elliptic functions  
${\scr E} := \frac{2}{\pi} \int_0^{\pi/2} (1-k\sin^2\psi)^{1/2}\, d\psi  
= F(-\frac{1}{2},\frac{1}{2}; 1; k)$ and ${\scr K} := \frac{2}{\pi}
\int_0^{\pi/2} (1-k\sin^2\psi)^{-1/2}\, d\psi 
= F(\frac{1}{2},\frac{1}{2}; 1; k)$, where $k := L^2/(r_0^2 + L^2)$
was introduced in Eq.~(\ref{5.27}). As indicated, the elliptic 
integrals can also be expressed in terms of hypergeometric
functions. These appear naturally in the course of averaging over the 
angles $\beta$ --- see Eq.~(\ref{A.28}). In fact, to obtain our final
expressions for the regularization parameters we have rationalized
their dependence on the hypergeometric functions by using the
recurrence relation of Eq.~(\ref{A.31}).  

As a final remark we note that the regularization parameters depend on 
$\Delta$ in two distinct ways. First, Eqs.~(\ref{1.30})--(\ref{1.32})  
show that the parameters $A_{(\mu)}$ are proportional to
$\mbox{sign}(\Delta)$ and therefore discontinuous across $\Delta 
= 0$; this behavior accounts for the discontinuity across $r=r_0$ of
each mode $\Phi_{(\mu)lm}$ of the retarded field. Second, the 
regularization parameters contain an implicit dependence on $\Delta$
that is not shown in Eqs.~(\ref{1.30})--(\ref{1.45}). Indeed, the
right-hand side of each equation should include terms that depend on
$\Delta$ but vanish in the limit $\Delta \to 0$; they are not
displayed precisely because they do not survive the limiting procedure 
described in Eq.~(\ref{5.6}).     
 
\begin{acknowledgments} 
We thank Leor Barack and Steve Detweiler for useful discussions that 
put us on the right track, and Bernard Whiting for his comments on an
earlier version of this manuscript. Part of this work was carried out
at the Aspen Center for Physics during the 2005 summer workshop LISA
Data: Analysis, Sources, and Science; we gratefully acknowledge the
Center and the workshop organizers for their support and kind
hospitality. This work was supported by the Natural Sciences and
Engineering Research Council of Canada.   
\end{acknowledgments} 

\appendix*
\section{Multipole coefficients} 

In this Appendix we collect results from the literature that are
required for the computation of the regularization parameters in
Sec.~V. Mostly we rely on the techniques developed in the paper by 
Detweiler, Messaritaki, and Whiting \cite{detweiler-etal:03}.  

\subsection{Decomposition of a scalar function in spherical harmonics} 

Let $f(\theta,\phi)$ be a scalar field on $S^2$. We decompose it in
spherical harmonics as 
\begin{equation}
f(\theta,\phi) = \sum_{lm} f_{lm} Y^{lm}(\theta,\phi).  
\label{A.1}
\end{equation}
The sum over the integer $l$ extends from $l = 0$ to $l = \infty$,
while the sum over the integer $m$ ranges from $m = -l$ to 
$m = l$. The field's spherical-harmonic modes are given by 
\begin{equation} 
f_{lm} = \int f(\theta,\phi) \bar{Y}^{lm}(\theta,\phi)\, d\Omega,  
\label{A.2}
\end{equation}
where an overbar indicates complex conjugation, and $d\Omega =
\sin\theta\, d\theta d\phi$ is an element of solid angle. 

We shall be interested in the value of $f$ at the special point
$(\theta = \frac{\pi}{2}, \phi = 0)$. This we express as 
\begin{equation} 
f({\textstyle \frac{\pi}{2}},0) = \sum_l f_l,  
\label{A.3}
\end{equation}
with 
\begin{equation} 
f_l = \sum_{m=-l}^l f_{lm} Y^{lm}({\textstyle \frac{\pi}{2}},0).  
\label{A.4}
\end{equation} 
The quantities $f_l$ associated with $f(\theta,\phi)$ play a
fundamental role below; we shall refer to them as the 
{\it multipole coefficients} of the function $f$.   

\subsection{Rotation of the angular coordinates} 

A convenient way to calculate the multipole coefficients $f_l$ is to
perform a rotation of the angular coordinates that maps the special
point $(\frac{\pi}{2},0)$ to the North pole of the new coordinate
system \cite{barack-ori:02, barack-ori:03a}. The rotation from the old
angles $(\theta,\phi)$ to the new angles $(\alpha,\beta)$ is defined
by the equations   
\begin{eqnarray} 
\sin\theta\cos\phi &=& \cos\alpha, 
\label{A.5} \\ 
\sin\theta\sin\phi &=& \sin\alpha \cos\beta,  
\label{A.6} \\ 
\cos\theta &=& \sin\alpha \sin\beta. 
\label{A.7}
\end{eqnarray} 
It is easy to see that this does indeed map the point $(\theta =
\frac{\pi}{2}, \phi = 0)$ to the point $(\alpha = 0, \beta = ?)$,
with $\beta$ undetermined. The special point is a singular point of
the new coordinate system, and as we shall see, this property is a
source of simplification in the computation of $f_l$. 

It is well-known that a rotation of the angular coordinates 
changes $Y_{lm}(\theta,\phi)$ into a mixture of functions 
$Y_{lm'}(\alpha,\beta)$ with $m'$ ranging from $-l$ to $l$; the
rotation mixes $m$ but leaves $l$ invariant. Because the quantity on
the left-hand-side of Eq.~(\ref{A.3}) is a scalar, whose value is
unchanged by the rotation, we may conclude that the rotation leaves
the multipole coefficients $f_l$ invariant. We shall make use of 
this simple fact to find an efficient way to compute these quantities.    

The function $f(\alpha,\beta)$ can be expanded in spherical harmonics
$Y^{lm}(\alpha,\beta)$ as in Eq.~(\ref{A.1}), and such a decomposition
implies that $f(0,?) = \sum_{lm} f_{lm} Y^{lm}(\alpha=0,\beta=?) 
= \sum_l \sqrt{(2l+1)/(4\pi)} f_{l0}$, after evaluating the spherical  
harmonics. Comparing this with Eq.~(\ref{A.3}) yields   
\begin{equation} 
f_l = \sqrt{\frac{2l+1}{4\pi}} 
f_{l0}[(\alpha,\beta)\ \mbox{decomposition}].  
\label{A.8}
\end{equation}
As indicated in Eq.~(\ref{A.8}), the multipole coefficient $f_l$ is 
proportional to the axisymmetric mode $f_{l0}$ of the function
$f(\alpha,\beta)$. This is given by 
\begin{equation} 
f_{l0} = \sqrt{\frac{2l+1}{4\pi}} \int f(\alpha,\beta)
P_l(\cos\alpha)\, d\cos\alpha\, d\beta, 
\label{A.9}
\end{equation}
where $P_l(\cos\alpha)$ is a Legendre polynomial. Equation (\ref{A.8})
therefore becomes 
\begin{equation} 
f_l = \frac{2l+1}{4\pi} \int f(\alpha,\beta) P_l(\cos\alpha)\, 
d\cos\alpha\, d\beta. 
\label{A.10}
\end{equation} 
If we let 
\begin{equation} 
f_l(\beta) := \frac{1}{2} (2l+1) \int_{-1}^{1} 
f(\alpha,\beta) P_l(\cos\alpha)\, d\cos\alpha, 
\label{A.11} 
\end{equation}
then Eq.~(\ref{A.10}) can be expressed as
\begin{equation} 
f_l = \bigl\langle f_l(\beta) \bigr\rangle, 
\label{A.12}
\end{equation} 
where 
\begin{equation} 
\bigl\langle f_l(\beta) \bigr\rangle = \frac{1}{2\pi} 
\int_0^{2\pi} f_l(\beta)\, d\beta 
\label{A.13}
\end{equation}
is the average of $f_l(\beta)$ over all angles $\beta$. Equations
(\ref{A.11}) and (\ref{A.12}) summarize the method by which the
multipole coefficients of the function $f(\theta,\phi)$ are computed.     

Equations (\ref{A.11})--(\ref{A.13}) can be given an alternative
interpretation that turns out to be useful for our purposes. Suppose
that we are presented with a scalar function $f(\alpha,\beta)$ and
that we wish to represent its dependence on $\alpha$ in terms of an
expansion in Legendre polynomials. We would write  
\begin{equation} 
f(\alpha,\beta) = \sum_l f_l(\beta) P_l(\cos\alpha), 
\label{A.14}
\end{equation}
and we would use the orthogonality properties of the Legendre
polynomials to express $f_l(\beta)$ as in Eq.~(\ref{A.11}). The
average of $f_l(\beta)$ over the angles $\beta$, as defined by
Eq.~(\ref{A.13}), would then give us the multipole coefficients
$f_l$. It is this interpretation that will be emphasized in the
rest of this Appendix.      

We pause here to remark that the calculational method described above  
to compute the multipole coefficients $f_l$ takes advantage of the
fact that the rotation of the coordinate system maps the special point 
$(\theta=\frac{\pi}{2},\phi=0)$ to the singular point
$(\alpha=0,\beta=?)$. It is the singular nature of the new coordinates
at the North pole which produces the equality of Eq.~(\ref{A.8}). This
method works well because $f(\theta,\phi)$ is a {\it scalar function}
of the angles: the rotation does not change the numerical value of the 
function. The method would not work as well for vectorial or tensorial
functions of the angles, because the map to a singular point would
alter the value of the functions by a singular factor. While this
difficulty can be averted \cite{barack-ori:02, barack-ori:03a}, this
introduces complications that can be avoided by choosing to deal with
scalar functions only. It is this observation that has motivated us to
work in terms of the scalars $\Phi_{(\mu)} 
:= \Phi_\alpha e^\alpha_{(\mu)}$ instead of the vector 
$\Phi_\alpha := \nabla_\alpha \Phi$.  

\subsection{Distance function $\rho(\alpha,\beta)$} 

The multipole coefficients that are required in Sec.~V are those
associated with negative and positive powers of the ``distance
function'' $\rho(\alpha,\beta)$, defined by Eq.~(\ref{5.28}), 
\begin{equation} 
\rho^2 := 2 (r_0^2 + L^2) \chi (\delta^2 + 1 - \cos\alpha). 
\label{A.15} 
\end{equation}
The symbols that appear in Eq.~(\ref{A.15}) are all introduced in 
Sec.~V D. We have $r_0$ denoting the current radial position of the
charged particle, and $L$ is the particle's (conserved) angular
momentum per unit mass. In addition,   
\begin{equation}
\delta^2 := \frac{E^2 r_0^4}{2(r_0^2+L^2)^2 (r_0-2M)^2} 
\frac{\Delta^2}{\chi}, 
\label{A.16}
\end{equation} 
in which $E$ denotes the particle's (conserved) energy per unit mass,
and   
\begin{equation} 
\Delta := r - r_0, 
\label{A.17}
\end{equation}
is the radial component of the displacement vector $w^\alpha =
x^\alpha - \bar{x}^\alpha$. And finally, 
\begin{equation} 
\chi := 1 - k\sin^2\beta, \qquad 
k := \frac{L^2}{r_0^2 + L^2} 
\label{A.18}
\end{equation}
contains the dependence of $\rho$ on the angle $\beta$. 

\subsection{Decomposition of $(\delta^2 + 1 - \cos\alpha)^{-n-1/2}$ in
  Legendre polynomials} 

The calculation of the regularization parameters described in Sec.~V
requires multipole coefficients for various powers of $\rho$, in a
context in which $\Delta$, and therefore $\delta$, is small. The
relevant powers of $\rho$ are $\rho^{-2n-1}$ with $n = -1, 0, 1, 2$,  
and so on. As was explained in the paragraph surrounding
Eq.~(\ref{A.14}), the starting point of a computation of multipole
coefficients is the decomposition of those selected powers of $\rho$
in Legendre polynomials. The most important piece of $\rho$ for these
decompositions is the factor $(\delta^2 + 1 - \cos\alpha)^{1/2}$, and
we therefore need expansions of the corresponding powers of this
quantity in Legendre polynomials.    

Quoting from Detweiler, Messaritaki, and Whiting
\cite{detweiler-etal:03}, we have     
\begin{equation} 
(\delta^2 + 1 - \cos\alpha)^{-n-1/2} = \sum_l A^n_l(\delta)
P_l(\cos\alpha), 
\label{A.19}
\end{equation} 
in terms of coefficients $A^n_l(\delta)$ that can be expanded in
powers of $\delta$. They obey the recurrence relation   
\begin{equation} 
A_l^{n+1} = -\frac{1}{(2n+1)\delta} \frac{d}{d\delta} 
A^n_l, 
\label{A.20}
\end{equation}
and special values are given by 
\begin{eqnarray}
A^0_l &=& \sqrt{2} - (2l+1)\delta + \frac{(2l+1)^2}{2\sqrt{2}}
\delta^2 - \frac{1}{3}l(l+1)(2l+1)\delta^3 + O(\delta^4), 
\label{A.21} \\ 
A^1_l &=& \frac{2l+1}{\delta} - \frac{(2l+1)^2}{\sqrt{2}} 
+ l(l+1)(2l+1) \delta + O(\delta^2), 
\label{A.22} \\ 
A^2_l &=& \frac{2l+1}{3\delta^3} - \frac{l(l+1)(2l+1)}{3\delta} 
+ O(\delta^0),  
\label{A.23} \\ 
A^3_l &=& \frac{2l+1}{5\delta^5} - \frac{l(l+1)(2l+1)}{15\delta^3} 
+ O(\delta^{-1}),   
\label{A.24} \\ 
A^4_l &=& \frac{2l+1}{7\delta^7} - \frac{l(l+1)(2l+1)}{35\delta^5} 
+ O(\delta^{-3}).    
\label{A.25}
\end{eqnarray} 
By induction from Eq.~(\ref{A.20}) we infer that 
\begin{equation} 
A^n_l = \frac{2l+1}{(2n-1)\delta^{2n-1}} 
\Bigl[ 1 + O(\delta^2) \Bigr], \qquad n \geq 2. 
\label{A.26}
\end{equation} 

We shall also need a decomposition for $(\delta^2 + 1 
- \cos\alpha)^{1/2}$. This is given by Eq.~(\ref{A.19}) with $n=-1$,  
and in this case the expansion coefficients are
\cite{detweiler-etal:03}  
\begin{equation} 
A^{-1}_l = -\frac{2\sqrt{2}}{(2l-1)(2l+3)} + O(\delta^2). 
\label{A.27} 
\end{equation} 

The results displayed in this subsection can be used in concert with 
Eq.~(\ref{A.15}) to calculate the Legendre coefficients
$[\rho^{-2n-1}]_l(\beta)$ associated with the functions 
$\rho^{-2n-1}(\alpha,\beta)$. The dependence on $\beta$ is contained   
in the factors of $\chi$ that are hidden in the definitions of
$\delta$ and $\rho$; refer back to Eqs.~(\ref{A.16}) and (\ref{A.18}).   

\subsection{Averaging over $\beta$} 

The next step in the calculation of the multipole coefficients 
$(\rho^{-2n-1})_l$ is to carry out the averaging over the angles 
$\beta$, as defined by Eq.~(\ref{A.13}). Because the dependence on 
$\beta$ is contained in $\chi$, what we need are expressions for
$\langle \chi^{-p} \rangle$, where $p$ is some (positive or negative)
number.    

Again quoting from Detweiler, Messaritaki, and Whiting
\cite{detweiler-etal:03}, we have  
\begin{equation} 
\langle \chi^{-p} \rangle = F_p 
:= F(p,{\textstyle \frac{1}{2}}; 1; k), 
\label{A.28}
\end{equation}
where $F(a,b;c;x)$ is the hypergeometric function, and where $k$ is
the constant defined by Eq.~(\ref{A.18}). Special cases of
Eq.~(\ref{A.28}) are   
\begin{equation} 
\langle \chi^{-1} \rangle = \frac{1}{\sqrt{1-k}} 
= \frac{\sqrt{r_0^2 + L^2}}{r_0} 
\label{A.29}
\end{equation}
and 
\begin{equation} 
\langle \chi \rangle = 1 - \frac{1}{2} k. 
\label{A.30}
\end{equation} 

The hypergeometric functions $F_p$ are linked by a recurrence relation   
displayed in Eq.~(15.2.10) of Ref.~\cite{abramowitz-stegun:72}. When 
specialized to our specific situation, this equation reads 
\begin{equation} 
F_{p+1} = \frac{p-1}{p(k-1)} F_{p-1} 
+ \frac{1 - 2p + (p - {\textstyle \frac{1}{2}})k}{p(k-1)} F_p. 
\label{A.31} 
\end{equation} 
Using this identity we find that the $F_p$'s that appear in our
expressions for $(\rho^{-2n-1})_l$ below --- see Eqs.~(\ref{A.36}) and 
(\ref{A.37}) --- can all be expressed in terms of $F_{1/2}$ and
$F_{-1/2}$. These, in turn, can be expressed in terms of complete
elliptic integrals: According to Eqs.~(17.3.1), (17.3.9), and
(17.3.10) of Ref.~\cite{abramowitz-stegun:72}, we have 
\begin{equation} 
F_{1/2} = {\scr K} := \frac{2}{\pi} K(k) := \frac{2}{\pi}
\int_0^{\pi/2} (1-k\sin^2\psi)^{-1/2}\, d\psi    
\label{A.32}
\end{equation}
and 
\begin{equation} 
F_{-1/2} = {\scr E} := \frac{2}{\pi} E(k) := \frac{2}{\pi}
\int_0^{\pi/2} (1-k\sin^2\psi)^{1/2}\, d\psi.     
\label{A.33}
\end{equation}
Our results in Sec.~V E are expressed in terms of the elliptic
integrals. 

We conclude this subsection with the remark that any function $g$ of
$\chi$ is necessarily periodic with period $\pi$ in the
interval $0 < \beta < 2\pi$; the function's behavior in the interval
$0 < \beta < \pi$ is replicated in the interval $\pi < \beta <
2\pi$. Furthermore, any such function is necessarily symmetric
(even) about $\beta = \pi/2$ in the first interval, and about $\beta = 
3\pi/2$ in the second interval. These properties are sufficient to
infer that  
\begin{equation} 
\langle g(\chi) \sin\beta \rangle = 
\langle g(\chi) \cos\beta \rangle = 
\langle g(\chi) \sin\beta \cos\beta \rangle = 0
\label{A.34}
\end{equation} 
for any function $g(\chi)$. 

\subsection{Final results: Multipole coefficients
  $(\rho^{-2n-1})_l$}   

The multipole coefficients of the functions
$\rho^{-2n-1}(\alpha,\beta)$ are denoted $(\rho^{-2n-1})_l$. They are
calculated by implementing the procedure described around
Eq.~(\ref{A.14}). Most of the required computations were already
performed in Secs.~4 and 5 of this Appendix; here we simply
collect the results and put it all together.     

By combining Eqs.~(\ref{A.15}), (\ref{A.16}), (\ref{A.19}), 
(\ref{A.21})--(\ref{A.27}), as well as Eqs.~(\ref{A.28}) and
(\ref{A.29}), we arrive at the following listing of multipole
coefficients:  
\begin{eqnarray}  
\Delta (\rho^{-3})_l &=& (l + {\textstyle \frac{1}{2}})  
\frac{r_0 - 2M}{E r_0^3} \mbox{sign}(\Delta) + O(\Delta), 
\label{A.35} \\ 
(\chi^{-p} \rho^{-1})_l &=& \frac{F_{p+1/2}}{\sqrt{r_0^2+L^2}} 
+ O(\Delta), 
\label{A.36} \\ 
(\chi^{-p} \rho)_l &=& -\frac{\sqrt{r_0^2+L^2} F_{p-1/2}}
{(l - {\textstyle \frac{1}{2}})(l + {\textstyle \frac{3}{2}})}  
+ O(\Delta^2).
\label{A.37}
\end{eqnarray} 
We also record the scaling relations
\begin{eqnarray}
\Delta (\chi^{-p} \rho^{-1})_l &=& O(\Delta), 
\label{A.38} \\ 
\Delta^2 (\chi^{-p} \rho^{-3})_l &=& O(\Delta), 
\label{A.39} \\ 
\Delta^4 (\chi^{-p} \rho^{-5})_l &=& O(\Delta), 
\label{A.40} \\ 
\Delta^7 (\chi^{-p} \rho^{-7})_l &=& O(\Delta^2), 
\label{A.41} \\ 
\Delta^{10} (\chi^{-p} \rho^{-9})_l &=& O(\Delta^3), 
\label{A.42}
\end{eqnarray}
which follow from the same set of equations. 

\bibliography{../bib/master} 
\end{document}